\documentclass[british,english]{paper}
\usepackage[T1]{fontenc}
\usepackage[latin9]{inputenc}
\usepackage{geometry}
\usepackage{color}
\definecolor{shadecolor}{rgb}{1, 0, 0}
\usepackage{babel}
\usepackage{refstyle}
\usepackage{framed}
\usepackage{mathrsfs}
\usepackage{amsthm}
\usepackage{amsmath}
\usepackage{amssymb}
\usepackage{esint}
\PassOptionsToPackage{normalem}{ulem}
\usepackage{ulem}
\usepackage{tocloft}
\usepackage[unicode=true,
 bookmarks=true,
 breaklinks=true,pdfborder={0 0 1},backref=false,colorlinks=true]
 {hyperref}
\definecolor{MyDarkBlue}{rgb}{0.2,0.2,0.6}
\definecolor{MyDarkRed}{rgb}{0.7,0.1,0.2}
\hypersetup{
 pdftitle={Lectures on Integrable Structures in Quantum Field Theory and Massive ODE/IM Correspondence},
 pdfauthor={Stefano Negro},
 citecolor=MyDarkRed,
 linkcolor=MyDarkBlue,
 urlcolor=MyDarkBlue
}
\usepackage[light]{iwona}
\makeatletter

\AtBeginDocument{\providecommand\subref[1]{\ref{sub:#1}}}
\providecommand{\tabularnewline}{\\}
\RS@ifundefined{subref}
  {\def\RSsubtxt{section~}\newref{sub}{name = \RSsubtxt}}
  {}
\RS@ifundefined{thmref}
  {\def\RSthmtxt{theorem~}\newref{thm}{name = \RSthmtxt}}
  {}
\RS@ifundefined{lemref}
  {\def\RSlemtxt{lemma~}\newref{lem}{name = \RSlemtxt}}
  {}

\numberwithin{equation}{section}
\numberwithin{figure}{section}

\usepackage{hyperref,amssymb}

\usepackage[font=small]{caption}
\usepackage{tikz}
\usetikzlibrary{plotmarks,calc,decorations, decorations.pathmorphing, patterns}
\usetikzlibrary{arrows}

\tikzstyle dynkin node=[very thick,shape=circle,draw,inner sep=0pt,minimum size=5mm]
\tikzstyle dynkin line=[very thick]
\tikzstyle inverse line=[gray,line width=1.46pt,line cap=round, dash pattern=on 0pt off 2\pgflinewidth]
\tikzstyle red phase=[red,decoration={snake,amplitude=0.1mm,segment length=1.6mm},decorate]
\tikzstyle blue phase=[blue,decoration={snake,amplitude=0.1mm,segment length=0.9mm},decorate]
\tikzstyle green phase=[green,decoration={snake,amplitude=0.1mm,segment length=0.9mm},decorate]
\tikzstyle brown phase=[brown,decoration={snake,amplitude=0.1mm,segment length=0.9mm},decorate]

\tikzstyle arrow=[thick,rounded corners=18pt,-latex]
\tikzstyle box=[draw,rounded corners,outer sep=4pt]
\tikzstyle B node=[outer sep=0pt]
\tikzstyle Q node=[inner sep=1pt,outer sep=0pt]

\makeatother

\begin{document}

\title{Lectures on Integrable Structures in Quantum Field Theory and Massive
ODE/IM Correspondence}

\date{06 July 2015}

\author{Stefano Negro}

\institution{{\footnotesize{}LPTENS, Ecole Normale Supérieure,}\\
{\footnotesize{}PSL Research University, Sorbonne Universités,}\\
{\footnotesize{}UPMC Univ. Paris 06, CNRS UMR 8549,}\\
{\footnotesize{}24 Rue Lhomond, 75005 Paris, France}\\[0.2cm]
\href{mailto:steff.negro@gmail.com}{\ttfamily steff.negro@gmail.com}\\
\href{mailto:negro@lpt.ens.fr}{\ttfamily negro@lpt.ens.fr}}
\maketitle
\vspace{2cm}
\begin{abstract}
This review was born as notes for a lecture given at the YRIS school
on integrability in Durham, in the summer of 2015. It deals with a
beautiful method, developed in the mid-nineties by V.V. Bazhanov,
S.L. Lukyanov and A.B. Zamolodchikov and, as such, called BLZ. This
method can be interpreted as a field theory version of the quantum
inverse scattering (QIS), also known as algebraic Bethe ansatz (ABA).
Starting with the case of conformal field theories (CFT) we show how
to build the field theory analogues of commuting transfer $T$ matrices
and Baxter $Q$-operators of integrable lattice models. These objects
contain the complete information of the integrable structure of the
theory, \textit{viz.} the integrals of motion, and can be used, as
we will show, to derive the thermodynamic Bethe ansatz (TBA) and non-linear
integral (NLIE) equations. This same method can be easily extended
to the description of integrable structures of certain particular
massive deformations of CFTs; these, in turn, can be described as
quantum group reductions of the quantum sine-Gordon model and it is
an easy step to include this last theory in the framework of BLZ approach.
Finally we show an interesting and surprising connection of the BLZ
structures with classical objects emerging from the study of classical
integrable models via the inverse scattering transform method. This
connection goes under the name of ODE/IM correspondence and we will
present it for the specific case of quantum sine-Gordon model only.\\

This review is part of a collection of papers \cite{Bomb_16,Levk_16,Loeb_16,Tong_16,Torr_16}
all of which were born out of lectures given at YRIS school in Durham.

\end{abstract}

\newpage{}
\tableofcontents{}

\section{Introduction}

The history of Integrable Systems is as old as that of Classical Mechanics
and the two were, for the largest part of 18th century, more or less
coinciding. Following the formulation of Isaac Newton\textquoteright s
laws of motion, for more than a century, eminent mathematicians and
physicists such as J. B. d\textquoteright Alembert, L. Euler, J. L.
Lagrange, C. G. J. Jacobi and sir W. R. Hamilton devoted many works
to the problem of finding exact solutions to Newton\textquoteright s
equations. These efforts brought about a striking amount of results,
which condensed in the theory of Lagrangian mechanics first and that
of Hamiltonian mechanics then, culminating in the first definition
of integrability, as given by J. E. É. Bour and J. Liouville. Although
the discovery of many new integrable systems quickly followed, by
the end of the 19th century a fundamental result of J. H. Poincaré
doused the excitement of the mathematical and physical community,
effectively deeming the integrable systems as exceptions amongst the
Hamiltonian ones. From that moment the theory of integrable system
laid more or less dormant until the second half of the '60, when the
idea of integrability resurfaced thanks to the efforts of C. S. Gardner,
J. M. Greene, M. D. Kruskal, R. Miura, P. D. Lax, L. D. Faddeev, V.
E. Zakharov and many other. From that pivotal half-decade, the theory
of integrable systems begun growing more and more, incorporating results
obtained in other branches of theoretical physics, first and foremost
Bethe's method for the study of quantum spin chains as well as Baxter's
approach to 2D statistical lattice models. The number of publications
devoted to the study of integrability grew steadily, especially after
the introduction of Conformal Field Theories by A. A. Belavin, A.
M. Polyakov and A. B. Zamolodchikov and the ``first superstring revolution''.
Finally a last breakthrough came just before the end of the century,
thanks to J. Maldacena: the ``AdS-CFT'' correspondence. This discovery
``opened the floodgates'' (to borrow the words of Polyakov) and
stimulated an impressive amount of work, especially in the field of
integrable models.

As the title explicitly reveals, this review deals with the analysis
of the integrable structures in field theories. What is meant with
this denomination are not simply the fundamental objects that are
seen appearing in all integrable theories: the integrals of motion
and their ``generating functions'', the $T$ and $Q$ operators.
The expression ``integrable structures'' encompasses the whole algebraic
skeleton which allows for the building of integrability to stand.
The study of integrable structures in field theory was first exhaustively
addressed to by V.V. Bazhanov, S.L. Lukyanov and A.B. Zamolodchikov
in a remarkable series of papers \cite{Bazh_Luky_AZam_96,Bazh_Luky_AZam_97-1,Bazh_Luky_AZam_99,Bazh_Luky_AZam_97-2}
and the goal of these 50 odd pages is essentially to go through their
work and present it in a uniform and coherent way, providing details
and insights in the definitions that we hope will help the reader
to understand this beautiful approach. It is, nonetheless, impossible
to include in a single review all the implications of the method introduced
by Bazhanov, Lukyanov and Zamolodchikov (hereafter addressed to as
the BLZ method), as his connections with the theory of quantum integrable
systems and CFTs are deep and widespread. For this reason we included
a list of references which will be addressed to in the text when a
certain topic will be simply cited.

Given the length and the complexity of the subject we decided to keep
the main body of the notes as clean as possible by separating heavier
calculations and in-depth analyses, not strictly essential to the
progression of the review, to boxed sections. These ``in-depth boxes''
are interleaved with the main body and the reader can, depending on
its necessities and the level of its knowledge, skip them without
missing anything fundamental about the method. We believe, however,
that they can be extremely useful in getting a deeper understanding
of the subject and of its many relations to other topics of integrability.

We would like to address a final word of caution to students and young
researchers first approaching this subject: do not feel discouraged
if you cannot grasp every aspect presented here. The BLZ method requires
the use of diverse advanced mathematical concepts and the computations
sketched here are often very technical and demanding; insisting on
understanding everything at a first reading would be foolish. Instead
we suggest multiple readings so that, at each step, it would be possible
to go through the concepts, references and calculations in more and
more detail. We especially suggest the readers willing to spend time
learning this method to go, at some point, through the computations
outlined here as this will often force them to explore the references
and think about the very meaning of the objects into play: the reward
will surely be a deeper and broader understanding of the concepts
exposed here.

This review is \foreignlanguage{british}{organised} as follows. In
the first section, after a brief review of Conformal Field Theories
(CFTs) and of classical KdV hierarchy, we will begin building the
integrability objects for the $c<1$ CFTs from scratch. We first introduce
the quantum transfer matrices $\mathbf{T}_{j}$ and show how they
can be interpreted as sort of generating functions for the quantum
integrals of motion. We will then broaden our view, \foreignlanguage{british}{generalising}
the algebraic setting and constructing the Baxter operators $\mathbf{Q}$.
While doing so we will also show how these objects can be used to
derive the useful TBA, Bethe Ansatz and NLIE equations, making a connection
to the other reviews in this volume. Following this will be a section
devoted to the extension of the previous results to the massive integrable
deformations of CFTs. This section contains a brief account on the
theory of integrable CFT deformations which can be skipped by the
readers already familiar with the concept. Finally, in the last section
we present a completely different yet, we believe, really interesting
approach to the construction of the integrable structures in the particular
case of sine-Gordon model. This method, bearing the name ODE/IM correspondence,
reveal an intimate and still poorly understood connection between
the theory of classical and quantum integrable models. Finally, given
the large number of parameters appearing in this review, we thought
it would be useful to collect the most relevant ones and the relationship
amongst them here, in Table \ref{tab:rel_par}.

\begin{table}

\begin{centering}
\begin{tabular}{|c|c|c|}
\hline 
Central charge & Conformal dimension & Spectral parameter\tabularnewline
\hline 
\hline 
$c$ & $h$ & $\lambda$\tabularnewline
\hline 
$\beta=\sqrt{\frac{1-c}{24}}-\sqrt{\frac{25-c}{24}}$ &  & $\theta=\left(1+\xi\right)\log\left(\lambda\right)$\tabularnewline
\hline 
$q=e^{\mathsf{i}\pi\beta^{2}}=-e^{-\mathsf{\frac{i\pi}{1+\xi}}}$ & $p\;:\; h=\left(\frac{p}{\beta}\right)^{2}+\frac{c-1}{24}$ & $y=\frac{\Gamma\left(1-\beta^{2}\right)}{\beta^{2}}\lambda$\tabularnewline
\hline 
$\xi=\frac{\beta^{2}}{1-\beta^{2}}$ &  & $\varkappa=-\mathsf{i}\frac{\pi}{\sin\left(\pi\beta^{2}\right)}\lambda$\tabularnewline
\hline 
\end{tabular}\protect\caption{Relationship amongst parameters\label{tab:rel_par}}

\par\end{centering}

\end{table}

\section{Integrable Structures of Conformal Field Theory\label{sec:Integrable-conformal}\index{Integrable Structures of Conformal Field Theory}}

The 2D CFTs are the perfect and probably best known example of exactly
solvable quantum field theories. From the year 1984, when the concept
of CFT was first introduced in an article of A.A. Belavin, A.M. Polyakov
and A.B. Zamolodchikov \cite{ABel_APol_AZam_84}, up to our days they
received a great deal of attention and most of their features are
now known, to the point of making them a self-contained theory which
is very often subject of advanced courses in theoretical physics.
The usual approach of these courses concerns what we might call a
representation-theoretical \foreignlanguage{british}{characterisation}
of CFTs, that is to say a description and classification of their
spectrum in terms of modules over the Virasoro algebra (or one of
its extensions). This point of view, which employs a wide array of
mathematical concepts, ranging from representation of infinite-dimensional
algebras to number theory, has the advantage of being extremely neat
and crystalline clear; however the concepts directly related to integrability
as we know them from classical and lattice models, such as integrals
of motion, Bethe ansatz equations and so on, do not seem to play a
primary role and remain hidden somewhere inside this elegant construction.
Such is its power that the legitimate question whether there is actually
any need to address to the integrable structures in CFTs arises. Anyhow,
this ``conventional'' point of view is limited to CFTs only and,
when dealing with, for example, their deformations one would like
methods closer to those employed in their lattice regularisation,
or in their classical limits, to be available. Looking in this direction,
there exists an alternative approach to CFTs, pioneered by Al.B.Zamolodchikov
\cite{AlZa_91}, where the spectrum is described in terms of scattering
states of a set of massless particles%
\footnote{Note that this characterisation is not unique! One can choose different
massless particle bases of the Hilbert space, with different particle
content and different scattering amplitudes. These different choices
reflect the possibility of reaching a certain CFT as massless limit
of different massive field theories.%
}. The fundamental object in this context is the factorisable \textit{massless
$S$-matrix}. This approach is closer to the integrable structures,
\textit{viz.} it is closer to a separated variables description. The
aim of this first part is to describe this integrable structure of
CFT by building the ``fundamental objects'': the operators $\mathbf{T}_{j}$
and $\mathbf{Q}_{\pm}$. These can be interpreted as the continuum
analogues of commuting transfer matrices and Baxter operators of integrable
lattice models and, in fact, it is wise to keep in mind this parallel
with lattice models, in order to understand the origin and meaning
of most formulae.

\subsection{Brief overview of CFT basic concepts\label{sub:Brief-overview-of}\index{Brief overview of CFT basic concepts}}

The goal of the following section is to recall the basics of 2D CFT
and to set up the notation, without any pretense of completeness.
It is intended for readers having already a good knowledge of the
subject; for those who are less familiar with it, there exists a plethora
of, often very good, reviews and books dealing with 2D CFT, offering
a wide range of different points of view. I suggest \cite{Gins_89}
for a straightforward introduction and the references therein for
a more in-depth study; the most daring might consider the ``Big Yellow
Book'' by P. Di Francesco, P. Mathieu and D. Sénéchal \cite{DiFr_Math_Sene}.

\paragraph{The Virasoro algebra}

A Conformal Field Theory (CFT) in $D$ Euclidean dimensions, that
is a local, isotropic field theory possessing no characteristic length
scale, is invariant under the global conformal group $SO(D+1,1)$,
a non-compact Lie group of dimension $\frac{1}{2}(D+1)(D+2)$. Although
bigger than the Galilei group $\mathbb{R}^{D}\rtimes SO(D)$, whose
dimension is $\frac{1}{2}D(D+1)$, it is still not sufficient to grant
integrability to the CFT: we need an infinite number of symmetries
to perform this task%
\footnote{The intuitive reason for that comes from Liouville's definition of
an integrable system. This states that a system is integrable iff
it possesses the same number of conserved quantities and degrees of
freedom. By Nöther theorem conserved charges correspond to symmetries
and so we need a set of symmetries having dimension equal to the degrees
of freedom of our system. But a field theory has ``infinite degrees
of freedom''! Hence we need an infinite number of symmetries to make
it integrable.%
} and that is exactly what we find if we look at the particular case
$D=2$. In fact when we consider the conformal transformations of
a plane, even if the special orthogonal group $SO(3,1)$ has dimension
$6$, we can find an infinite number of conformal coordinate transformations:
the holomorphic mappings from the complex plane (or part of it) onto
itself. This discrepancy between the finiteness of the conformal group
and the infinite amount of independent conformal coordinate transformations
is easily resolved by remarking that most of these last are not globally
well-defined. The set of infinitesimal conformal transformations form
an infinite dimensional local algebra, the renowned \textit{Virasoro
algebra}%
\footnote{Actually the algebra of infinitesimal conformal mappings is the Witt
algebra. However in the quantum theory this symmetry is anomalous
and the Witt algebra gets extended to $\textit{Vir}$ by the addition
of a \textit{central charge} $c$, also called \textit{conformal anomaly}.%
} $\textit{Vir}$, which exponentiate to the \emph{Virasoro group}
$\widehat{\textrm{diff}S^{1}}$ (the centrally extended group of diffeomorphisms
of the unit circle) \cite{Witt_88,Pres_Sega}. This last contains,
as a subgroup, the Möbius group $SL(2,\mathbb{C})\big/\mathbb{Z}_{2}\sim SO(3,1)$
of global conformal transformations. Since a local field theory should
be sensitive to local symmetries, even if the related transformations
are not globally defined, the behaviour of a $(1+1)$-dimensional
is indeed constrained by the full algebra $Vir$. It is precisely
the local conformal invariance that, being infinite-dimensional, allows
for exact solutions of 2D CFTs to exist. The Virasoro algebra $\textit{Vir}$
is generated by the operators $\left\{ L_{n}\right\} _{n\in\mathbb{Z}}$
, obeying the famous commutation relations
\[
[L_{n},\, L_{m}]=(n-m)\, L_{n+m}+\frac{c}{12}n\,(n^{2}-1)\,\delta_{n,-m}\;,
\]
where $c$ is a number called central charge or conformal anomaly.

Let us now consider a CFT on a flat Euclidean plane, having coordinates
$(x,y)$. The presence of scale invariance means that we are dealing
with massless theories. Were we considering a Minkowski geometry $(x^{0},x^{1})$,
it would be then natural to describe the system using light-cone coordinates
$(x^{+},x^{-})=(x^{1}+x^{0},x^{1}-x^{0})$, so that left/right-moving
massless fields depend uniquely on, respectively, $x^{+}$ or $x^{-}$.
Here, in the same spirit, it turns out to be extremely useful to introduce
the complex coordinates $(w,\overline{w})=(x+\mathsf{i}y,x-\mathsf{i}y)$
and the notion of left/right-moving fields turns into that of purely
holomorphic/antiholomorphic Euclidean fields. The algebra of symmetries
of a CFT will thus be the direct sum of two Virasoro algebras: $\textit{Vir}\oplus\overline{\textit{Vir}}$.
From now on we will consider the complex coordinates $w$ and $\bar{w}$
to be independent, so that $\textit{Vir}\oplus\overline{\textit{Vir}}$
naturally acts on $\mathbb{C}^{2}$ and we can treat each term in
the direct sum independently and effectively work only with holomorphic
fields. When the time comes to compute physical quantities we will
``remember'' to add the antiholomorphic contributions and impose
the ``real slice'' condition $\bar{w}=w^{\ast}$.

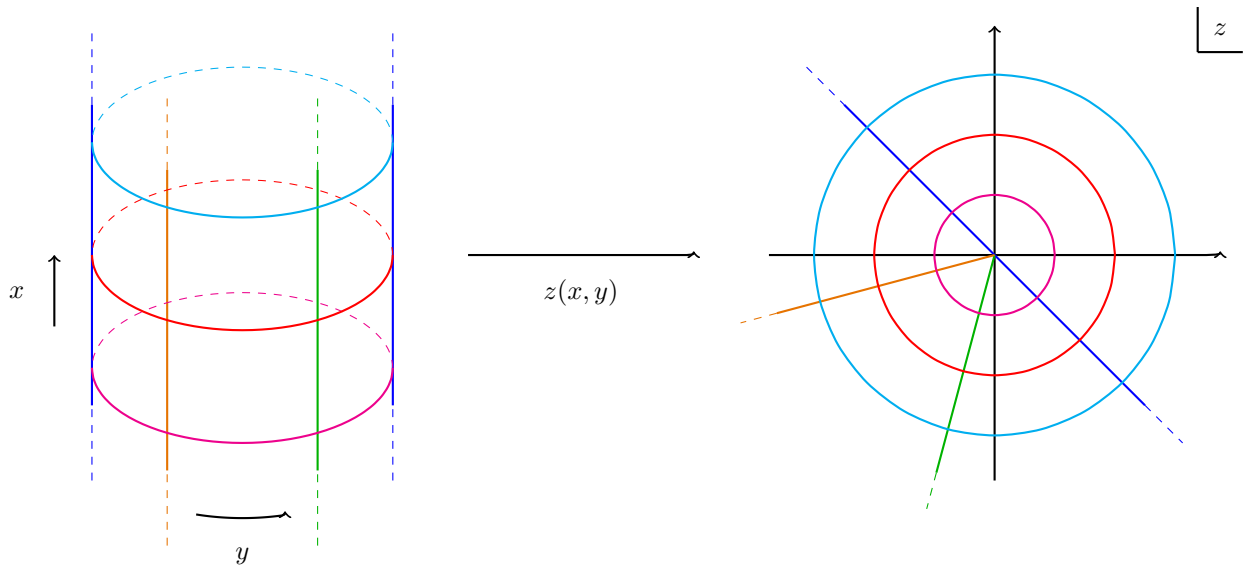
\begin{figure}[t]
\begin{centering}
\begin{tikzpicture}

\draw[scale=1,domain=0:4,smooth,variable=\x, color=blue,thick,-] plot ({-7},{\x});
\draw[scale=1,domain=0:4,smooth,variable=\x, color=blue,thick,-] plot ({-3},{\x});
\draw[scale=1,domain=4:5,smooth,variable=\x, color=blue,dashed,-] plot ({-7},{\x});
\draw[scale=1,domain=4:5,smooth,variable=\x, color=blue,dashed,-] plot ({-3},{\x});
\draw[scale=1,domain=-1:0,smooth,variable=\x, color=blue,dashed,-] plot ({-7},{\x});
\draw[scale=1,domain=-1:0,smooth,variable=\x, color=blue,dashed,-] plot ({-3},{\x});

\draw[scale=1,domain=0:4,smooth,variable=\x, color=orange!90!black,thick,-] plot ({-6},{\x-0.866});
\draw[scale=1,domain=0:4,smooth,variable=\x, color=green!70!black,thick,-] plot ({-4},{\x-0.866});
\draw[scale=1,domain=4:5,smooth,variable=\x, color=orange!90!black,dashed,-] plot ({-6},{\x-0.866});
\draw[scale=1,domain=4:5,smooth,variable=\x, color=green!70!black,dashed,-] plot ({-4},{\x-0.866});
\draw[scale=1,domain=-1:0,smooth,variable=\x, color=orange!90!black,dashed,-] plot ({-6},{\x-0.866});
\draw[scale=1,domain=-1:0,smooth,variable=\x, color=green!70!black,dashed,-] plot ({-4},{\x-0.866});

\draw[scale=1,domain=0:3.141,smooth,variable=\x, color=magenta,dashed] plot ({2*cos(\x r)-5},{sin(\x r)+0.5});
\draw[scale=1,domain=3.141:6.283,smooth,variable=\x, color=magenta,thick] plot ({2*cos(\x r)-5},{sin(\x r)+0.5});
\draw[scale=1,domain=0:3.141,smooth,variable=\x, color=red,dashed] plot ({2*cos(\x r)-5},{sin(\x r)+2});
\draw[scale=1,domain=3.141:6.283,smooth,variable=\x, color=red,thick] plot ({2*cos(\x r)-5},{sin(\x r)+2});
\draw[scale=1,domain=0:3.141,smooth,variable=\x, color=cyan,dashed] plot ({2*cos(\x r)-5},{sin(\x r)+3.5});
\draw[scale=1,domain=3.141:6.283,smooth,variable=\x, color=cyan,thick] plot ({2*cos(\x r)-5},{sin(\x r)+3.5});

\draw[scale=1,domain=1.05:1.95,smooth,variable=\x, color=black,thick,->] plot ({-7.5},{\x});
\draw[scale=1,domain=4.4:5,smooth,variable=\x, color=black,thick,->] plot ({2*cos(\x r)-5},{sin(\x r)-0.5});
\node[] at (-5,-2) {$y$}; \node[] at (-8,1.5) {$x$};

\draw[scale=1,domain=3:6,smooth,variable=\x, color=black,thick,->] plot ({\x-5},{2});
\node[] at (4.5-5,1.5) {$z(x,y)$};

\draw[scale=1,domain=2:8,smooth,variable=\x, color=black,thick,->] plot ({\x},{2});
\draw[scale=1,domain=-1:5,smooth,variable=\x, color=black,thick,->] plot ({5},{\x});
\draw[scale=1,domain=-2:2,smooth,variable=\x, color=blue,thick,-] plot ({\x+5},{-\x+2});
\draw[scale=1,domain=0:3,smooth,variable=\x, color=green!70!black,thick,-] plot ({-(\x)*(0.258)+5},{-(\x)*0.965+2});
\draw[scale=1,domain=0:3,smooth,variable=\x, color=orange!90!black,thick,-] plot ({-(\x)*0.965+5},{-(\x)*(0.258)+2});
\draw[scale=1,domain=-2.5:-2,smooth,variable=\x, color=blue,dashed,-] plot ({\x+5},{-\x+2});
\draw[scale=1,domain=2:2.5,smooth,variable=\x, color=blue,dashed,-] plot ({\x+5},{-\x+2});
\draw[scale=1,domain=3:3.5,smooth,variable=\x, color=green!70!black,dashed,-] plot ({-(\x)*(0.258)+5},{-(\x)*0.965+2});
\draw[scale=1,domain=3:3.5,smooth,variable=\x, color=orange!90!black,dashed,-] plot ({-(\x)*0.965+5},{-(\x)*(0.258)+2});
\draw[scale=1,domain=0:6.283,smooth,variable=\x, color=magenta,thick] plot ({0.8*cos(\x r)+5},{0.8*sin(\x r)+2});
\draw[scale=1,domain=0:6.283,smooth,variable=\x, color=red,thick] plot ({1.6*cos(\x r)+5},{1.6*sin(\x r)+2});
\draw[scale=1,domain=0:6.283,smooth,variable=\x, color=cyan,thick] plot ({2.4*cos(\x r)+5},{2.4*sin(\x r)+2});
\draw[scale=1,domain=-0.3:0.3,smooth,variable=\x, color=black,thick,-] plot ({7.7},{\x+5});
\draw[scale=1,domain=-0.3:0.3,smooth,variable=\x, color=black,thick,-] plot ({\x+8},{4.7});

\node[] at (8,5) {$z$};
\end{tikzpicture}\protect\caption{Map from the cylinder $(x,y)$ to the complex $z$-plane.\label{fig:Map_cyl_plane}}

\par\end{centering}

\end{figure}

Since we are dealing with massless fields, we must pay attention to
infrared divergencies. For this reason we invest one of the dimensions,
say $y$, with the rôle of spatial dimension and compactify it on
a unit circle: $y+2\pi\equiv y$. This procedure defines our theory
on a cylinder $\mathbb{R}\times S^{1}$. Next we can perform the following
conformal map
\[
w\longrightarrow z\doteq e^{w}\equiv e^{x+\mathsf{i}y}\;,
\]
which ``squashes'' the cylinder on the $z$-complex plane as shown
pictorially in Figure \ref{fig:Map_cyl_plane}. It is easy to see
that the ``time'' direction $x$ is mapped in the radial one $\rho=\sqrt{z\overline{z}}$,
while the ``space'' direction $y$ is sent in the angular one $\phi=\frac{1}{2\mathsf{i}}\log\left(\frac{z}{\overline{z}}\right)$.
The infinite past and infinite future $x=\pm\infty$ are sent to the
points $0$ and $\infty$, respectively, of the $z$-plane. The usual
procedure of quantisation in this setup is called \textit{radial quantisation}\textit{\emph{.}}

The subalgebra of $\textit{Vir}\times\overline{\textit{Vir}}$ generated
by $\left\{ L_{i},\overline{L}_{i}\right\} _{i=-1}^{1}$ is associated
with the global conformal group $SL(2,\mathbb{C})\big/\mathbb{Z}_{2}$
and is anomaly free. It is useful for characterising physical states.
In fact suppose that we are working, as we will, with eigenstates
of the operators $L_{0}$ and $\bar{L}_{0}$ and denote the eigenvalues,
called \emph{conformal weights, as} $h$ and $\bar{h}$. Consider
the following two particular operators of $SL(2,\mathbb{C})\big/\mathbb{Z}_{2}$:
\begin{itemize}
\item $L_{0}+\overline{L}_{0}$: on the cylinder it generates the translations
along the time direction and gets mapped, on the plane, to the generator
of dilatations $(z,\overline{z})\rightarrow\gamma(z,\overline{z})$.
In radial quantisation, it corresponds to the Hamiltonian of the system;
\item $\mathsf{i}(L_{0}-\overline{L}_{0})$: on the cylinder it generates
the translations along the space direction and gets mapped, on the
plane, to the generator of rotations $(z,\overline{z})\rightarrow(e^{\mathsf{i}\alpha}z,e^{-\mathsf{i}\alpha}\overline{z})$.
In radial quantisation, it corresponds to the momentum of the system.
\end{itemize}
The eigenvalues of these two operators are the \emph{scaling dimension}
$\Delta\doteq h+\bar{h}$ and the \emph{conformal spin} $s\doteq h-\bar{h}$.
In the context of radial quantisation they correspond to the energy
and the momentum of the state.

\paragraph{The energy-momentum tensor}

The energy-momentum tensor $T^{\mu\,\nu}$ is defined as the conserved
current associated to the invariance of the system with respect to
coordinate transformations $\epsilon_{\mu}(x)$:
\[
\delta_{\epsilon}\mathscr{A}=\frac{1}{2}\int d^{2}x\, T^{\mu\,\nu}(x)\,\left(\partial_{\mu}\epsilon_{\nu}(x)+\partial_{\nu}\epsilon_{\mu}(x)\right)\;,
\]
where $\mathscr{A}$ is the action of the system. The energy-momentum
tensor is symmetric in its indices $T^{\mu\,\nu}=T^{\nu\,\mu}$ and
for a CFT it is traceless $T_{\;\mu}^{\mu}=0$ (this is actually valid
in any dimension $D$). If $D=2$, then, we only have one component
for each chirality:
\[
T_{\textrm{plane}}(z)\doteq-2\pi T^{z\, z}\;,\qquad\overline{T}_{\textrm{plane}}(\overline{z})\doteq-2\pi T^{\overline{z}\,\overline{z}}\;.
\]
These two components are the generating functions of the Virasoro
generators
\[
T_{\textrm{plane}}(z)=\sum_{n\in\mathbb{Z}}z^{-n-2}L_{n}\;,\qquad\overline{T}_{\textrm{plane}}(\overline{z})=\sum_{n\in\mathbb{Z}}\overline{z}^{-n-2}\overline{L}_{n}\;,
\]
which, in turn, can be expressed in terms of $T$ by means of Cauchy
theorem:
\[
L_{n}=\frac{1}{2\pi\mathsf{i}}\ointop_{0}dz\, z^{n+1}T_{\textrm{plane}}(z)\;,\qquad\overline{L}_{n}=\frac{1}{2\pi\mathsf{i}}\ointop_{0}d\overline{z}\,\overline{z}^{n+1}\overline{T}_{\textrm{plane}}(\overline{z})\;.
\]
These formulae allow to build the energy-momentum tensor also for
those CFT who do not possess an action (or if that action is not known).
In the following we will be considering CFTs defined on the cylinder
and is useful to have an expression of the energy-momentum tensor
in this geometry: 
\[
T(w)=-\frac{c}{24}+\sum_{n\in\mathbb{Z}}e^{\mathsf i n w}L_{-n}\;,\qquad\overline{T}(\overline{w})=-\frac{c}{24}+\sum_{n\in\mathbb{Z}}e^{-\mathsf i n\overline{w}}L_{-n}\;.
\]
The corresponding expressions of the Virasoro generators take the
form
\[
L_{n}=\frac{1}{2\pi}\intop_{0}^{2\pi}dw\, e^{\mathsf i n w}T(w)\;,\qquad\overline{L}_{n}=\frac{1}{2\pi}\intop_{0}^{2\pi}d\overline{w}\, e^{-\mathsf i n \overline{w}}\overline{T}(\overline{w})\;.
\]
It is useful to remark here that on the cylinder $T(w+2\pi)=T(w)$
and $\overline{T}(\overline{w}+2\pi)=\overline{T}(\overline{w})$.

\paragraph{Primary fields}

Let us consider a transformation of coordinates $(z,\overline{z})\rightarrow(\omega(z),\overline{\omega}(\overline{z}))$;
any field in our CFT which transforms as follows
\[
\varphi_{h,\overline{h}}(z,\overline{z})\rightarrow\varphi_{h,\overline{h}}'\left(\omega(z),\overline{\omega}(\overline{z})\right)=\left(\frac{d\omega}{dz}\right)^{-h}\left(\frac{d\overline{\omega}}{d\overline{z}}\right)^{-\overline{h}}\varphi_{h,\overline{h}}(z,\overline{z})\;,
\]
is named \textit{primary} field, while the quantities $h$ and $\overline{h}$
are the holomorphic and anti-holomorphic conformal weights introduced
above. All the fields which are not primary will be called \textit{descendants}.
The energy-momentum tensor is an example of a particular descendant
field, called \textit{quasi-primary} as it transforms as a primary
only under global conformal transformations:
\[
T(z)\rightarrow T'(\omega(z))=\left(\frac{d\omega}{dz}\right)^{-h}T(z)+\frac{c}{12}\left\{ z;\omega(z)\right\} \;,
\]
where $\left\{ z;\omega(z)\right\} $ is the Schwartzian derivative
of $\omega(z)$, which vanishes iff $\omega\in SL(2,\mathbb{C})\big/\mathbb{Z}_{2}$.
A consequence of the form of primary fields and energy-momentum tensor
is the particular operator product expansion (OPE) that they satisfy:
\[
T(z)\varphi_{h,\overline{h}}(z',\overline{z}')\sim\frac{h}{(z-z')^{2}}\varphi_{h,\overline{h}}(z',\overline{z}')+\frac{1}{z-z'}\partial_{z'}\varphi_{h,\overline{h}}(z',\overline{z}')\;,
\]
\[
T(z)T(z')\sim\frac{c/2}{(z-z')^{4}}+\frac{2}{(z-z')^{2}}T(z')+\frac{1}{z-z'}\partial_{z'}T(z')\;.
\]

\paragraph{Hilbert space}

The Hilbert space $\mathscr{H}_{ph}$ of our CFT is built on some
vacuum state $\left|\,0\,\right\rangle $, which must be invariant
under $SL(2,\mathbb{C})\big/\mathbb{Z}_{2}$ and satisfy
\[
L_{n}\left|\,0\,\right\rangle =0\;,\qquad\overline{L}_{n}\left|\,0\,\right\rangle =0\;,\quad\forall n\geq-1\;,
\]
which is a consequence of the request that $T(z)\left|\,0\,\right\rangle $
and $\overline{T}(\overline{z})\left|\,0\,\right\rangle $ be well-defined
as $(z,\overline{z})\rightarrow(0,0)$%
\footnote{This request contain in itself that of invariance with respect to
$SL(2,\mathbb{C})\big/\mathbb{Z}_{2}$.%
}. The action of a primary field on this vacuum generates eigenstates
of the Hamiltonian:
\[
\left|\, h,\overline{h}\,\right\rangle \doteq\varphi_{h,\overline{h}}(0,0)\left|\,0\,\right\rangle \;.
\]
The fact that these are eigenstates is easily obtained from the OPE
properties of the primary fields:
\[
[L_{0},\varphi_{h,\overline{h}}(0,0)]=\frac{1}{2\pi\mathsf{i}}\ointop_{0}dz\, z\, T(z)\,\varphi_{h,\overline{h}}(0,0)=h\varphi_{h,\overline{h}}(0,0)\;.
\]
Similar properties are valid for other operators $L_{n}$ and one
finds
\[
\begin{array}{ccc}
L_{0}\left|\, h,\overline{h}\,\right\rangle =h\left|\, h,\overline{h}\,\right\rangle  &  & L_{n}\left|\, h,\overline{h}\,\right\rangle =0\\
 & \;;\qquad\\
\overline{L}_{0}\left|\, h,\overline{h}\,\right\rangle =\overline{h}\left|\, h,\overline{h}\,\right\rangle  &  & \overline{L}_{n}\left|\, h,\overline{h}\,\right\rangle =0
\end{array}\;,\quad\forall n>0\;.
\]
So these states, which we call primary like the fields generating
them, are highest-weight vectors for $\textit{Vir}\times\overline{\textit{Vir}}$.
We can thus generate a subset $\mathcal{V}_{h}\otimes\mathcal{V}_{\overline{h}}$
of the Hilbert space by the free action of $\left\{ L_{n},\overline{L}_{n}\right\} _{n=-\infty}^{-1}$
on the primary state $\left|\, h,\overline{h}\,\right\rangle $:
\[
\mathcal{V}_{a}\equiv\mathcal{V}_{h_{a}}\doteq\left\{ L_{-k_{1}}L_{-k_{2}}\ldots L_{-k_{n}}\left|\, h,\overline{h}\,\right\rangle \;\Big\backslash\;1\leq k_{1}\leq\cdots\leq k_{n}\;\Big\backslash\;\forall n\geq0\right\} \;.
\]
The vector field $\mathcal{V}_{a}$ is closed under the action of
$\textit{Vir}$ and it's called \textit{Verma module}.

The Hilbert space of a CFT is embedded in some suitable way (we will
not cover this topic here) into $\mathscr{H}_{ch}\otimes\overline{\mathscr{H}}_{ch}$,
where $\mathscr{H}_{ch}\doteq\bigoplus_{a}\mathcal{V}_{a}$ is the
space of ``right-chiral'' states. The index $a$ of the direct sum
runs on the admissible conformal dimension of our CFT; this number
depends on $c$ and is usually infinite. There are however some notable
exceptions where the structure of the Verma modules becomes degenerate
and the number of allowed highest weights becomes finite. Let us list
three important categories of CFTs:
\begin{itemize}
\item \textbf{Unitary non degenerate}: when $c\geq1$, the conformal dimension
may take a continuum of positive values; for any of these, $\mathcal{V}_{h}$
is unitary. An example in this class is the free boson, corresponding
to $c=1$;
\item \textbf{Minimal models}: when
\[
c=1-6\frac{(m-m')^{2}}{m\, m'}\;,\quad m\perp m'\;,\; m,m'\in\mathbb{N}\big\slash\left\{ 0\right\} \;,
\]
the allowed conformal dimensions are restricted to take a finite number
of discrete values, indexed by two integers $(r,s)$:
\[
h_{r,s}=\frac{(m\, r-m's)^{2}-(m-m')^{2}}{4\, m\, m'}\;,\quad\begin{array}{c}
1\leq r<m'\\
\\
1\leq s<m
\end{array}\;.
\]
These models are denoted as $\mathcal{M}_{m,m'}$. An example in this
class is the Yang-Lee model, corresponding to the choice $m=5$ and
$m'=2$, meaning $c=-\frac{22}{5}$;
\item \textbf{Unitary minimal models}: amongst the above models, the unitary
ones are those with $m'=m+1$ and they usually are denoted as $\mathcal{M}_{m}$.
An example in this class is the Ising model, corresponding to $m=3$
and $c=\frac{1}{2}$.
\end{itemize}

\paragraph{Integrals of motion}

The first step towards the unveiling of the integrable structures
of CFT was performed by R. Sasaki and I. Yamanaka \cite{Sasa_Yama_88}.
They considered a CFT on a cylinder and the algebra $\mathcal{U}(\textit{Vir})$
generated by the energy-momentum tensor $T(w)$ along with composite
fields built as (normal ordered) powers of $T(w)$ and its derivatives.
What they found is that there exists an infinite dimensional abelian
subalgebra $\mathcal{I}\subset\mathcal{U}(\textit{Vir})$ spanned
by ``local integrals of motion'' (IM):
\[
\mathcal{I}=\left\{ \mathbf{I}_{2k-1}\right\} _{k=1}^{\infty}\;,\quad\mathbf{I}_{2k-1}\doteq\intop_{0}^{2\pi}\frac{dw}{2\pi}T_{2k}(w)\;,\quad[I_{2k-1},I_{2l-1}]=0\;,
\]
where the cylinder radius is fixed at $R=1$. The densities are some
regularised polynomials of $T(w)$ and its derivatives; for example
\begin{equation}
T_{2}(w)\doteq T(w)\;,\quad T_{4}(w)\doteq\,:T^{2}(w):\;,\quad T_{6}(w)\doteq\,:T^{3}(w):+\frac{c+2}{12}\,:\left(\partial T(w)\right)^{2}:\;,\label{eq:IMdensities}
\end{equation}
where the regularised product is defined as
\[
:T^{2}(w):\,\doteq\ointop_{w}\frac{dw'}{2\pi\mathsf{i}}\frac{\mathscr{T}\left(T(w')T(w)\right)}{w'-w}\;,
\]
and $\mathscr{T}$ is the ``chronological product''%
\footnote{This is the corresponding on the cylinder of the radial ordering on
the $z$-plane.%
}
\[
\mathscr{T}\left(A(w')B(w)\right)=\begin{cases}
A(w')B(w) & \textrm{if}\;\Im(w)>\Im(w')\\
B(w)A(w') & \textrm{if}\;\Im(w')>\Im(w)
\end{cases}\;.
\]
Although there is no known closed formula for the densities $T_{2k}(w)$,
they are uniquely determined by the requirement of commutativity of
local IMs and by the \textit{spin assignment} rule:
\[
\ointop_{w}\frac{dw'}{2\pi\mathsf{i}}(w'-w)\mathscr{T}\left(T(w')T_{2k}(w)\right)=2k\, T_{2k}(w)\;,
\]
which can be simply implemented by requiring $T_{2k}(w)$ to be a
polynomial of total grade $2k$ and assigning grade $2$ to $T$ and
grade $1$ to derivatives. The first few IMs are ``easily'' computed
\begin{eqnarray}
\mathbf{I}_{1} & = & L_{0}-\frac{c}{24}\;,\qquad\textrm{This is the (chiral part of the) Hamiltonian!}\;,\label{eq:Integr_CFT_1}\\
\mathbf{I}_{3} & = & 2\sum_{n=1}^{\infty}L_{-n}L_{n}+L_{0}^{2}-\frac{c+2}{12}L_{0}+c\frac{5c+22}{4\times6!}\;,\label{eq:Integr_CFT_3}\\
\mathbf{I}_{5} & = & \!\!\!\!\!\!\!\!\!\!\!\!\sum_{n_{1}+n_{2}+n_{3}=0}\!\!\!\!\!\!\!\!\!\!:L_{n_{1}}L_{n_{2}}L_{n_{3}}:\,+\sum_{n=1}^{\infty}\left[\frac{c+11}{6}n^{2}-1-\frac{c}{4}\right]L_{-n}L_{n}+\frac{3}{2}\sum_{n=1}^{\infty}L_{1-2n}L_{2n-1}+\nonumber \\
 &  & \qquad\qquad\quad-\frac{c+4}{8}L_{0}^{2}+5(c+2)\frac{3c+20}{4\times6!}L_{0}-5c\frac{3c+14}{4\times9!}(7c+68)\label{eq:Integr_CFT_5}
\end{eqnarray}

\subsection{Brief overview of classical KdV\label{sub:(Very)-brief-overview}\index{(Very) brief overview of classical KdV}}

In this section we will briefly present some very basic concepts and
facts about the classical KdV hierarchy. The reader interested in
this topic can find a good starting point for the study of classical
integrability in the review \cite{Torr_16}; for a more advanced read
we suggest the beautiful book \cite{Babe_Bern_Talo}.

Why being concerned with the classical KdV? The reason is very simple.
It is known \cite{Sasa_Yama_88} that the CFTs with $c<1$ are, in
some sense, a quantum version of the classical KdV; in fact if we
consider the ``classical limit'' $c\rightarrow-\infty$ and perform
the following substitutions:
\[
T(w)\rightarrow-\frac{c}{6}U(w)\;,\qquad[\;,\;]\rightarrow\frac{6\pi}{\mathsf{i}\, c}\{\;,\;\}_{P}\;,
\]
where $\{\;,\;\}_{P}$ are the Poisson brackets, the Virasoro algebra
reduces to the following Poisson algebra
\[
\{U(w),U(w')\}_{P}=2\left(U(w)+U(w')\right)\delta'(w-w')+\delta'''(w-w')\;,
\]
which is known to describe the second Hamiltonian structure of KdV,
provided the Hamiltonian is chosen amongst the classical IM:
\begin{eqnarray*}
I_{1}^{\textrm{cl}} & = & \intop_{0}^{2\pi}\frac{dw}{2\pi}U(w)\\
I_{3}^{\textrm{cl}} & = & \intop_{0}^{2\pi}\frac{dw}{2\pi}U^{2}(w)\\
I_{5}^{\textrm{cl}} & = & \intop_{0}^{2\pi}\frac{dw}{2\pi}\left[U^{3}(w)-\frac{1}{2}\left(\partial_{w}U(w)\right)^{2}\right]\\
 & \cdots & \;.
\end{eqnarray*}
We choose the field $U(w)$ to be periodic $U(w+2\pi)=U(w)$, just
like $T(w)$. These classical IMs, which form a commutative Poisson
algebra $\{I_{2k-1},I_{2l-1}\}_{P}=0$, are clearly the classical
versions of the operators (\ref{eq:Integr_CFT_1}-\ref{eq:Integr_CFT_5}).
Different choices of Hamiltonian bring us to different equation of
motion:
\begin{eqnarray}
I_{1}^{\textrm{cl}} & : & \partial_{t_{1}}U=\partial_{w}U\nonumber \\
I_{3}^{\textrm{cl}} & : & \partial_{t_{3}}U=\partial_{w}^{3}U+6U\,\partial_{w}U\qquad\quad\textrm{The "canonical" KdV}\label{eq:KdV_tower}\\
I_{5}^{\textrm{cl}} & : & \partial_{t_{5}}U=-\partial_{w}^{5}U-2U\,\partial_{w}^{3}U+5\partial_{w}U\partial_{w}^{2}U+20U^{2}\partial_{w}U\nonumber \\
 & \cdots & \;.\nonumber 
\end{eqnarray}
This infinite sequence of partial differential equations is called
\textit{KdV hierarchy} and it can be shown to be equivalent to a description
of the isospectral deformations of the following second order differential
operator depending on a \emph{spectral parameter} $\lambda$:
\begin{equation}
L(w\vert\lambda)\doteq\partial_{w}^{2}+U(w)-\lambda^{2}\;,\label{eq:Linear_Lax_KdV}
\end{equation}
called \textit{Lax operator}. The connection between this operator
and the tower of differential equations (\ref{eq:KdV_tower}) relies
on the existence of an infinite set of operators $M_{2n-1}(w)$, such
that
\[
\frac{d}{dt_{2n-1}}L(w\vert\lambda)=\left[M_{2n-1}(w),L(w\vert\lambda)\right]\;\Longleftrightarrow\textrm{KdV equation associated to }I_{2n-1}^{\textrm{cl}}\textrm{ is satisfied.}
\]
For example the canonical KdV equation is obtained from the operator
\[
M_{3}(w)=4\partial_{w}^{3}+6U(w)\partial_{w}+3U'(w)\;.
\]
Associated to each Lax operator, there exists a differential equation,
called usually \emph{auxiliary equation} (or \emph{system} if one
has to deal with matrix Lax operators). In our case the equation has
the form (from here on the prime $'$ will denote differentiation)
\[
L(w\vert\lambda)\psi(w\vert\lambda)=\psi''(w)-\left(\lambda^{2}-U(w)\right)\psi(w)=0\;.
\]
This is a second order differential equation and, as such, possesses
two linearly independent solutions $\psi_{1}(w\vert\lambda)$ and
$\psi_{2}(w\vert\lambda)$. Very important characteristics of differential
equations are the \emph{monodromy properties} of the solutions; these
can be encoded in the \emph{monodromy matrix}, defined as
\[
\left(\psi_{1}(w\vert\lambda)\,,\,\psi_{2}(w\vert\lambda)\right)\mathbf{M}(\lambda)=\left(\psi_{1}(w+2\pi\vert\lambda)\,,\,\psi_{2}(w+2\pi\vert\lambda)\right)\;.
\]
Out of the monodromy matrix is then possible to define the $T$\emph{-function}.
This is a central object in integrable systems and is defined most
simply as the trace of $\mathbf{M}$
\begin{equation}
\mathbf{T}(\lambda)\doteq\textit{tr}\,\mathbf{M}(\lambda)\;.\label{eq:KdVTop}
\end{equation}
Although an explicit expression of $\mathbf{T}$ can be complicated
to obtain, we can express it as an asymptotic series at large $\lambda$:
\begin{equation}
\frac{1}{2\pi}\log\left[\mathbf{T(\lambda)}\right]\underset{\lambda\rightarrow\infty}{\sim}\lambda\left[1-\sum_{n=1}^{\infty}c_{n}I_{2n-1}^{\textrm{cl}}\lambda^{-2n}\right]\;,\label{eq:KdV_asympt_T}
\end{equation}
where $c_{1}=1/2$ and $c_{n}=\frac{(2n-3)!!}{2^{n}\, n!}$.

\begin{framed}%

\paragraph*{WKB expansion of the Lax auxiliary equation}

\begin{flushleft}
It is instructive to compute explicitly the expression of the monodromy
matrix. In order to do so we need to find a representation of the
solutions to the differential equation
\[
\psi''(w)=\left(\lambda^{2}-U(w)\right)\psi(w)\;,
\]
and a standard procedure which allows us to do so is the WKB method
\cite{Bend_Orsz}. The first step consists in introducing a small
parameter $\epsilon^{2}$ in front of the second derivative:
\[
\epsilon^{2}\psi''(w)=\left(\lambda^{2}-U(w)\right)\psi(w)\;,
\]
and search for solutions of the form
\[
\psi(w)\sim\exp\left[\frac{1}{\epsilon}S(w)+A_{0}(w)+\sum_{n=1}^{\infty}\epsilon^{n}A_{n}(w)\right]\;,\textrm{as}\;\epsilon\rightarrow0\;,
\]
where the sign $\sim$ reminds us that the right-hand side is an asymptotic
series. By inserting this form in the differential equation and isolating
each power of $\epsilon$, we find 
\begin{align*}
\epsilon^{0}\;: & \quad S'(w)^{2}=\lambda^{2}-U(w)\quad\Rightarrow\quad S(w)=\pm\intop_{w_{0}}^{w}\sqrt{\lambda^{2}-U(w')}dw'\;,\\
\epsilon^{1}\;: & \quad S''(w)+2S'(w)A_{0}'(w)=0\quad\Rightarrow\quad A_{0}(w)=k-\frac{1}{4}\log\left(\lambda^{2}-U(w)\right)\;,\\
\epsilon^{n}\;: & \quad2S'(w)A_{n-1}'(w)+A_{n-2}''(w)+\sum_{k=0}^{n-2}A_{k}'(w)A_{n-k}'(w)=0\;,\quad\forall n>1\;.
\end{align*}
This is a triangular system of differential equations, which allows
us to obtain the $n$-th term by the simple integration of a first-order
differential equation. What's more, each even order equation happens
to be the difference of total derivatives; for example
\[
\epsilon^{3}\;:\qquad A_{2}'(w)=\partial_{w}\left[\frac{\left(\partial_{w}\log S'(w)\right)^{2}-2\partial_{w}^{2}\log S'(w)}{16\, S'(w)^{2}}\right]\;.
\]
The odd-order equations, on the other hand, are proper first-order
differential equations, as an example, the first term reads
\[
\epsilon^{2}\;:\qquad A_{1}(w)=\intop_{w_{0}}^{w}\frac{2S'''(w')S'(w')-3S''(w')^{2}}{8S'(w')^{3}}dw'\;.
\]
Now, in order to obtain an expression for $\mathbf{M}$, we need to
see what happens to our solution when we shift $w\rightarrow w+2\pi$.
This is easily computed remembering that $U(w+2\pi)=U(w)$, so that
\[
A_{2n}(w+2\pi)=A_{2n}(w)\;,\qquad\forall n\geq0\;,
\]
while
\[
A_{2n-1}(w+2\pi)=\intop_{w_{0}}^{w+2\pi}\mathcal{A}_{2n-1}\left[U(w')\right]dw'=\intop_{w_{0}-2\pi}^{w_{0}}\mathcal{A}_{2n-1}\left[U(w')\right]dw'+\intop_{w_{0}}^{w}\mathcal{A}_{2n-1}\left[U(w')\right]dw'\;,
\]
so that
\[
A_{2n-1}(w+2\pi)=A_{2n-1}(w)+\intop_{0}^{2\pi}\mathcal{A}_{2n-1}\left[U(w')\right]dw'\;,
\]
where $\mathcal{A}_{2n-1}\left[U(t)\right]$ is some functional of
$U(t)$ and $A_{-1}(w)\equiv S(w)$. From this we easily infer that
the monodromy matrix is diagonal with eigenvalues
\[
\exp\left[\pm\intop_{0}^{2\pi}\left(\sqrt{\lambda^{2}-U(w')}+\sum_{n=1}^{\infty}\mathcal{A}_{2n-1}\left[U(w')\right]\right)dw'\right]\;.
\]
The first two terms in the large-$\lambda$ expansion of $\mathbf{T}$
(\ref{eq:KdV_asympt_T}) come from the expansion of the square root
for large $\lambda$,
\[
\intop_{0}^{2\pi}\frac{dw'}{2\pi}\sqrt{\lambda^{2}-U(w')}\underset{\lambda\rightarrow\infty}{\sim}\intop_{0}^{2\pi}\frac{dw'}{2\pi}\lambda\left(1-\frac{U(w')}{2\lambda^{2}}-\frac{U(w')^{2}}{8\lambda^{4}}\ldots\right)=\lambda\left(1-\frac{1}{2\lambda^{2}}I_{1}^{\textrm{cl}}-\frac{1}{8\lambda^{4}}I_{3}^{\textrm{cl}}\right)\;.
\]
On the other hand, higher order terms require the computation of more
and more $A_{n}(w)$ in the WKB expansion; as a simple example, let
us take in consideration $A_{1}$
\[
\mathcal{A}_{1}(w)=-\frac{4U''(w)(\lambda^{2}-U(w))-5U'(w)^{2}}{32(\lambda^{2}-U(w))^{\frac{5}{3}}}\underset{\lambda\rightarrow\infty}{\sim}-\frac{U''(w)}{8\lambda^{3}}-\frac{5U'(w)^{2}+6U(w)U''(w)}{32\lambda^{5}}\;.
\]
When integrating the above term between $0$ and $2\pi$, all total
derivatives vanish and we can perform integration by parts, so that
the coefficient of $\lambda^{-5}$ reads, as expected
\[
-\frac{1}{16}\intop_{0}^{2\pi}\frac{dw'}{2\pi}\left(U(w')^{3}-\frac{1}{2}U'(w')^{2}\right)=-\frac{1}{16}I_{5}^{\textrm{cl}}\;.
\]

\par\end{flushleft}\end{framed}

\begin{flushleft}
We have thus shown that the $T$-function (\ref{eq:KdVTop}) serves
as a sort of generating function for the classical IMs (\ref{eq:KdV_asympt_T}).
However we can do more, much more: in fact we can construct an infinite
tower of Poisson commuting T-functions! This is a consequence of a
deep connection between KdV hierarchy and the Lie algebra $sl(2)$
as we are going to briefly hint at. A close look at the differential
operator (\ref{eq:Linear_Lax_KdV}) shows that it can be factorised:
\[
L(w\vert\lambda)=\left(\partial_{w}+\phi'(w)\right)\left(\partial_{w}-\phi'(w)\right)-\lambda^{2}\;,
\]
with the field $\phi(w)$ being the \emph{Miura transform} of $U(w)$
\cite{Miur_68}:
\begin{equation}
-U(w)=\left(\phi'(w)\right)^{2}+\phi''(w)\;,\label{eq:Miura_trans}
\end{equation}
having canonical Poisson brackets
\[
\left\{ \phi(w),\phi(w')\right\} _{P}=\epsilon(w-w')\;,\qquad\epsilon(x)=n\;,\ 2\pi n<x\leq2\pi(n+1)\;.
\]
Note that since the field $U(w)$ is periodic, the Miura field has
to be taken, in full generality, quasiperiodic
\[
\phi(w+2\pi)=\phi(w)+2\pi\mathsf{i}p\;.
\]
We can now reduce the second order differential equation $L(w\vert\lambda)\psi(w\vert\lambda)=0$
to a system of first order equations:
\[
\begin{cases}
\left(\partial_{w}-\phi'(w)\right)\psi(w\vert\lambda)=\lambda\tilde{\psi}(w\vert\lambda)\\
\left(\partial_{w}+\phi'(w)\right)\tilde{\psi}(w\vert\lambda)=\lambda\psi(w\vert\lambda)
\end{cases}\;,
\]
which can be written in matrix form
\[
\left(\partial_{w}-\phi'(w)\sigma^{3}-\lambda\sigma^{1}\right)\Psi(w\vert\lambda)=0\;,
\]
with $\sigma^{i}$ being the Pauli matrices:
\[
\sigma^{1}=\left(\begin{array}{cc}
0 & 1\\
1 & 0
\end{array}\right)\;,\quad\sigma^{2}=\left(\begin{array}{cc}
0 & -\mathsf{i}\\
\mathsf{i} & 0
\end{array}\right)\;,\quad\sigma^{3}=\left(\begin{array}{cc}
1 & 0\\
0 & -1
\end{array}\right)\;.
\]
Now comes the generalisation: since the connection between $L$ and
the hierarchy of KdV equations (\ref{eq:KdV_tower}) uniquely relies
on the commutation relations between $L$ and the operators $M_{2n-1}$,
we can think of defining an abstract Lax operator 
\[
\mathscr{L}(w\vert\lambda)\doteq\partial_{w}-\phi'(w)\, H-\lambda(E+F)\;,
\]
where $H$, $E$ and $F$ are the generators of $sl(2)$ Lie algebra:
\[
[H,E]=2E\;,\quad[H,F]=-2F\;,\quad[E,F]=2H\;.
\]
The commutation properties of this operator are exactly the same as
those of $L$ and, moreover, it reduces to this last when the $2$-dimensional
representation of $sl(2)$ is chosen. We thus expect the monodromy
properties of this operator to encode information on the KdV hierarchy.
Now, let $\pi_{j}$, with $j\in\frac{1}{2}\mathbb{N}$, denote the
$(2j+1)$-dimensional representation of $sl(2)$, such that $\pi_{j}\left[H\right]=\textrm{diag}\left(2j,2j-2,\ldots,-2j+2,-2j\right)$.
Consider the matrix equation
\[
\pi_{j}\left[\mathscr{L}(w\vert\lambda)\right]\Psi_{j}(w\vert\lambda)=0\;,
\]
where $\Psi_{j}(w\vert\lambda)$ is a $(2j+1)$-dimensional vector,
and let us repeat what has been done just above. In order to obtain
a nice form of the solution to this equation, we rewrite it as follows
(we omit $\pi_{j}$ in the next few equations, for clarity):
\[
\left(\partial_{w}-\phi'(w)H\right)\Psi(w)=e^{\phi(w)H}\partial_{w}e^{-\phi(w)H}\Psi(w)=\lambda\left(E+F\right)\Psi(w)\;,
\]
where the first passage is allowed, since $H$ is diagonal. Now define
$\tilde{\Psi}(w)\doteq e^{-\phi(w)H}\Psi(w)$, so that it satisfies
the equation
\[
\partial_{w}\tilde{\Psi}(w)=\lambda e^{-\phi(w)H}\left(E+F\right)e^{\phi(w)H}\tilde{\Psi}(w)=\lambda\left(e^{-2\phi(w)}E+e^{2\phi(w)}F\right)\tilde{\Psi}(w)\;,
\]
where we used the property of any Lie algebra element $A$: $e^{\alpha H}Ae^{-\alpha H}=e^{\alpha\textrm{ad}_{H}(A)}A$,
where the adjoint action is defined by $[H,A]=\textrm{ad}_{H}(A)\, A$.
The general solution of a first-order matrix equation can be written
as a \textit{path-ordered exponential}\textit{\emph{:}}
\[
\tilde{\Psi}(w)=\mathscr{P}\exp\left[\lambda\intop_{0}^{w}dw'\left(e^{-2\phi(w')}E+e^{2\phi(w')}F\right)\right]\Psi^{0}\;,
\]
with $\Psi^{0}$ being an arbitrary constant vector, representing
the integration constants. A path-ordered exponential $\mathscr{P}\exp\left[\int_{0}^{w}a(w')dw'\right]$
is defined as the following series expansion
\[
\mathscr{P}\exp\left[\int_{0}^{w}a(w')dw'\right]=\sum_{n=0}^{\infty}\frac{1}{n!}\intop_{0}^{w}\cdots\intop_{0}^{w}\mathscr{P}\left[a(w_{1}')\cdots a(w_{n}')\right]dw_{1}'\cdots dw_{n}'\;,
\]
and the path-ordering $\mathscr{P}$ forces an ordering of decreasing
argument from left to right:
\[
\mathscr{P}\left[a(w_{1})a(w_{2})\right]=\begin{cases}
a(w_{1})a(w_{2}) & w_{1}>w_{2}\\
a(w_{2})a(w_{1}) & w_{1}<w_{2}
\end{cases}\;.
\]
Re-expressing $\tilde{\Psi}$ in terms of $\Psi$, we obtain
\[
\Psi_{j}(w\vert\lambda)=\pi_{j}\left\{ e^{\phi(w) H}\mathscr{P}\exp\left[\lambda\intop_{0}^{w}dw'\left(e^{-2\phi(w')}E+e^{2\phi(w')}F\right)\right]\right\} \Psi_{j}^{0}\;,
\]

\par\end{flushleft}

Now, for each representation $\pi_{j}$ we can define a monodromy
matrix:
\begin{equation}
\mathbf{M}_{j}(\lambda)=\pi_{j}\left\{ e^{2\pi\mathsf{i}p H}\mathscr{P}\exp\left[\lambda\intop_{0}^{2\pi}dw\left(e^{-2\phi(w)}E+e^{2\phi(w)}F\right)\right]\right\} \;,\label{eq:KdV_monodromy}
\end{equation}
and a corresponding $L$-matrix%
\footnote{In the literature this object is sometimes called Lax matrix.%
}
\begin{equation}
\mathbf{L}_{j}(\lambda)\doteq\pi_{j}\left[e^{-\pi\mathsf{i}p H}\right]\mathbf{M}_{j}(\lambda)\;.\label{eq:KdV_Lmat}
\end{equation}
This last matrix can be shown to satisfy the $r$-matrix Poisson relation:
\begin{equation}
\left\{ \mathbf{L}_{j}(\lambda)\overset{\otimes}{,}\mathbf{L}_{j'}(\lambda')\right\} _{P}=[\mathbf{r}_{j\, j'}(\lambda/\lambda'),\,\mathbf{L}_{j}(\lambda)\otimes\mathbf{L}_{j'}(\lambda')]\;,\label{eq:r_mat_pois}
\end{equation}
where the $r$-matrix is defined as
\[
\mathbf{r}_{j\, j'}(\lambda)\doteq\left(\pi_{j}\otimes\pi_{j'}\right)\left[\mathbf{r}(\lambda)\right]\;,\quad\mathbf{r}(\lambda)\doteq\frac{\lambda+\lambda^{-1}}{\lambda-\lambda^{-1}}\frac{H\otimes H}{2}+\frac{2}{\lambda-\lambda^{1}}\left(E\otimes F+F\otimes E\right)\;.
\]
The $r$-matrix Poisson algebra tells us immediately that the quantities%
\footnote{The operator denoted here with $\mathbf{T}$ are not to be confused
with the densities of IM (\ref{eq:IMdensities}) introduced above.%
}
\[
\mathbf{T}_{j}(\lambda)=\textit{tr}\,\mathbf{M}_{j}(\lambda)\;,
\]
are in involution with respect with the Poisson brackets
\[
\left\{ \mathbf{T}_{j}(\lambda),\mathbf{T}_{j'}(\lambda')\right\} _{P}=0\;,
\]
and are expected to generate the classical IMs in their asymptotic
limit. Note that $\mathbf{T}_{\frac{1}{2}}(\lambda)=\mathbf{T}(\lambda)$.

We will not be showing the explicit proof of the relation (\ref{eq:r_mat_pois}),
however we wish to close this section suggesting an approach to the
computation which we think gives an intuitive interpretation of the
form of $\mathbf{M}_{j}$. The core of this approach resides in the
following expansion of the path-ordered exponential as a ``continuum
limit'' of an ordered product:
\[
\mathscr{P}\exp\left[\intop_{0}^{2\pi}a(w')dw\right]=\lim_{N\rightarrow\infty}e^{a(w_{N})\Delta w}e^{a(w_{N-1})\Delta w}\cdots e^{a(w_{0})\Delta w}\;,
\]
where $w_{j}=j\Delta w$ and $\Delta w=\frac{2\pi}{N}$. This expression
allows us to write $\mathbf{L}_{j}$ as (here too we omit $\pi_{j}$
for clarity)
\[
\mathbf{L}(\lambda)=e^{\pi\mathsf{i}p H}\lim_{N\rightarrow\infty}\mathbf{l}(w_{N}\vert\lambda)\mathbf{l}(w_{N-1}\vert\lambda)\cdots\mathbf{l}(w_{0}\vert\lambda)\;,
\]
where the matrices $\mathbf{l}_{j}$ are
\[
\mathbf{l}(w\vert\lambda)=\sum_{n=0}^{\infty}\frac{\lambda^{n}\Delta w^{n}}{n!}\left(e^{-2\phi(w)}E+e^{2\phi(w)}F\right)^{n}\;.
\]
It is now sufficient to show that $\mathbf{l}$ satisfies the relation
(\ref{eq:r_mat_pois}), for any value of $w$ and $\lambda$. This
is reminiscent of the approach to lattice models, in which we have
a matrix $\mathbf{l}$ on each site and the full transfer matrix of
the system is built as a trace of the product of these matrices for
each site. In fact we can interpret the matrix $\mathbf{L}(\lambda)$
as the ``continuum limit'' of the monodromy matrix of a lattice
model, where $e^{\pi\mathsf{i}pH}$ plays the role of twist.

\subsection{The quantum monodromy matrix and the $T$-operators\label{sub:The-quantum-monodromy}\index{The quantum monodromy matrix and the T -operators}}

Let us now concentrate on our goal: we are going to reproduce in the
$c<-2$ CFTs%
\footnote{The restriction to this domain will be clearer later.%
} what has been sketched above for the classical KdV hierarchy. Namely
we are going to address the problem of simultaneous diagonalisation
of the local IMs (\ref{eq:Integr_CFT_1}-\ref{eq:Integr_CFT_5}) with
a method that can be interpreted as a version of the Quantum Inverse
Scattering (QIS) \cite{Bogo_Izer_Kore} for field theories. Just as
it was suggested in the previous section, all the object we are going
to introduce have a counterpart in lattice models and it is a good
idea to keep in mind this parallelism. On the other hand these objects
will be the quantised version of those introduced above for the classical
KdV hierarchy and we are going to use the same symbols to denote them.
Note that, from now on, we will consider the right chirality only.

In order to proceed to the construction of the objects $\mathbf{T}_{j}$,
we first need the quantum version of Miura transformation: \textit{the
Feigin-Fuchs free field representation} \cite{Feig_Fuch_82}
\begin{equation}
-\beta^{2}T(w)=:\left(\varphi'(w)\right)^{2}:\,+(1-\beta^{2})\varphi''(w)+\frac{\beta^{2}}{24}\;,\qquad\beta\doteq\sqrt{\frac{1-c}{24}}-\sqrt{\frac{25-c}{24}}\;,\label{eq:Feigin_Fuchs}
\end{equation}
where $\varphi(w)$ is a free field
\[
\varphi(w)=\mathsf{i} Q+\mathsf{i} P w+\sum_{n\neq0}\frac{a_{-n}}{n}e^{\mathsf{i} n w}\;,
\]
and the normal ordering $:\cdot:$ consist in placing the $a_{n}$
oscillators in increasing $n$ from left to right. The operators $Q$,
$P$ and $\left\{ a_{n}\right\} _{n\neq0}$ generate an Heisenberg
algebra:
\[
[Q,P]=\frac{\mathsf{i}}{2}\beta^{2}\;,\qquad[a_{n},a_{m}]=\frac{n}{2}\beta^{2}\delta_{n+m,0}\;.
\]
It is easy to see that this transformation becomes exactly (\ref{eq:Miura_trans})
as $c\rightarrow-\infty$. With this expression for $T(w)$ we are
able to give a description of the Hilbert space $\mathscr{H}_{ch}$
in terms of Fock spaces $\mathscr{F}_{p}$, defined as highest-weight
modules over the Heisenberg algebra; the highest-weight vector $\left|\, p\,\right\rangle \in\mathscr{F}_{p}$
obeys to the following relations
\[
P\left|\, p\,\right\rangle =p\left|\, p\,\right\rangle \;,\qquad a_{n}\left|\, p\,\right\rangle =0\;,\;\forall n>0\;.
\]
The Fock space thus defined is isomorphic to the Verma module $\mathcal{V}_{h}$,
where
\[
h=\left(\frac{p}{\beta}\right)^{2}+\frac{c-1}{24}\;,
\]
and we can describe the Hilbert space as
\[
\mathscr{H}_{ch}=\bigoplus_{a}\mathscr{F}_{a}\;,\qquad\mathscr{F}_{a}\equiv\mathscr{F}_{p_{a}}\;,
\]
where the direct sum runs over the values of $p$ corresponding to
the allowed Virasoro highest-weights. These Fock spaces are naturally
graded under the action of $L_{0}$:
\[
\mathscr{F}_{p}=\bigoplus_{\ell=0}^{\infty}\mathscr{F}_{p}^{(\ell)}\;,\qquad L_{0}\mathscr{F}_{p}^{(\ell)}=(h+\ell)\mathscr{F}_{p}^{(\ell)}\;.
\]
With some simple algebraic manipulation we can express the Virasoro
generators $\left\{ L_{n}\right\} $ in terms of the Heisenberg algebra
as
\begin{eqnarray*}
\beta^{2}L_{n} & = & \beta^{2}\frac{c-1}{24}+2\sum_{j\neq0,n}a_{j}a_{n-j}+a_{n}\left(2P-n(1-\beta^{2})\right)\;,\\
\beta^{2}L_{0} & = & \beta^{2}\frac{c-1}{24}+2\sum_{j=1}^{\infty}a_{-j}a_{j}+P^{2}\;.
\end{eqnarray*}
Since in theory we know how to express the local IMs $\left\{ \mathbf{I}_{2k-1}\right\} $
in terms of the local densities $T_{2k}(w)$, the formers can be re-expressed
in terms of polynomials in the free field $\varphi(w)$ and its derivatives:
\[
\mathbf{I}_{2k-1}=(-1)^{k}\beta^{-2k}\intop_{0}^{2\pi}\frac{dw}{2\pi}\left[\,:\left(\varphi'(w)\right)^{2k}:\,+\underset{\textrm{Higher derivatives of }\varphi(w)}{\underbrace{\cdots\cdots\cdots\cdots\cdots}}\right]\;.
\]
As it is evident from their definition (remember the spin assignment
request), each term in a local IM, as complicated as it might be,
is nevertheless a product of operators $L_{n_{i}}$, where the sum
of indices vanishes: $\sum_{i}n_{i}=0$. As a consequence $[L_{0},\mathbf{I}_{2k-1}]=0$
and the local IMs act invariantly on the level subspaces $\mathscr{F}_{p}^{(\ell)}$.
The full diagonalisation of the integrals of motion is thus reduced
to their diagonalisation on each level subspace, which requires a
finite number of algebraic manipulations; these, however, become rapidly
extremely involved and so far the result is known only for some simple
cases, e.g. for the vacuum $\left|p\right\rangle $
\begin{eqnarray*}
I_{1}^{(\textrm{vac})}(h,c) & = & h-\frac{c}{24}\;,\\
I_{3}^{(\textrm{vac})}(h,c) & = & h^{2}-\frac{c+2}{12}h+c\frac{5c+22}{4\times6!}\;,\\
I_{5}^{(\textrm{vac})}(h,c) & = & h^{3}-\frac{c+4}{8}h^{2}+5(c+2)\frac{3c+20}{4\times6!}h-5c(3c+14)\frac{7c+68}{4\times9!}\;,\\
 & \cdots
\end{eqnarray*}
where $\mathbf{I}_{2k-1}\left|\, p\,\right\rangle =I_{2k-1}^{(\textrm{vac})}\left|\, p\,\right\rangle $.

In the setting provided by the Fock description of the Hilbert space,
we can easily follow the footprints of \subref{(Very)-brief-overview}
and define quantum counterparts of the monodromy matrices (\ref{eq:KdV_monodromy})
and of the $L$-matrices (\ref{eq:KdV_Lmat}). In order to do so,
we consider the quantum enveloping algebra $\mathcal{U}_{q}\left(sl(2)\right)$
\cite{Kuli_Resh_Skly_81,Fuch} generated by the elements $E$, $F$
and $H$:
\[
[H,E]=2E\;,\quad[H,F]=-2F\;,\quad[E,F]=\frac{q^{H}-q^{-H}}{q-q^{-1}}\;,
\]
with
\[
q\doteq e^{\mathsf{i}\pi\beta^{2}}\;,
\]
and let $\pi_{j}$ denote the $(2j+1)$-dimensional representation
of this algebra. The ``quantum monodromy matrices'' are then defined
as the following operator-valued matrices%
\footnote{Note that these objects can be informally interpreted as the monodromy
matrices of the solution $\Psi_{j}$ to the operator-matrix equation
$\pi_{j}\left[\mathscr{L}(w\vert\lambda)\right]\Psi_{j}(w\vert\lambda)=0$,
where $\mathscr{L}(w\vert\lambda)=\partial-\varphi'(w)\, H-\lambda\left(q^{\frac{H}{2}}E+q^{-\frac{H}{2}}F\right)$.
Here one has to take care of normal ordering when defining the corresponding
of $\tilde{\Psi}=:e^{-\varphi}:\Psi$. This gives rise to the presence
of vertex operators in the path-ordered integral.%
}
\begin{equation}
\mathbf{M}_{j}(\lambda)\doteq\pi_{j}\left\{ e^{2\pi \mathsf{i}P H}\mathscr{P}\exp\left[\lambda\intop_{0}^{2\pi}dw\left(V_{-}(w)q^{\frac{H}{2}}E+V_{+}(w)q^{-\frac{H}{2}}F\right)\right]\right\} \;,\label{eq:quantum_monodromy}
\end{equation}
where
\[
V_{\pm}(w)\doteq\,:e^{\pm2\varphi(w)}:\,=\exp\left[\pm2\sum_{n=1}^{\infty}\frac{a_{-n}}{n}e^{\mathsf{i}nw}\right]e^{\pm2\mathsf{i}\left(Q+Pw\right)}\exp\left[\mp2\sum_{n=1}^{\infty}\frac{a_{n}}{n}e^{-\mathsf{i}nw}\right]\;,
\]
are called vertex operators; they have conformal dimension $\beta^{2}$
and act on the Fock spaces by shifting the highest weight:
\[
V_{\pm}(w)\;:\quad\mathscr{F}_{p}\longrightarrow\mathscr{F}_{p\pm\beta^{2}}\;.
\]
The reason why in the path ordered exponent the combinations $q^{\frac{H}{2}}E$
and $q^{-\frac{H}{2}}F$ appear is related to the fact that the correct
construction of $\mathbf{M}_{j}$ should start from the quantum affine
enveloping algebra $\mathcal{U}_{q}\left(\widehat{sl(2)}\right)$;
we will return to this point in \subref{Baxter-Q-operators}.

The $L$-operators are defined in the same way as in the classical
case:
\begin{equation}
\mathbf{L}_{j}(\lambda)\doteq\pi_{j}\left[e^{-\pi \mathsf{i}P H}\right]\mathbf{M}_{j}(\lambda)\;.\label{eq:quantum_L}
\end{equation}
Both the quantum monodromy matrices and the $L$-operators are $(2j+1)\times(2j+1)$
matrices whose elements are operators acting on the space
\[
\hat{\mathscr{F}}_{p}\doteq\bigotimes_{n=-\infty}^{\infty}\mathscr{F}_{p+n\beta^{2}}\;.
\]
They have to be understood as power series in $\lambda$:
\[
\mathbf{L}_{j}(\lambda)=\pi_{j}\left[e^{\pi\mathsf{i}P H}\sum_{k=0}^{\infty}\lambda^{k}\intop_{\underset{w_{1}\geq\cdots\geq w_{k}}{0}}^{2\pi}dw_{1}\cdots dw_{k}\mathcal{K}(w_{1})\cdots\mathcal{K}(w_{k})\right]\;,
\]
where we introduced
\begin{equation}
\mathcal{K}(w)\doteq V_{-}(w)q^{\frac{H}{2}}E+V_{+}(w)q^{-\frac{H}{2}}F\;.\label{eq:Numerator}
\end{equation}
These series converge for any $\lambda$ if%
\footnote{This is most easily inferred from the fact that the vertex operators
$V_{\pm}$ have conformal dimension $\beta^{2}$ and, thus, $V_{+}(w)V_{-}(w')\sim\left(w-w'\right)^{-2\beta^{2}}\left(1+O(w-w')\right)$.
So, for the integrals to converge we must impose $\beta^{2}<\frac{1}{2}$,
which is equivalent to $c<-2$.%
} $-\infty<c<-2$; outside this region the definition of $\mathbf{M}_{j}$
and $\mathbf{L}_{j}$ necessitate a proper regularisation. In these
notes we will limit ourselves to the cases $c<-2$.

The operators $\mathbf{L}_{j}$ are tailored in such a way that the
following $RLL$ relation is satisfied
\begin{equation}
\mathbf{R}_{j\, j'}(\lambda/\lambda')\left(\mathbf{L}_{j}(\lambda)\otimes\mathbb{I}\right)\left(\mathbb{I}\otimes\mathbf{L}_{j'}(\lambda')\right)=\left(\mathbb{I}\otimes\mathbf{L}_{j'}(\lambda')\right)\left(\mathbf{L}_{j}(\lambda)\otimes\mathbb{I}\right)\mathbf{R}_{j\, j'}(\lambda/\lambda')\;,\label{eq:RLL_relation}
\end{equation}
where $\mathbf{R}_{j\, j'}(\lambda)$ is the trigonometric $R$-matrix
of $\mathcal{U}_{q}\left(sl(2)\right)$, acting on $\pi_{j}\otimes\pi_{j'}$.
The fundamental-fundamental case $j=j'=\frac{1}{2}$ reads as follows%
\footnote{Note that this is the same exact matrix as for the 6-vertex model
\cite{Kiri_Resh_87}!%
}
\begin{equation}
\mathbf{R}_{\frac{1}{2}\frac{1}{2}}(\lambda)=\begin{pmatrix}\frac{\lambda}{q}-\frac{q}{\lambda}\\
 & \lambda-\lambda^{-1} & q^{-1}-q\\
 & q^{-1}-q & \lambda-\lambda^{-1}\\
 &  &  & \frac{\lambda}{q}-\frac{q}{\lambda}
\end{pmatrix}\;.\label{eq:R_fund_fund}
\end{equation}
In order to check the validity of the RLL relation it is possible
to adopt a brute force method, that is discretise the $\mathscr{P}$
exponential and compute the two sides of the relation, or interpret
$\mathbf{L}_{j}$ and $\mathbf{R}_{j\, j'}$ as particular realisations
of universal objects of the algebra $\mathcal{U}_{q}\left(\widehat{sl(2)}\right)$;
this last approach is sketched in \subref{Baxter-Q-operators}.

We can finally define the ``quantum transfer matrices'' as traces
of the quantum monodromy matrices (\ref{eq:quantum_monodromy}):
\begin{equation}
\mathbf{T}_{j}(\lambda)\doteq\textrm{tr}_{\pi_{j}}\left(\mathbf{M}_{j}(\lambda)\right)\;.\label{eq:quantum_transfer}
\end{equation}
As a direct consequence of the $RLL$ relation, these matrices form
a commuting family:
\[
[\mathbf{T}_{j}(\lambda),\mathbf{T}_{j'}(\lambda')]=0\;,
\]
moreover they commute with the operator $P$ and, as such, act invariantly%
\footnote{An informal way to see this is to notice that the operators $\mathcal{K}(w)$
are traceless; moreover the only terms in $\mathbf{T}_{j}$ having
non-vanishing trace are those containing an equal number of operators
$E$ and $F$. Thus, only products of elements of the type $V_{-}(w)V_{+}(w')$
appear and these act on Fock spaces as $\mathscr{F}_{p}\rightarrow\mathscr{F}_{p+\beta^{2}}\rightarrow\mathscr{F}_{p}$.%
} on each $\mathscr{F}_{p}$. Finally through some tedious computation
\cite{Bazh_Luky_AZam_97-2} it is possible to show that, with the
definition (\ref{eq:quantum_transfer}), the quantum transfer matrices
commute with all the local IMs%
\footnote{Note that the proof is limited to some low order IM; a full proof
of the commutativity is still lacking.%
}:
\[
[\mathbf{T}_{j}(\lambda),\mathbf{I}_{2k-1}]=0\;,
\]
which means that the level subspaces $\mathscr{F}_{p}^{(\ell)}$ are
the eigenspaces of $\mathbf{T}_{j}(\lambda)$.

\begin{figure}[h]
\begin{centering}
\begin{tikzpicture}

\draw[scale=1,domain=0:3,smooth,variable=\x, color=blue,thick,-] plot ({-5.5},{\x});
\draw[scale=1,domain=0:3,smooth,variable=\x, color=green!70!black,thick,-] plot ({\x-7},{1.5});

\node[] at (-8,1.5) {$\mathbf{l}(w_j)\;:$};
\node[] at (-5.5,4) {$w_j$};

\draw[scale=1,domain=0:3,smooth,variable=\x, color=blue,thick,-] plot ({1},{\x});
\draw[scale=1,domain=0:3,smooth,variable=\x, color=blue,thick,-] plot ({2},{\x});
\draw[scale=1,domain=0:3,smooth,variable=\x, color=blue,thick,-] plot ({4},{\x});
\draw[scale=1,domain=0:3,smooth,variable=\x, color=blue,thick,-] plot ({5},{\x});
\draw[scale=1,domain=0:6,smooth,variable=\x, color=green!70!black,thick,-] plot ({\x},{1.5});
\draw[scale=1,domain=0:1,smooth,variable=\x, color=green!70!black,dashed,-] plot ({\x-1},{1.5});
\draw[scale=1,domain=6:7,smooth,variable=\x, color=green!70!black,dashed,-] plot ({\x},{1.5});

\node[] at (-1.5,1.55) {$\mathbf{L}\;:$};
\node[] at (1,4) {$w_N$}; \node[] at (2,4) {$w_{N-1}$};
\node[] at (3,4) {$\cdots$}; \node[] at (3,1) {$\cdots$};
\node[] at (3,2) {$\cdots$}; \node[] at (4,4) {$w_1$};
\node[] at (5,4) {$w_0$};

\end{tikzpicture}\protect\caption{Graphical representation of operators $\mathbf{l}(w_{j})$ and $\mathbf{L}$;
the horizontal green line represents the auxiliary space, while the
vertical blue ones correspond to the quantum spaces.\label{fig:Graphical-Ls}}

\par\end{centering}

\end{figure}
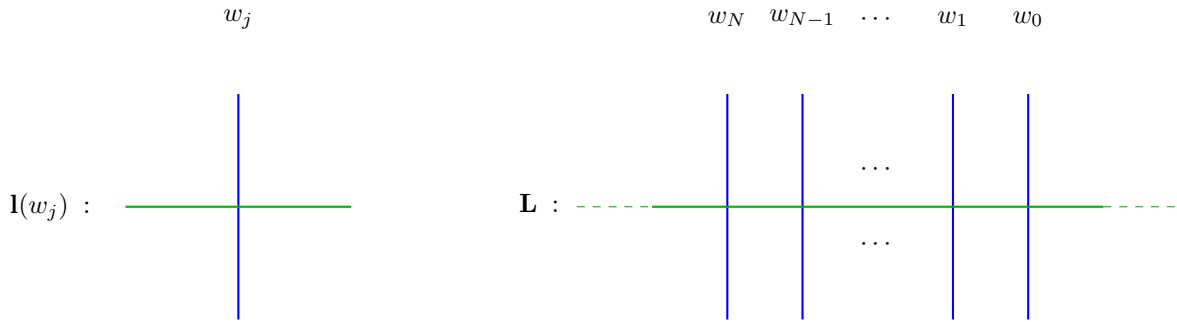

Before explicitly presenting the simple case of $\pi_{j}=\pi_{\frac{1}{2}}$,
we wish to underline the connection with lattice models. Just as in
the classical case, we can express the $\mathbf{L}$-operators as
a continuum limit of a product:
\[
\mathbf{L}(\lambda)=e^{\pi\mathsf{i}PH}\lim_{N\rightarrow\infty}\mathbf{l}(w_{N}\vert\lambda)\mathbf{l}(w_{N-1}\vert\lambda)\cdots\mathbf{l}(w_{0}\vert\lambda)\;.
\]
Here the ``local'' operators $\mathbf{l}$ are expressed as
\[
\mathbf{l}(w\vert\lambda)=\exp\left[\lambda\mathcal{K}(w)\Delta w\right]\sim1+\lambda\mathcal{K}(w)\Delta w\;.
\]
The form of $\mathcal{K}(w)$ is exactly that which we would expect
from a lattice model:
\[
\mathcal{K}(w)=\sum_{j=\pm}V_{j}(w)\omega_{j}\;,
\]
where $\omega_{\pm}$ are generators of the $\mathcal{U}_{q}\left(\widehat{sl(2)}\right)$
algebra in matrix realisation (see \subref{Baxter-Q-operators} for
more details), and $V_{\pm}$ are a vertex operator realisation of
the same algebra. In fact the operators
\[
V_{0}(w)=\sqrt{2}\partial_{w}\varphi(w)\;,\qquad V_{\pm}(w)=:e^{\pm2\varphi(w)}\;,
\]
satisfy the $\widehat{sl(2)}$ subalgebra at level $1$ \cite{DiFr_Math_Sene}.
So we can interpret $\mathbf{l}$ as being the tensor product of two
operators acting on two different spaces: one, corresponding to the
matrices $\omega_{j}$, is the auxiliary space; the other, corresponding
to the vertex operators $V_{j}$, is the quantum space. A pictorial
representation is given in Figure \ref{fig:Graphical-Ls}. We wish
to stress that this connection is by no means mathematically precise,
but rather an intuitive interpretation of the physical meaning of
the operators introduced above.

\begin{framed}%

\paragraph*{The basic representation}

All that has been said until now is rather general and abstract. In
order to make things more concrete, let us concentrate on the simplest
amongst the quantum transfer matrices, namely $\mathbf{T}(\lambda)\equiv\mathbf{T}_{\frac{1}{2}}(\lambda)$.
The $2$-dimensional representation of $\mathcal{U}_{q}\left(sl(2)\right)$
can be chosen as
\[
\pi_{\frac{1}{2}}\left[H\right]=\begin{pmatrix}1 & 0\\
0 & -1
\end{pmatrix}\;,\quad\pi_{\frac{1}{2}}\left[E\right]=\begin{pmatrix}0 & 1\\
0 & 0
\end{pmatrix}\;,\quad\pi_{\frac{1}{2}}\left[F\right]=\begin{pmatrix}0 & 0\\
1 & 0
\end{pmatrix}\;,
\]
and the operator $\mathbf{T}(\lambda)$ can thus be written as a power
series in $\lambda^{2}$ (due to the tracelessness of the matrices
$E$ and $F$):
\begin{equation}
\mathbf{T}(\lambda)=2\cos\left(2\pi P\right)+\sum_{n=1}^{\infty}\lambda^{2n}\mathbf{G}_{2n}\;,\label{eq:nonlocal_int_in_T}
\end{equation}
where
\begin{eqnarray}
\mathbf{G}_{2n} & \doteq & q^{n}\!\!\!\!\!\!\!\!\intop_{\underset{w_{1}\geq\cdots\geq w_{2n}}{0}}^{2\pi}\!\!\!\!\!\!\!\! dw_{1}\cdots dw_{2n}\;\, e^{2\pi\mathsf{i}P}V_{-}(w_{1})V_{+}(w_{2})\cdots V_{-}(w_{2n-1})V_{+}(w_{2n})+\nonumber \\
 &  & \qquad\qquad\qquad+e^{-2\pi\mathsf{i}P}V_{+}(w_{1})V_{-}(w_{2})\cdots V_{+}(w_{2n-1})V_{-}(w_{2n})\;,\label{eq:nonlocal_IMs}
\end{eqnarray}
are operators commuting amongst themselves and with the local IMs
\[
[\mathbf{G}_{2n},\mathbf{G}_{2m}]=0\;,\qquad[\mathbf{G}_{2n},\mathbf{I}_{2k-1}]=0\;,
\]
and, for this reason, are called \textit{non-local integrals of motion}.
Just as their local counterpart, they act invariantly on each level
subspace $\mathscr{F}_{p}$ and, in particular, the highest weight
vector $\left|\, p\,\right\rangle $ is one of their common eigenstates:
\[
\mathbf{G}_{2n}\left|\, p\,\right\rangle =G_{2n}^{(\textrm{vac})}(p)\left|\, p\,\right\rangle \;.
\]
The vacuum eigenvalues can be calculated rather straightforwardly
from the definition (\ref{eq:nonlocal_IMs}):
\begin{eqnarray*}
 &  & G_{2n}^{(\textrm{vac})}(p)=\!\!\!\!\!\!\!\!\!\!\!\!\!\!\!\!\intop_{\underset{w_{1}\geq\tilde{w}_{1}\geq\cdots\geq w_{n}\geq\tilde{w}_{n}}{0}}^{2\pi}\!\!\!\!\!\!\!\!\!\!\!\!\!\!\!\! dw_{1}d\tilde{w}_{1}\cdots dw_{n}d\tilde{w}_{n}\prod_{j>i\geq1}^{n}\left[4\sin\left(\frac{w_{i}-w_{j}}{2}\right)\sin\left(\frac{\tilde{w}_{i}-\tilde{w}_{j}}{2}\right)\right]^{2\beta^{2}}\times\\
 &  & \qquad\qquad\qquad\times\prod_{i,j=1}^{n}\left[2\sin\left(\frac{w_{i}-\tilde{w}_{j}}{2}\right)\right]^{-2\beta^{2}}2\cos\left[2p\left(\pi+\sum_{j=1}^{n}(\tilde{w}_{j}-w_{j})\right)\right]\;,
\end{eqnarray*}
and when $n=1$, this expression greatly simplifies to
\[
G_{2}^{(\textrm{vac})}(p)=\intop_{0}^{2\pi}dw\intop_{0}^{w}d\tilde{w}\frac{2\cos\left(2\pi p+2p(\tilde{w}-w)\right)}{\left[2\sin\left(\frac{w-\tilde{w}}{2}\right)\right]^{2\beta^{2}}}=\frac{4\pi^{2}\Gamma\left(1-\beta^{2}\right)}{\Gamma\left(1-2p-\beta^{2}\right)\Gamma(1+2p-\beta^{2})}\;.
\]
In order to obtain these results one has to use the following property
of vertex operators:
\[
\left\langle p\right|V_{\epsilon}(w)V_{\tilde{\epsilon}}(\tilde{w})\left|p\right\rangle =e^{-2\mathsf{i}p\left(\tilde{\epsilon}w+\epsilon\tilde{w}\right)}\left[2\sin\left(\frac{w-\tilde{w}}{2}\right)\right]^{-2\epsilon\tilde{\epsilon}\beta^{2}}\;,\qquad\epsilon,\tilde{\epsilon}=\pm1\;,
\]
and the Wick theorem.

All the operators $\mathbf{T}_{j}(\lambda)$ are entire functions
of $\lambda^{2}$, possessing an essential singularity at infinity
due to the accumulation of zeroes on the negative $\lambda^{2}$-axis.
This can be shown by comparison with a result obtained in \cite{Fend_Lesa_Sale_95},
where a series similar to (\ref{eq:nonlocal_int_in_T}) was analysed
and shown to converge in the whole complex plane, defining an entire
function with an essential singularity at infinity. As it turns out,
the coefficients of this series are larger in absolute value than
(\ref{eq:nonlocal_IMs}), meaning that $\mathbf{T}$ is entire as
well. Finally the entirety of the operators $\mathbf{T}_{j}$ directly
descends from this result, thanks to the $T$-system we are going
to present just below (\ref{eq:T_system}).

We are interested in obtaining an asymptotic series expansion of $\mathbf{T}(\lambda)$
since, recalling equation (\ref{eq:KdV_asympt_T}), we expect the
integrals of motion to appear there as coefficients. In fact, as pointed
out just above, it is possible to examine the discretised version
of $\mathbf{M}_{\frac{1}{2}}(\lambda)$. Then, by means of standard
Algebraic Bethe Ansatz (ABA), and, subsequently, taking the continuum
limit back to $\mathbf{M}_{\frac{1}{2}}(\lambda)$ one obtains the
following expression for the quantum transfer matrix:
\[
\mathbf{T}(\lambda)=\boldsymbol{\Lambda}(q\lambda)+\boldsymbol{\Lambda}^{-1}(q^{-1}\lambda)\;,
\]
where
\[
\log\boldsymbol{\Lambda}(q\lambda)\underset{\underset{\left|\arg\left(\lambda\right)\right|<\pi}{\left|\lambda\right|\rightarrow\infty}}{\sim}m\lambda^{1+\xi}-\sum_{k=1}^{\infty}C_{k}\mathbf{I}_{2k-1}\lambda^{(1+\xi)(1-2k)}\;,\qquad\xi\doteq\frac{\beta^{2}}{1-\beta^{2}}\;.
\]
The constants in the asymptotic expansion are
\begin{eqnarray}
m & = & 2\sqrt{\pi}\frac{\Gamma\left(\frac{1}{2}-\frac{\xi}{2}\right)}{\Gamma\left(1-\frac{\xi}{2}\right)}\left[\Gamma\left(\frac{1}{1+\xi}\right)\right]^{1+\xi}\;,\label{eq:m-parameter}\\
C_{k} & = & \frac{1+\xi}{k!}\left(\frac{\pi\xi}{1+\xi}\right)^{k}\left(\frac{2}{C_{0}}\frac{\Gamma\left(\frac{1}{2}-\frac{\xi}{2}\right)}{\Gamma\left(1-\frac{\xi}{2}\right)}\right)^{2k-1}\frac{\Gamma\left[(1+\xi)(k-\frac{1}{2})\right]}{\Gamma\left[1+(k-\frac{1}{2})\xi\right]}\;.\label{eq:C-parameter}
\end{eqnarray}
Notice how in this asymptotic expansion, $\lambda$ appears with fractional
powers $1+\xi$; this might seem surprising, until one remembers that
the vertex operators $V_{\pm}(w)$ carry a conformal dimension $h_{V}=\beta^{2}=\frac{\xi}{1+\xi}$.
We can thus think to the spectral parameter $\lambda$ as carrying
an anomalous dimension $\left[\lambda\right]=\left[\textrm{length}\right]^{-\frac{1}{1+\xi}}$.
The explicit computations which yield these results are rather lengthy
and we will not present them here. Nonetheless we encourage the interested
reader to delve into them, starting from the above cited paper \cite{Fend_Lesa_Sale_95}.\end{framed}

\subsection{T-system, Y-system and Thermodynamic Bethe Ansatz equations\label{sub:T-system,-Y-system-and}}

Let us return to the analysis of the quantum monodromy matrices $\mathbf{T}_{j}(\lambda)$
associated highest dimensional representations of $\mathcal{U}_{q}\left(sl(2)\right)$.
It is easily deduced from their definition that them too are power
series in $\lambda^{2}$:
\[
\mathbf{T}_{j}(\lambda)=\frac{\sin\left(2(2j+1)\pi P\right)}{\sin\left(2\pi P\right)}+\sum_{n=1}^{\infty}\lambda^{2n}\mathbf{G}_{2n}^{(j)}\;.
\]
The surprising fact about these expansions is that they are deeply
interrelated; in fact the non-local IMs $\mathbf{G}_{2n}^{(j)}$ with
$j>\frac{1}{2}$ can all be written as polynomials in $\mathbf{G}_{2n}^{(\frac{1}{2})}\equiv\mathbf{G}_{2n}$,
e.g.
\begin{eqnarray}
\mathbf{G}_{2}^{(j)} & = & A_{j}(2\pi P,\pi\beta^{2})\mathbf{G}_{2}\;,\nonumber \\
\mathbf{G}_{4}^{(j)} & = & A_{j}(2\pi P,2\pi\beta^{2})\mathbf{G}_{4}+B_{j}(2\pi P,\pi\beta^{2})\mathbf{G}_{2}\;,\label{eq:higher_Gs}\\
 & \cdots & \;,\nonumber 
\end{eqnarray}
where
\begin{eqnarray*}
A_{j}(a,b) & \doteq & \frac{1}{4\sin a\sin b}\left[\frac{\sin\left[(2j+1)(a-b)\right]}{\sin\left(a-b\right)}-\frac{\sin\left[(2j+1)(a+b)\right]}{\sin\left(a+b\right)}\right]\;,\\
B_{j}(a,b) & \doteq & \frac{1}{16\sin a\sin b\sin2b}\left[\frac{\sin\left[(2j+1)(a-2b)\right]}{\sin\left(a-b\right)\sin\left(a-2b\right)}+\frac{\sin\left[(2j+1)(a+2b)\right]}{\sin\left(a+b\right)\sin\left(a+2b\right)}+\right.\\
 &  & \qquad\qquad\qquad\qquad\left.-2\cos b\frac{\sin\left[(2j+1)a\right]}{\sin\left(a-b\right)\sin\left(a+b\right)}\right]\;.
\end{eqnarray*}
These polynomial relations suggest that there might exist algebraic
relation between quantum transfer matrices belonging to different
representations $\pi_{j}$. This is indeed the case, as the operators
$\mathbf{T}_{j}$ satisfy the following system of finite-difference
functional equations
\begin{equation}
\mathbf{T}_{j}(q^{\frac{1}{2}}\lambda)\mathbf{T}_{j}(q^{-\frac{1}{2}}\lambda)=1+\mathbf{T}_{j+\frac{1}{2}}(\lambda)\mathbf{T}_{j-\frac{1}{2}}(\lambda)\;,\label{eq:T_system}
\end{equation}
known as \textit{$T$-system} or \textit{Hirota bilinear equations}
\cite{Kuni_Naka_Suzu_94,Kric_Lipa_Wieg_Zabr_97}. This system of equations
is a direct consequence of the $RLL$ relation (\ref{eq:RLL_relation})
and can be obtained by using a procedure called $R$-matrix fusion%
\footnote{In fact, while the expressions (\ref{eq:higher_Gs}) can be obtained
by brute force computation, we prefer to consider it as a consequence
of (\ref{eq:T_system}).%
}, well known in lattice theory \cite{Kuli_Resh_Skly_81}. It is worth
noticing that equations (\ref{eq:RLL_relation}), (\ref{eq:T_system})
and the $R$-matrix (\ref{eq:R_fund_fund}) are essentially the same
as the corresponding ones in the integrable $XXZ$ model \cite{Kiri_Resh_87}.
This, clearly, is not just a coincidence as the underlying algebraic
structure of the latter is the same as that of the CFTs we are studying
here; this structure knows nothing about the discrete or continuous
nature of the system and is thus expected that the equations arising
from purely algebraic considerations (such as the $RLL$ relation
above or, as we will see, the $TQ$ equation) have the same structure,
no matter what is the model under study. The information on the different
nature of the models will be contained then in the analytical properties
of the objects involved in these relations. These considerations will
be precious later, when we will extend this setting to massive theories.

For generic values of the central charge $c$, we have an infinite
hierarchy of quantum transfer matrices which, thanks to the system
(\ref{eq:T_system}), can all be expressed in terms of the fundamental
one $\mathbf{T}(\lambda)$:
\begin{eqnarray*}
\mathbf{T}_{1}(\lambda) & = & \mathbf{T}(q^{\frac{1}{2}}\lambda)\mathbf{T}(q^{-\frac{1}{2}}\lambda)-1\;,\\
\mathbf{T}_{\frac{3}{2}}(\lambda) & = & \mathbf{T}(q\lambda)\mathbf{T}(\lambda)\mathbf{T}(q^{-1}\lambda)-\mathbf{T}(q^{\frac{1}{2}}\lambda)-\mathbf{T}(q^{-\frac{1}{2}}\lambda)\;,\\
 & \cdots & \;.
\end{eqnarray*}
With some algebraic effort, we can also recast the $T$-system in
the following form
\begin{equation}
\mathbf{T}(\lambda)\mathbf{T}_{j}(q^{\frac{2j+1}{2}}\lambda)=\mathbf{T}_{j-\frac{1}{2}}(q^{\frac{2j+2}{2}}\lambda)+\mathbf{T}_{j+\frac{1}{2}}(q^{\frac{2j}{2}}\lambda)\;.\label{eq:T_syst_alternative}
\end{equation}
Let us now introduce the $Y$-operators as follows \cite{Klum_Pear_92}:
\[
\mathbf{Y}_{j}(\theta)\doteq\mathbf{T}_{j-\frac{1}{2}}(\lambda)\mathbf{T}_{j+\frac{1}{2}}(\lambda)\;,\qquad\lambda^{1+\xi}=e^{\theta}\;,
\]
with the convention $\mathbf{T}_{0}=1$ and $\mathbf{T}_{-\frac{1}{2}}=0$;
then it is easily showed that they satisfy the $Y$-system equations
\begin{equation}
\mathbf{Y}_{j}^{+}\mathbf{Y}_{j}^{-}=\left(1+\mathbf{Y}_{j+\frac{1}{2}}\right)\left(1+\mathbf{Y}_{j-\frac{1}{2}}\right)\;,\label{eq:Y_system_gen}
\end{equation}
where we have introduced the short-hand notation for shifts: $\mathbf{Y}^{\pm}\doteq\mathbf{Y}(\theta\pm\mathsf{i}\pi\frac{\xi}{2})$.
This last infinite system of finite difference equation can be further
recast in an infinite set of non-linear integral equations, known
as \emph{Thermodynamic Bethe Ansatz equations} whose general form
is the following
\begin{equation}
\epsilon_{j}^{(\ell)}(\theta)=\mathcal{Z}^{(\ell)}(\theta)-\sum_{k}\intop_{-\infty}^{\infty}d\theta'\,\varphi_{j\, k}\left(\theta-\theta'\right)\log\left[1+e^{-\epsilon_{k}^{(\ell)}(\theta')}\right]\;,\label{eq:TBA_gen}
\end{equation}
where the \emph{pseudo-energies} $\epsilon_{j}^{(\ell)}$ are the
logarithms of the $Y$-operators eigenvalues:
\[
\epsilon_{j}^{(\ell)}(\theta)\doteq\log\left[Y_{j}^{(\ell)}(\theta)\right]\;,\qquad\mathbf{Y}_{j}(\theta)\left|\ell\right\rangle =Y_{j}^{(\ell)}(\theta)\left|\ell\right\rangle \;,
\]
and $\ell$ labels the eigenstate under consideration. The function
$\mathcal{Z}^{(\ell)}(\theta)$ is called \emph{driving term} and
depends on the particular eigenstate, while the kernel $\varphi_{j\, k}(\theta)$
only depends on the algebraic structure of the $Y$-system. The procedure
to go from (\ref{eq:Y_system_gen}) to (\ref{eq:TBA_gen}) is intuitively
simple, however one has to take great care to the analytic properties
of the functions involved. More specifically one has to know the asymptotic
behaviour of the $Y$-functions, which will be encoded into the function
$\mathcal{Z}^{(\ell)}$. Moreover the presence of poles and zeroes
in the functions $Y_{j}^{(\ell)}$ might create a great deal of problems.
All these questions are addressed in the \cite{Tong_16} and we recommend
interested readers to refer to that review.

\begin{framed}%

\paragraph*{Truncation and the minimal models $\mathcal{M}_{2,2n+1}$ TBA}

The relations we derived just above, the $T$-system, the $Y$-system
and the TBA equations, although very simple-looking and fancy, still
consists of an infinity of equations for an infinite set of functions;
for this reason, dealing with them is, to use an euphemism, complicated.
However there are situations in which the number of equations and
functions involved reduce to a finite number; this fenomenon is called
\emph{truncation}. The parallel we traced above with the lattice model
helps us identify these cases: it is known that, for some particular
values of the parameters, the $XXZ$ system can be reduced to the
$RSOS$ model \cite{Andr_Baxt_Forr_84} and the $T$-system collapses
to a finite set of equations for a finite number of functions \cite{Baxt_Pear_82,Bazh_Resh_89}.
This phenomenon of truncation in $XXZ$ can be traced back to a purely
algebraic fact: when $q$ is a $N$-th root of unity the $(N+1)$-dimensional
representation $\pi_{\frac{N}{2}}$ of $\mathcal{U}_{q}\left(sl(2)\right)$
becomes reducible, while all the representations with $\frac{1}{2}\leq j<\frac{N}{2}$
remain irreducible. In particular
\[
\pi_{\frac{N}{2}}=\pi_{\frac{N}{2}-1}\oplus\varrho_{N}^{+}\oplus\varrho_{N}^{-}\;,
\]
where $\varrho_{N}^{\pm}$ are two particular one-dimensional representations
such that
\[
\varrho_{N}^{\pm}\left[E\right]=\varrho_{N}^{\pm}\left[F\right]=0\;,\qquad\varrho_{N}^{\pm}\left[H\right]=\pm N\;.
\]
Being purely algebraic and, as such, blind to the particular theory
overlying the algebra structure, we expect the phenomenon of truncation
to happen for CFTs as well. Indeed, considering the decomposition
above, and applying it to the abstract definition \ref{eq:quantum_monodromy}
we immediately obtain that
\[
\mathbf{T}_{\frac{N}{2}}(\lambda)=2\cos\left(2\pi NP\right)+\mathbf{T}_{\frac{N}{2}-1}(\lambda)\;,
\]
which makes (\ref{eq:T_system}) a closed set of equations for the
operators $\left\{ \mathbf{T}_{j}\right\} _{j=0}^{\frac{N}{2}-\frac{1}{2}}$.

In this case too is convenient to introduce the $Y$-operators, with
a slight modification with respect to the general case, due to the
finiteness of the system:
\begin{eqnarray*}
\mathbf{Y}_{j}(\theta) & \doteq & \mathbf{T}_{j-\frac{1}{2}}(\lambda)\mathbf{T}_{j+\frac{1}{2}}(\lambda)\;,\quad j=\frac{1}{2},1,\ldots,\frac{N}{2}-1\;,\\
\mathbf{Y}_{0}(\theta) & \doteq & 0\;,\\
\overline{\mathbf{Y}}(\theta) & \doteq & \mathbf{T}_{\frac{N}{2}-1}\;.
\end{eqnarray*}
The $Y$-system is then immediately seen to be as follows
\begin{eqnarray}
\mathbf{Y}_{j}^{+}\mathbf{Y}_{j}^{-} & = & \left(1+\mathbf{Y}_{j-\frac{1}{2}}\right)\left(1+\mathbf{Y}_{j+\frac{1}{2}}\right)\;,\qquad j=\frac{1}{2},1,\ldots,\frac{N}{2}-\frac{3}{2}\;,\nonumber \\
\mathbf{Y}_{\frac{N}{2}-1}^{+}\mathbf{Y}_{\frac{N}{2}-1}^{-} & = & \left(1+\mathbf{Y}_{\frac{N}{2}-\frac{3}{2}}\right)\left(1+e^{2\pi \mathsf{i}NP}\overline{\mathbf{Y}}\right)\left(1+e^{-2\pi \mathsf{i}NP}\overline{\mathbf{Y}}\right)\;,\label{eq:DN_Y_system}\\
\overline{\mathbf{Y}}^{+}\overline{\mathbf{Y}}^{-} & = & \left(1+\mathbf{Y}_{\frac{N}{2}-1}\right)\;.\nonumber 
\end{eqnarray}
This $Y$-system is called of type $D_{N}$ \cite{AlZa_91-1} as it
can be nicely encoded in the Dynkin diagram of said type. Indeed let
us associate each $Y$-operator with a node of a graph and draw lines
between these whenever the corresponding $Y$s appear in the same
equation. What we obtain is the diagram shown in the picture below:
a Dynkin diagram of type $D_{N}$.

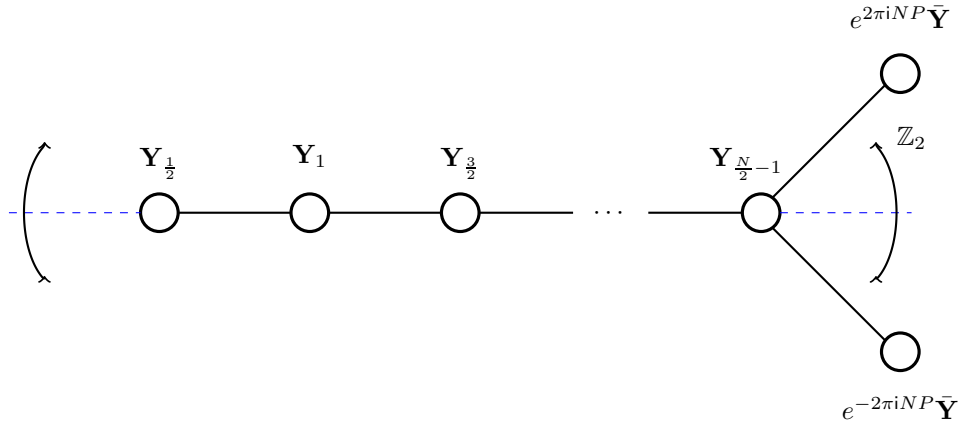
\begin{center}
\begin{tikzpicture}

\node[dynkin node] at (0,0){};
\draw[scale=1,domain=0.25:1.75,smooth,variable=\x, color=black,thick,-] plot ({\x},{0});
\node[dynkin node] at (2,0){};
\draw[scale=1,domain=2.25:3.75,smooth,variable=\x, color=black,thick,-] plot ({\x},{0});
\node[dynkin node] at (4,0){};
\draw[scale=1,domain=4.25:5.5,smooth,variable=\x, color=black,thick,-] plot ({\x},{0});
\node[] at (6,0) {$\cdots$};
\draw[scale=1,domain=6.5:7.75,smooth,variable=\x, color=black,thick,-] plot ({\x},{0});
\node[dynkin node] at (8,0){};
\draw[scale=1,domain=8.25:9.75,smooth,variable=\x, color=black,thick,-] plot ({\x-0.1},{\x-8.05});
\draw[scale=1,domain=8.25:9.75,smooth,variable=\x, color=black,thick,-] plot ({\x-0.1},{-\x+8.05}); \node[dynkin node] at (9.85,1.85){};
\node[dynkin node] at (9.85,-1.85){};

\node[] at (0,0.65) {$\mathbf Y_{\frac{1}{2}}$};
\node[] at (2,0.75) {$\mathbf Y_1$};
\node[] at (4,0.65) {$\mathbf Y_{\frac{3}{2}}$};
\node[] at (7.8,0.65) {$\mathbf Y_{\frac{N}{2}-1}$};
\node[] at (9.85,2.6) {$e^{2\pi\mathsf{i}NP}\bar{\mathbf Y}$};
\node[] at (9.85,-2.6) {$e^{-2\pi\mathsf{i}NP}\bar{\mathbf Y}$};

\draw[scale=1,domain=-2:-0.25,smooth,variable=\x, color=blue,dashed,-] plot ({\x},{0});
\draw[scale=1,domain=8.25:10,smooth,variable=\x, color=blue,dashed,-] plot ({\x},{0});
\draw[scale=1,domain=-1.1:1.1,smooth,variable=\x, color=black,thick,<->] plot ({0.5*cos(\x r)+9.3},{sin(\x r)});
\draw[scale=1,domain=2.04:4.24,smooth,variable=\x, color=black,thick,<->] plot ({0.5*cos(\x r)-1.3},{sin(\x r)});
\node[] at (10,1) {$\mathbb{Z}_2$};

\end{tikzpicture}
\par\end{center}

Note that for $q$ to be a root of unity, we must require $\beta^{2}$
to be a rational number, say $\beta^{2}=\frac{m'}{m}$ and, by virtue
of (\ref{eq:Feigin_Fuchs}), the central charge becomes
\[
c=13-6\left(\beta^{-2}+\beta^{2}\right)=1-6\frac{(m-m')^{2}}{m\, m'}\;,
\]
identifying our CFT as the minimal model $\mathcal{M}_{m,m'}$. Thanks
to this identification we now realize that the truncation of the $T$-system
is a clear reflection of the finite number of primary fields of these
CFTs.

We can get a further simplification of (\ref{eq:DN_Y_system}) by
considering the particular minimal models $\mathcal{M}_{2,2n+3}$
\[
c=1-3\frac{(2n+1)^{2}}{2n+3}\;,\quad\beta^{2}=\frac{2}{2n+3}\;,\quad p_{k}=\frac{2k-2n-3}{2(2n+3)}\;,\quad k=1,\ldots,n+1\;,
\]
for which the Ka\v{c} table is composed of a single line of $2(n+1)$
boxes, symmetric along the middle. This symmetry reflects itself in
the $T$-system and the following relation is valid
\[
\mathbf{T}_{n+\frac{1}{2}-1}(\lambda)=\mathbf{T}_{j}(\lambda)\;,\quad j=0,\frac{1}{2},\ldots,n+\frac{1}{2}\;\Longrightarrow\;\mathbf{T}_{n+\frac{1}{2}}(\lambda)=1\;,
\]
thanks to this which, the $Y$-system simplifies to
\begin{eqnarray*}
\mathbf{Y}_{j}^{+}\mathbf{Y}_{j}^{-} & = & \left(1+\mathbf{Y}_{j-\frac{1}{2}}\right)\left(1+\mathbf{Y}_{j+\frac{1}{2}}\right)\;,\qquad j=\frac{1}{2},1,\ldots,n\\
\mathbf{Y}_{n+\frac{1}{2}-j} & = & \mathbf{Y}_{j}\;,
\end{eqnarray*}
which corresponds to the Dynkin diagram of $A_{2n}$ type \cite{AlZa_91-1,Klas_Melz_92},
depicted in the following figure.

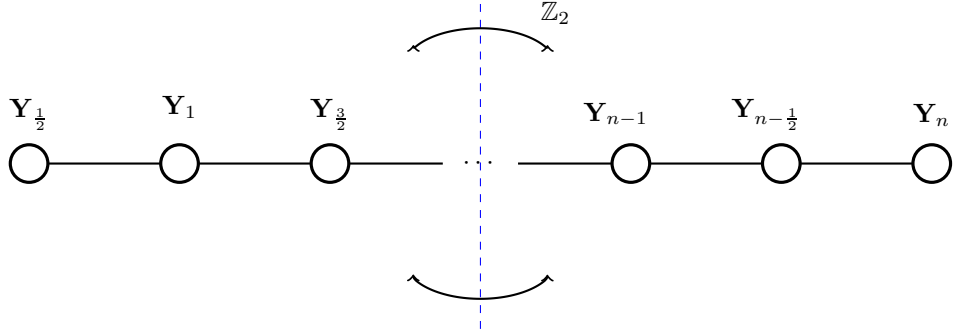
\begin{center}
\begin{tikzpicture}

\node[dynkin node] at (0,0){};
\draw[scale=1,domain=0.25:1.75,smooth,variable=\x, color=black,thick,-] plot ({\x},{0});
\node[dynkin node] at (2,0){};
\draw[scale=1,domain=2.25:3.75,smooth,variable=\x, color=black,thick,-] plot ({\x},{0});
\node[dynkin node] at (4,0){};
\draw[scale=1,domain=4.25:5.5,smooth,variable=\x, color=black,thick,-] plot ({\x},{0});
\node[] at (6,0) {$\cdots$};
\draw[scale=1,domain=6.5:7.75,smooth,variable=\x, color=black,thick,-] plot ({\x},{0});
\node[dynkin node] at (8,0){};
\draw[scale=1,domain=8.25:9.75,smooth,variable=\x, color=black,thick,-] plot ({\x},{0});
\node[dynkin node] at (10,0){};
\draw[scale=1,domain=10.25:11.75,smooth,variable=\x, color=black,thick,-] plot ({\x},{0});
\node[dynkin node] at (12,0){};
\node[] at (0,0.65) {$\mathbf Y_{\frac{1}{2}}$};
\node[] at (2,0.75) {$\mathbf Y_1$};
\node[] at (4,0.65) {$\mathbf Y_{\frac{3}{2}}$};
\node[] at (7.8,0.65) {$\mathbf Y_{n-1}$};
\node[] at (9.8,0.65) {$\mathbf Y_{n-\frac{1}{2}}$};
\node[] at (12,0.65) {$\mathbf Y_n$};
\draw[scale=1,domain=-2.2:2.2,smooth,variable=\x, color=blue,dashed,-] plot ({6},{\x});
\draw[scale=1,domain=0.47:2.67,smooth,variable=\x,color=black,thick,<->] plot ({cos(\x r)+6},{0.5*sin(\x r)+1.3}); \draw[scale=1,domain=3.61:5.81,smooth,variable=\x, color=black,thick,<->] plot ({cos(\x r)+6},{0.5*sin(\x r)-1.3});
\node[] at (7,2) {$\mathbb{Z}_2$};
\end{tikzpicture}
\par\end{center}

Let us now focus on the particular eigenvalues $Y_{j}^{\textrm{gr.st}}(\theta)$
corresponding to the ground state $\left|\, p_{n+1}\,\right\rangle $
(this is the state with lowest $L_{0}$ eigenvalue). We know that
the functions $T_{j}^{\textrm{gr.st.}}(\lambda)$ are entire functions
of $\lambda^{2}$, with asymptotic behaviour
\[
T_{j}^{\textrm{gr.st.}}(\lambda)\underset{\lambda\rightarrow\infty}{\sim}m_{j}\,\lambda^{1+\xi}\;,\qquad m_{j}=\frac{2m}{\pi}\cot\left(\frac{\pi}{2}\xi\right)\sin\left(\pi j\xi\right)\;,
\]
with $m$ given in (\ref{eq:m-parameter}). This information is sufficient
to pass from the $Y$-system to the TBA equations:
\begin{equation}
\varepsilon_{j}(\theta)=\pi m_{j}e^{\theta}-\sum_{j'}\intop_{-\infty}^{\infty}\frac{d\theta'}{2\pi}\varphi_{j\, j'}(\theta-\theta')\log\left[1+e^{-\varepsilon_{j'}(\theta')}\right]\;,\label{eq:massless_TBA}
\end{equation}
where we introduced the \textit{pseudo-energies} $\varepsilon_{j}$
as
\[
Y_{j}^{\textrm{gr.st.}}(\theta)=e^{\epsilon_{j}(\theta)}\;.
\]
The kernel $\varphi_{j\, j'}(\theta)$ is defined from the equation
\[
\left(1-\varphi\right)^{-1}=1-s\,\hat{I}\;,\Rightarrow\;\varphi_{j\, k}(\theta)=-s(\theta)\,\hat{I}_{j\, k}+\sum_{i}\hat{I}_{j\, i}\intop_{-\infty}^{\infty}s(\theta'-\theta)\varphi_{i\, k}(\theta')d\theta'\;,
\]
with $\hat{I}_{j\, k}$ being the incidence matrix of $A_{2n}$ and
$s(\theta)=\frac{1}{\xi\cosh\left(\theta/\xi\right)}$ the inverse
of the shift operator: $s^{-1}\;:\; f\rightarrow f^{+}+f^{-}$. Taking
the Fourier transform of the above relation, with some effort, is
possible to show that the kernel $\varphi$ can be expressed as the
logarithmic derivative of the ``massless S-matrix'' $S_{j\, j'}(\theta)$
\cite{Klas_Melz_92}
\[
\varphi_{j\, j'}(\theta)\doteq-\mathsf{i}\partial_{\theta}\log\left(S_{j\, j'}(\theta)\right)\;,
\]
whose explicit form is known:
\[
S_{j\, j'}(\theta)=F_{j+j'}(\theta)F_{\left|j-j'\right|}(\theta)\!\!\!\!\!\!\prod_{k=1}^{2\min(j,j')-1}\!\!\!\!\!\! F_{\left|j-j'\right|+k}^{2}(\theta)\;,\quad F_{j}(\theta)\doteq\frac{\sinh\theta+\mathsf{i}\sin\left(\pi j\xi\right)}{\sinh\theta-\mathsf{i}\sin\left(\pi j\xi\right)}\;.
\]
Notice that this matrix appears directly from the algebraic properties
of the truncated $Y$-system. It is a consequence of the internal
consistency of this setting that $S_{j\, j'}$ happens to be exactly
the two-particle element of the factorisable scattering matrix proposed
in \cite{Freu_Klass_Melz_89} for the $S$-matrix\footnotemark description
of minimal models of the type $\mathcal{M}_{2,2n+1}$. This little
``miracle'' gives us a strong confirmation of the correctness of
the BLZ approach.\end{framed}\footnotetext{More information on the -matrix approach to integrable models can be found in \cite{Bomb_16}.}

\subsection{Baxter Q-operators\label{sub:Baxter-Q-operators}}

The construction of the $Q$-operators follows very closely that of
the $T$-operators presented above. Like these last they are defined
as traces of some particular monodromy matrix built out of vertex
operators and the generators of some algebra. The difference between
the two stands exactly in the choice of the algebra. For the construction
of $Q$-operators it turns out that we need the quantum oscillator
algebra $\textrm{osc}_{q}$ generated by $\left\{ \mathcal{H},\mathcal{E}_{+},\mathcal{E}_{-}\right\} $
with commutation relations
\[
[\mathcal{H},\mathcal{E}_{\pm}]=\pm2\mathcal{E}_{\pm}\;,\qquad q\mathcal{E}_{+}\mathcal{E}_{-}-q^{-1}\mathcal{E}_{-}\mathcal{E}_{+}=\frac{1}{q-q^{-1}}\;.
\]
The appearance of this algebra might seems strange as, at first sight,
it does not seem to be related to the $sl(2)$ algebraic structure
we have been using to construct everything else. Truth is, $\textrm{osc}_{q}$
and $sl(2)$ really are intimately related and the following in-depth
box explains this relation. We encourage the reader not familiar with
this fact to go through this explanation to better understand the
profound relation between the $T$- and $Q$-operators.

\begin{framed}%

\paragraph*{Quantum affine $sl(2)$ and universal operators}

In many cases, the right way to delve deeper in the core of a theory
is to generalise the mathematical setting; this not only opens the
way for further achievements but almost always cleans up the table
and bring about a great simplification of the structures: complicating
to clarify. It turns out that the most natural starting point for
the construction of the $L$- and $T$-operators is a slight generalisation
of the algebra $\mathcal{U}_{q}\big(su(2)\big)$: the quantum Ka\v{c}-Moody
affine algebra $\mathcal{U}_{q}\left(\widehat{sl(2)}\right)$. Using
this as a starting point we will obtain in one fell swoop a natural
description of both $T$- and $Q$-operators, displaying explicitly
their deep connection, as well as a setting in which the algebraic
relations introduced in the previous section can be easily demonstrated.
Let us thus introduce the algebra $\mathcal{U}_{q}\left(\widehat{sl(2)}\right)$:
it generated by the six elements $\left\{ x_{i},y_{i},h_{i}\right\} _{i=0}^{1}$
which satisfy the commutation relations 
\begin{eqnarray*}
[h_{i},x_{j}]=-a_{i\, j}x_{j}\;, & [h_{i},h_{j}]=0\qquad\qquad\, & ,\; i,j=0,1\;,\\
{}[h_{i},y_{j}]=a_{i\, j}y_{j}\,\,\;\;, & [y_{i},x_{j}]=\delta_{i\, j}\frac{q^{h_{i}}-q^{h_{j}}}{q-q^{-1}} & ,\; i,j=0,1\;,
\end{eqnarray*}
and the quantum Serre relations
\begin{eqnarray*}
x_{i}^{3}x_{j}-[3]_{q}x_{i}^{2}x_{j}x_{i}+[3]_{q}x_{i}x_{j}x_{i}^{2}-x_{j}x_{i}^{3}=0\;,\\
y_{i}^{3}y_{j}-[3]_{q}y_{i}^{2}y_{j}y_{i}+[3]_{q}y_{i}y_{j}y_{i}^{2}-y_{j}y_{i}^{3}=0\quad,
\end{eqnarray*}
where we define the $q$-numbers as
\[
[n]_{q}\doteq\frac{q^{n}-q^{-n}}{q-q^{-1}}\;,\qquad[n]_{q}\underset{q\rightarrow1}{\longrightarrow}n\;.
\]
The matrix $a_{i\, j}$ is the Cartan matrix of the affine algebra
$\widehat{su(2)}$:
\[
a_{i\, j}=\begin{pmatrix}2 & -2\\
-2 & 2
\end{pmatrix}_{i\, j}\;.
\]
In order to be consistently defined this algebra necessitate the further
introduction of the grade operator $d$
\begin{eqnarray*}
[d,h_{0}]=[d,h_{1}]=0\;, & [d,x_{1}]=x_{1}\;\;,\\
{}[d,x_{0}]=[d,y_{0}]=0\;, & [d,y_{1}]=-y_{1}\;,
\end{eqnarray*}
and the central charge (obviously \uline{not} the same thing as
the central charge $c$ of the CFT!) $k=h_{0}+h_{1}$. The algebra
thus defined is a \textit{quasitriangular Hopf algebra} \cite{Fuch,Drin_87,Loeb_16}
whose co-multiplication
\[
\Delta\;:\quad\mathcal{U}_{q}\left(\widehat{sl(2)}\right)\longrightarrow\mathcal{U}_{q}\left(\widehat{sl(2)}\right)\otimes\mathcal{U}_{q}\left(\widehat{sl(2)}\right)\;,
\]
is defined by its action on the generators:
\begin{eqnarray*}
 &  & \Delta(x_{i})=x_{i}\otimes1+q^{-h_{i}}\otimes x_{i}\;,\quad\Delta(y_{i})=y_{i}\otimes q^{h_{i}}+1\otimes y_{i}\;,\\
 &  & \Delta(h_{i})=h_{i}\otimes1+1\otimes h_{i}\;,\quad\quad\;\,\Delta(d)=d\otimes1+1\otimes d\;.
\end{eqnarray*}
There exists a second possible choice for comultiplication:
\[
\Delta'\doteq\sigma\circ\Delta\;,\qquad\sigma\left(A\otimes B\right)=B\otimes A\;,\;\forall A,B\in\mathcal{U}_{q}\left(\widehat{sl(2)}\right)\;,
\]
and the property of quasitriangularity makes sure \cite{Khor_Tols_92}
there exists an object called universal $R$-matrix
\[
\mathcal{R}\in\mathcal{B}_{+}\otimes\mathcal{B}_{-}\;,
\]
intertwining between these two co-multiplications
\begin{equation}
\Delta'(A)\mathcal{R}=\mathcal{R}\Delta(A)\;,\qquad\forall A\in\mathcal{U}_{q}\left(\widehat{sl(2)}\right)\;.\label{eq:R_Intertwining}
\end{equation}
Here $\mathcal{B}_{+}$ and $\mathcal{B}_{-}$ are the Borel subalgebras
of $\mathcal{U}_{q}\left(\widehat{sl(2)}\right)$ generated by the
elements $\left\{ h_{0},h_{1},y_{0},y_{1}\right\} $ and $\left\{ h_{0},h_{1},x_{0},x_{1}\right\} $,
respectively. This universal $R$-matrix satisfies the Yang-Baxter
equation
\[
\mathcal{R}^{1\,2}\mathcal{R}^{1\,3}\mathcal{R}^{2\,3}=\mathcal{R}^{2\,3}\mathcal{R}^{1\,3}\mathcal{R}^{1\,2}\;,\qquad\begin{cases}
\mathcal{R}^{1\,2}=\mathcal{R}\otimes1\\
\mathcal{R}^{2\,3}=1\otimes\mathcal{R}\\
\mathcal{R}^{1\,3}=\left(\sigma\otimes1\right)\mathcal{R}^{2\,3}
\end{cases}\;,
\]
as is obtained straightforwardly from the action of the comultiplication
\[
\left(\Delta\otimes1\right)\mathcal{R}=\mathcal{R}^{1\,3}\mathcal{R}^{2\,3}\;,\qquad\left(1\otimes\Delta\right)\mathcal{R}=\mathcal{R}^{1\,3}\mathcal{R}^{1\,2}\;,
\]
and the relation (\ref{eq:R_Intertwining}). An explicit form of the
universal $R$-matrix for $\mathcal{U}_{q}\left(\widehat{sl(2)}\right)$
can be found in \cite{Khor_Tols_92-1}.

Using the generators introduced above, we can build the following
abstract operator
\[
\mathcal{L}\doteq e^{\mathsf{i}\pi Ph}\mathscr{P}\exp\left[\intop_{0}^{2\pi}\mathcal{K}(w)dw\right]\;,\qquad\mathcal{K}(w)\doteq V_{-}(w)y_{0}+V_{+}(w)y_{1}\;,
\]
where the operators $P$ and $V_{\pm}$ are the same ones we defined
in Sect. \ref{sub:The-quantum-monodromy} and we set $h_{0}=-h_{1}=h$,
which corresponds to choosing the central charge $k$ to be zero.
It is evident that $\mathcal{L}\in\mathcal{B}_{+}$ and, with some
simple computation, one can show that
\[
\Delta\left(\mathcal{L}\right)=\left(\mathcal{L}\otimes1\right)\left(1\otimes\mathcal{L}\right)\;,\qquad\Delta'\left(\mathcal{L}\right)=\left(1\otimes\mathcal{L}\right)\left(\mathcal{L}\otimes1\right)\;.
\]
For this reason, the $RLL$ relation follows automatically from the
definition of $\mathcal{L}$ and $\mathcal{R}$
\begin{equation}
\mathcal{R}\left(\mathcal{L}\otimes1\right)\left(1\otimes\mathcal{L}\right)=\left(1\otimes\mathcal{L}\right)\left(\mathcal{L}\otimes1\right)\mathcal{R}\;.\label{eq:RLL_univ}
\end{equation}

This is an extremely important relation; being completely abstract,
contains in itself the $RLL$ relation for any possible representation
of $\mathcal{U}_{q}\left(\widehat{sl(2)}\right)$. In particular if
we are able to map $\mathcal{L}$ into $\mathbf{L}_{j}$, then the
(\ref{eq:RLL_relation}) will be automatically demonstrated: we need
to find an homomorphism between $\mathcal{U}_{q}\left(\widehat{sl(2)}\right)$
and $\mathcal{U}_{q}\left(sl(2)\right)$. More precisely we want a
whole family of homomorphisms, since we need to introduce the spectral
parameter $\lambda$ which, in this abstract setting, is absent. These
homomorphisms actually exist and are named \textit{evaluation representations}:
\[
\textrm{ev}_{\lambda}\;:\;\mathcal{U}_{q}\left(\widehat{sl(2)}\right)\rightarrow\mathcal{U}_{q}\left(sl(2)\right)\;,\quad\begin{cases}
x_{0}\mapsto\lambda^{-1}F\, q^{-\frac{H}{2}}\\
y_{0}\mapsto\lambda q^{\frac{H}{2}}E\\
h_{0}\mapsto H
\end{cases}\;,\;\begin{cases}
x_{1}\mapsto\lambda^{-1}E\, q^{\frac{H}{2}}\\
y_{1}\mapsto\lambda q^{-\frac{H}{2}}F\\
h_{1}\mapsto-H
\end{cases}\;,
\]
where $\left\{ E,F,H\right\} $ are the usual generators of $\mathcal{U}_{q}\left(sl(2)\right)$.
It is immediate to verify that
\[
\pi_{j}(\lambda)\left[\mathcal{L}\right]=\mathbf{L}_{j}(\lambda)\;,\qquad\pi_{j}(\lambda)\doteq\pi_{j}\circ\textrm{ev}_{\lambda}\;,
\]
while it is less obvious but still verifiable that, starting from
the general definition of \cite{Khor_Tols_92-1}, we obtain 
\[
\left(\pi_{j}(\lambda)\otimes\pi_{j'}(\lambda')\right)\left[\mathcal{R}\right]=\rho_{j\, j'}\left(\lambda/\lambda'\right)\mathbf{R}_{j\, j'}\left(\lambda/\lambda'\right)\;,
\]
with $\rho_{j\, j'}(\lambda)$ being an uninteresting scalar factor.
In this light the operators $\mathbf{T}_{j}(\lambda)$ are nothing
but a specific representation of the following more general ``universal
T-operator''
\begin{equation}
\mathcal{T}\doteq\textit{tr}_{\mathcal{U}_{q}\left(\widehat{sl(2)}\right)}\left[e^{\pi\mathsf{i}Ph}\mathcal{L}\right]\;,\label{eq:abstract_T}
\end{equation}
and one can think of defining new operators from $\mathcal{T}$, by
choosing different representations of $\mathcal{U}_{q}\left(\widehat{sl(2)}\right)$.
Any two of these operators will commute amongst themselves, as a direct
consequence of the universal $RLL$ relation (\ref{eq:RLL_univ}),
and this is precisely a property that we want for our $Q$-operators.
However with the family $\pi_{j}(\lambda)$, we have exhausted all
the finite dimensional evaluation representations: we need to look
elsewhere. As mentioned above, the correct choice of algebra for the
construction of $Q$-operators happens to be $\textrm{osc}_{q}$.
Even though it might look rather different from $\mathcal{U}_{q}\left(\widehat{sl(2)}\right)$,
it is an easy exercise to show that the two following maps 
\[
\omega_{\lambda}^{\pm}\;:\;\mathcal{U}_{q}\left(\widehat{sl(2)}\right)\rightarrow\textrm{osc}_{q}\;,\qquad\begin{cases}
h\mapsto\pm\mathcal{H}\\
y_{0}\mapsto\lambda\mathcal{E}_{\pm}\\
y_{1}\mapsto\lambda\mathcal{E}_{\mp}
\end{cases}\;,
\]
are homomorphisms of $\mathcal{B}_{+}$ into the quantum oscillator
algebra $\textrm{osc}_{q}$.\end{framed}

Consider now any representation $\rho$ of $\textrm{osc}_{q}$, such
that the following object
\[
Z(p)\doteq\textit{tr}_{\rho}\left[e^{2\pi\mathsf{i}p\mathcal{H}}\right]\;,
\]
exists and do not vanish for $\Im(p)<0$. Then we can construct the
following operators
\[
\mathbf{L}_{\pm}(\lambda)\doteq\rho_{\pm}(\lambda)\left[\mathcal{L}\right]\equiv\rho\left\{ e^{\pm\pi\mathsf{i}P\mathcal{H}}\mathscr{P}\exp\left[\lambda\intop_{0}^{2\pi}dw\left(V_{-}(w)q^{\pm\frac{\mathcal{H}}{2}}\mathcal{E}_{\pm}+V_{+}(w)q^{\mp\frac{\mathcal{H}}{2}}\mathcal{E}_{\mp}\right)\right]\right\} \;,
\]
where
\[
\rho_{\pm}(\lambda)\doteq\rho\circ\omega_{\lambda}^{\pm}\;,
\]
and the corresponding realisation of $\mathcal{T}$%
\footnote{Note that in this case we need to add a regularising factor to the
trace. This is due to infinite dimensionality of the algebra $\textrm{osc}_{q}$.%
}
\[
\mathbf{A}_{\pm}(\lambda)\doteq Z^{-1}(\pm P)\textit{tr}_{\rho}\left[e^{\pm\pi\mathsf{i}P\mathcal{H}}\mathbf{L}_{\pm}(\lambda)\right].
\]
Notice how, allowing analytic continuation from the $\Im(p)<0$ half-plane,
these operators enjoy the symmetry
\[
\mathbf{A}_{\pm}(\lambda)\underset{\left(p,\varphi(w)\right)\rightarrow\left(-p,-\varphi(w)\right)}{\longrightarrow}\mathbf{A}_{\mp}(\lambda)\;.
\]
These operators, as much as the other we introduced, have to be understood
as power series in $\lambda^{2}$ (since, here too, the odd terms
vanish under the trace)
\begin{equation}
\mathbf{A}_{\pm}(\lambda)=1+\sum_{n=1}^{\infty}\sum_{\underset{\sum_{i}\sigma_{i}=0}{\left\{ \sigma_{i}=\pm1\right\} _{i=1}^{2n}}}\lambda^{2n}a_{2n}\left(\sigma_{1},\ldots,\sigma_{2n}\vert\pm P\right)J_{2n}\left(\mp\sigma_{1},\mp\sigma_{2},\ldots,\mp\sigma_{2n}\right)\;,\label{eq:A_exp_zero}
\end{equation}
where we introduced the two functions
\begin{eqnarray*}
 &  & J_{2n}(\sigma_{1},\ldots,\sigma_{2n})\doteq q^{n}\intop_{\underset{w_{1}\geq\cdots\geq w_{2n}}{0}}^{2\pi}dw_{1}\cdots dw_{2n}V_{\sigma_{1}}(w_{1})\cdots V_{\sigma_{2n}}(w_{2n})\;,\\
 &  & a_{2n}(\sigma_{1},\ldots,\sigma_{2n}\vert P)\doteq Z^{-1}(P)\textit{tr}_{\rho}\left[e^{2\pi\mathsf{i}P\mathcal{H}}\mathcal{E}_{\sigma_{1}}\cdots\mathcal{E}_{\sigma_{2n}}\right]\;.
\end{eqnarray*}
The really interesting fact about this decomposition is that the dependance
on $\mathcal{U}_{q}\left(\widehat{sl(2)}\right)$ is contained entirely
in the coefficients $a_{2n}$ which turn out to be uniquely determined
by the commutation relations of the generators and the cyclic properties
of the trace; as such these do \uline{not} depend on the chosen
representation $\rho$ of $\textrm{osc}_{q}$! Clearly the coefficients
$J_{2n}$ are closely related to the non-local integrals of motion
$\mathbf{G}_{2n}$, however a more neat expansion of the operators
$\mathbf{A}_{\pm}$ is the following
\begin{equation}
\log\left(\mathbf{A}_{\pm}(\lambda)\right)=-\sum_{n=1}^{\infty}y^{2n}\mathbf{H}_{2n}\;,\qquad y\doteq\frac{\Gamma\left(1-\beta^{2}\right)}{\beta^{2}}\lambda\;,\label{eq:A_in_lambda_square}
\end{equation}
where the $\mathbf{H}_{2n}$ are a set of non-local integrals of motion
alternative to the $\mathbf{G}_{2n}$ ones and, obviously, algebraically
related to these, e.g.
\begin{equation}
\mathbf{H}_{1}=\frac{\beta^{4}\Gamma\left(\beta^{2}\right)}{4\pi\Gamma\left(1-\beta^{2}\right)\sin\left(2\pi P+\pi\beta^{2}\right)}\mathbf{G}_{1}\;.\label{eq:H1_in_G1}
\end{equation}

Finally we can introduce the Baxter $Q$ operators as
\[
\mathbf{Q}_{\pm}(\lambda)\doteq\lambda^{\pm2\frac{P}{\beta^{2}}}\mathbf{A}_{\pm}(\lambda)\;.
\]
Just as the operators $\mathbf{T}_{j}(\lambda)$, they act invariantly
on $\mathscr{F}_{p}$, which is an immediate consequence of the representation
(\ref{eq:A_exp_zero}) of $\mathbf{A}_{\pm}$. Below is a list of
the properties of $Q$ operators, which descend from from their definition
as representations of $\mathcal{T}$ (\ref{eq:abstract_T}) and from
the structure of the representations involved:
\begin{enumerate}
\item they commute amongst themselves and with all the $T$-operators
\[
[\mathbf{Q}_{\pm}(\lambda),\mathbf{Q}_{\pm}(\lambda')]=[\mathbf{Q}_{\pm}(\lambda),\mathbf{Q}_{\mp}(\lambda')]=[\mathbf{Q}_{\pm}(\lambda),\mathbf{T}_{j}(\lambda')]=0\;.
\]
Consequently they commute with all the IMs, local and nonlocal
\[
[\mathbf{Q}_{\pm}(\lambda),\mathbf{I}_{2k-1}]=[\mathbf{Q}_{\pm}(\lambda),\mathbf{G}_{2n}]=[\mathbf{Q}_{\pm}(\lambda),\mathbf{H}_{2n}]=0\;;
\]

\item they satisfy the Baxter $T-Q$ relation
\[
\mathbf{T}(\lambda)\mathbf{Q}_{\pm}(\lambda)=\mathbf{Q}_{\pm}(q\lambda)+\mathbf{Q}_{\pm}(q^{-1}\lambda)\;.
\]
This relation is a second order finite-difference equation whose ``potential''
$\mathbf{T}(\lambda)$ is a periodic function of $\log\left(\lambda^{2}\right)$;
for this reason the two solutions $\mathbf{Q}_{+}$ and $\mathbf{Q}_{-}$
can be interpreted as Bloch wave solutions to the $T-Q$ relation.
\item they satisfy the quantum wronskian relation
\begin{equation}
\mathbf{Q}_{+}(q^{\frac{1}{2}}\lambda)\mathbf{Q}_{-}(q^{-\frac{1}{2}}\lambda)-\mathbf{Q}_{+}(q^{-\frac{1}{2}}\lambda)\mathbf{Q}_{-}(q^{\frac{1}{2}}\lambda)=2\mathsf{i}\sin\left(2\pi P\right)\;.\label{eq:Quantum_Wronskian}
\end{equation}
This relation guarantees the independence of the functions $\mathbf{Q}_{+}$
and $\mathbf{Q}_{-}$, solutions to the $T-Q$ relation
\item Wronskian expression of $\mathbf{T}_{j}$
\begin{equation}
2\mathsf{i}\sin\left(2\pi P\right)\mathbf{T}_{j}(\lambda)=\mathbf{Q}_{+}(q^{j+\frac{1}{2}}\lambda)\mathbf{Q}_{-}(q^{-j-\frac{1}{2}}\lambda)-\mathbf{Q}_{+}(q^{-j-\frac{1}{2}}\lambda)\mathbf{Q}_{-}(q^{j+\frac{1}{2}}\lambda)\;.\label{eq:T_Wronskian}
\end{equation}

\end{enumerate}
\begin{framed}%

\paragraph*{The asymptotic of $\mathbf{A}_{\pm}$: a simple case}

A particularly simple situation is when $2p=N$, for some integer
$N$; then the quantum wronskian (\ref{eq:Quantum_Wronskian}) vanishes,
meaning that the $Q$-functions $\mathbf{Q}_{+}$ and $\mathbf{Q}_{-}$
are linearly dependent. In particular rather simple direct computation
shows that $\mathbf{A}_{+}(\lambda)\Big\vert_{2p=N}=\mathbf{A}_{-}(\lambda)\Big\vert_{2p=N}=\mathbf{A}^{(N)}(\lambda)$,
where this last operator can be written as
\[
\mathbf{A}^{(N)}(\lambda)\doteq\sum_{n=0}^{\infty}\frac{\varkappa^{2n}}{(n!)^{2}}q^{n}\mathscr{P}_{w}\left.\left\{ \left[\intop_{0}^{2\pi}\frac{dw}{2\pi}V_{+}(w)\right]^{n}\left[\intop_{0}^{2\pi}\frac{dw}{2\pi}V_{-}(w)\right]^{n}\right\} \right|_{2p=N}\;,
\]
with
\[
\varkappa\doteq-\mathsf{i}\frac{\pi\lambda}{\sin\left(\pi\beta^{2}\right)}\;,
\]
and the symbol $\mathscr{P}_{w}$ orders the factors from left to
right in decreasing $w$ order. The vacuum eigenvalue of this operator
$\mathbf{A}^{(N)}(\lambda)\left|\, p=\frac{N}{2}\,\right\rangle =A_{N}^{(\textrm{vac})}(\lambda)\left|\, p=\frac{N}{2}\,\right\rangle $
coincides with the one-dimensional Coulomb gas partition function
\begin{eqnarray*}
 &  & A_{N}^{(\textrm{vac})}(\lambda)\equiv\mathscr{Z}_{N}(\varkappa)\doteq\sum_{n=0}^{\infty}\frac{\varkappa^{2n}}{(n!)^{2}}\intop_{0}^{2\pi}\left(\prod_{\ell=1}^{n}\frac{dw_{\ell}d\tilde{w}_{\ell}}{4\pi^{2}}\right)e^{\mathsf{i}N\sum_{\ell=1}^{n}(\tilde{w}_{\ell}-w_{\ell})}\times\\
 &  & \qquad\qquad\qquad\qquad\times\frac{\prod_{i\neq j}\left|4\sin\left(\frac{w_{i}-w_{j}}{2}\right)\sin\left(\frac{\tilde{w}_{i}-\tilde{w}_{j}}{2}\right)\right|^{2\beta^{2}}}{\prod_{i,j}\left|2\sin\left(\frac{w_{i}-\tilde{w}_{j}}{2}\right)\right|^{2\beta^{2}}}\;,
\end{eqnarray*}
which can be shown \cite{Fend_Lesa_Sale_95} to define an entire function
of $\varkappa^{2}$, with asymptotic behaviour
\[
\log\left(\mathscr{Z}_{N}(\varkappa)\right)\underset{\varkappa^{2}\rightarrow\infty}{\sim}\varkappa^{\frac{1}{1-\beta^{2}}}\;.
\]
In fact with some effort one can show that the same beaviour is valid
for all the eigenvalues of $\mathbf{A}^{(N)}(\lambda)$, meaning that
\begin{equation}
\log\left(\mathbf{A}_{\pm}(\lambda)\right)\Big\vert_{2p=N}\;\underset{\lambda^{2}\rightarrow-\infty}{\sim}M\left(-\lambda^{2}\right)^{\frac{1}{2-2\beta^{2}}}\;.\label{eq:A_asymptotic}
\end{equation}
Although the entirety in $\lambda^{2}$ of operators $\mathbf{A}_{\pm}(\lambda)$
can be demonstrated in general, the above asymptotic behavior is explicitly
demonstrated only for $2p=N\in\mathbb{Z}$. Nonetheless it is reasonable
to assume that this behaviour is valid for any value of $p$ as we
shall do.\end{framed}

\subsection{Bethe ansatz and non-linear integral equation\label{sub:Destri-de-Vega-equation}}

In this section we will concentrate on the eigenvalue $Q(\lambda)\equiv Q_{+}^{\alpha}(\lambda)$
of $\mathbf{Q}_{+}(\lambda)$ on the state $\left|\,\alpha\,\right\rangle \in\mathscr{F}_{p}$;
similar considerations can be obtained for the eigenvalues of $\mathbf{Q}_{-}(\lambda)$.

Let us denote $T(\lambda)$ and $A(\lambda)$ the eigenvalues of $\mathbf{T}(\lambda)$
and $\mathbf{A}_{+}(\lambda)$, respectively, on the state $\left|\,\alpha\,\right\rangle $.
Then the following two equations descend directly from Baxter $T-Q$
relation
\begin{eqnarray*}
 &  & T(\lambda)Q(\lambda)=Q(q\lambda)+Q(q^{-1}\lambda)\;,\\
 &  & T(\lambda)A(\lambda)=e^{2\pi\mathsf{i}p}A(q\lambda)+e^{-2\pi ip}A(q^{-1}\lambda)\;.
\end{eqnarray*}
As we will shortly see, provided the analytic properties of the functions
$A$ and $T$, these equations impose severe restrictions on the allowed
solutions. Let us recall the properties of $A$ and $T$ we agree
on
\begin{itemize}
\item Analyticity: both functions $A(\lambda)$ and $T(\lambda)$ are entire
in $\lambda^{2}\in\mathbb{C}$;
\item Asymptotic behaviour:
\begin{eqnarray*}
 &  & A(\lambda)\underset{\lambda^{2}\rightarrow-\infty}{\sim}\exp\left[M\left(-\lambda^{2}\right)^{\frac{1}{2-2\beta^{2}}}\right]\;,\\
 &  & T(\lambda)\underset{\underset{\left|\arg\left(\lambda^{2}\right)\right|<\pi}{\left|\lambda^{2}\right|\rightarrow\infty}}{\sim}\exp\left[m\left(\lambda^{2}\right)^{\frac{1}{2-2\beta^{2}}}\right]\;,
\end{eqnarray*}
where $m$ is given by the formula (\ref{eq:m-parameter}) and $M$
is presented below in (\ref{eq:M-parameter}).
\item Location of zeroes%
\footnote{We will not make use of the knowledge about the zeroes of $T(\lambda)$.%
}: the zeroes $\left\{ \lambda_{k}^{2}\right\} _{k=0}^{\infty}$ of
$A(\lambda)$ are either real or pairs of complex conjugates. For
any eigenvalue $A(\lambda)$, the number of zeroes on the positive
real $\lambda^{2}$-axis accumulate towards $+\infty$, while the
number of other zeroes remains finite. For the vacuum, if $2p>-\beta^{2}$,
the only zeroes are those on the positive real $\lambda^{2}$-axis.
We avoid those values of the highest weight (e.g. $2p=-\beta^{2}-n$)
for which $\lambda_{0}^{2}=0$.
\end{itemize}
These properties allow us to use Hadamard factorisation theorem: if
$0<\beta^{2}<\frac{1}{2}$ then the asymptotic behaviour of $A(\lambda)$
tells us that its order $\rho_{A}$, as a function $\lambda^{2}$,
is $\frac{1}{2}<\rho_{A}<1$, meaning that we can write the very simple
product
\[
A(\lambda)=\prod_{k=0}^{\infty}\left(1-\frac{\lambda^{2}}{\lambda_{k}^{2}}\right)\;,\qquad A(0)=1\;.
\]

Now, let us take the ``$T-A$ relation'', which we can write as
\[
e^{2\pi\mathsf{i}p}T(\lambda)\frac{A(\lambda)}{A(q^{-1}\lambda)}=\mathfrak{a}(\lambda)+1\;,\qquad\mathfrak{a}(\lambda)\doteq e^{4\pi\mathsf{i}p}\frac{A(q\lambda)}{A(q^{-1}\lambda)}\;,
\]
and evauate it at $\lambda^{2}=\lambda_{k}^{2}$, recalling that $T(\lambda)$
is devoid of singularities at finite $\lambda^{2}$; what we obtain
is an infinite set of coupled algebraic equations of Bethe ansatz-type%
\footnote{For more information on Bethe Ansatz method of solution for integrable
models see \cite{Levk_16}%
}
\[
\mathfrak{a}(\lambda_{k})=-1\quad\Longrightarrow\quad\prod_{\ell=0}^{\infty}\frac{\lambda_{\ell}^{2}-q^{2}\lambda_{k}^{2}}{\lambda_{\ell}^{2}-q^{-2}\lambda_{k}^{2}}=-e^{-4\pi\mathsf{i}p}\;,\qquad\forall k\in\mathbb{N}^{0}\;.
\]
As always the infinity of equations and variables at hand might sound
slightly scary, however there exists a beautiful procedure which allow
us to ``resum'' these equations turning them into a single non-linear
integral equation (NLIE) paired with a finite set of Bethe ansatz
equations for a finite set of variables. The introduction of this
method in the context of QFT is due to C. Destri and H.J. de Vega
\cite{Dest_deVe_95}, although the non-linear integral equation first
appeared in a work of A. Klümper, M.T. Batchelor and P.A. Pearce \cite{Klum_Batc_Pear_91},
where it was used to compute the central charge of $6$- and $19$-vertex
models. We will not present the derivation of the equations, as it
follows the same exact lines of the original article; we limit ourselves
to displaying the result:
\begin{equation}
\begin{cases}
\mathsf{i}\log\left(\mathfrak{a}(\theta)\right)=-2\pi\frac{p}{\beta^{2}}+2M\cos\left(\pi\frac{\beta^{2}}{2-2\beta^{2}}\right)e^{\theta}+\\
\qquad\qquad\quad\quad+\mathsf{i}\sum_{a}\log\left(S(\theta-\theta_{a})\right)-2\mathcal{G}\star\Im\left[\log\left(1+\mathfrak{a}(\theta-\mathsf{i}0)\right)\right]\\
\mathfrak{a}(\theta_{a})=-1
\end{cases}\;,\label{eq:DDV_equation}
\end{equation}
where, as before, $\lambda^{1+\xi}\equiv\lambda^{\frac{1}{1-\beta^{2}}}=e^{\theta}$
and $\left\{ \theta_{a}\right\} _{a}$ corresponds to the set of those
zeroes $\left\{ \lambda_{a}^{2}=e^{2\theta_{a}(1-\beta^{2})}\right\} _{a}$
which lie outside the real positive $\lambda^{2}$-axis. We used the
sign $\star$ to denote the convolution of two functions:
\[
f\star g(\theta)\doteq\intop_{-\infty}^{\infty}d\theta'f(\theta-\theta')g(\theta')\equiv\intop_{-\infty}^{\infty}d\theta'f(\theta')g(\theta-\theta')\;,
\]
 and we introduced the kernel
\[
\mathcal{G}(\theta)\doteq\delta(\theta)+\frac{1}{2\pi\mathsf{i}}\partial_{\theta}\log\left(S(\theta)\right)\;,
\]
and the function
\[
S(\theta)\doteq\exp\left[-\mathsf{i}\intop_{-\infty}^{\infty}\frac{d\nu}{\nu}\sin\left(\nu\theta\right)\frac{\sinh\left(\pi\nu\frac{1+\xi}{2}\right)}{\cosh\left(\pi\frac{\nu}{2}\right)\sinh\left(\pi\nu\frac{\xi}{2}\right)}\right]\;,
\]
which coincides precisely with the soliton-soliton scattering amplitude
for the sine-Gordon model \cite{AZam_AlZa_79}. Given a solution $\mathfrak{a}(\lambda)$
of the NLIE (\ref{eq:DDV_equation}) above, one can recover the function
$A(\lambda)$ with the following formula
\begin{eqnarray}
 &  & \log\left(A(\lambda)\right)=-\mathsf{i}\intop_{-\infty-\mathsf{i}}^{\infty-\mathsf{i}}d\nu\frac{g(\nu+\mathsf{i}0)}{\cosh\left(\pi\frac{\nu+\mathsf{i}0}{2}\right)\sinh\left(\pi\xi\frac{\nu+\mathsf{i}0}{2}\right)}\left(-\lambda^{2}\right)^{\mathsf{i}\nu\frac{1+\xi}{2}}\;,\nonumber \\
 &  & g(\nu)=\intop_{-\infty}^{\infty}\frac{d\theta}{2\pi}\Im\left[\log\left(1+\mathfrak{a}(\theta-\mathsf{i}0)\right)\right]e^{-\mathsf{i}\nu\theta}\;.\label{eq:g_in_terms_of_a}
\end{eqnarray}
It is possible to use this formulae in the case of the vacuum eigenvalue
$A^{(\textrm{vac})}(\lambda)$ in the limit $p\rightarrow\infty$
to compute the exact form of the coefficient $M$ in (\ref{eq:A_asymptotic}).
This turns out to be
\begin{equation}
M=\frac{1}{\sqrt{\pi}}\Gamma\left(\frac{\xi}{2}\right)\Gamma\left(\frac{1-\xi}{2}\right)\left(\Gamma\left(\frac{1}{1+\xi}\right)\right)^{1+\xi}\;.\label{eq:M-parameter}
\end{equation}
We will not present the computations here and refer the interested
reader to the original article \cite{Bazh_Luky_AZam_97-1}.

\section{Integrable structures of massive integrable field theories\label{sec:Integrable-massive}}

Having extracted and analysed the integrable structures of conformal
field theories, a natural question arises: are these results ``exportable''
in massive field theories? The answer, at least for what concerns
theories obtained as integrable deformations of CFTs, is positive.
Actually, as it turns out, this extension is rather straightforward:
the equations keep the same exact form they have in the massless case.
As we already noticed above, this is expected since the algebraic
structure governing massless theories survives unscathed to the integrable
deformation. On the other hand, the analytic properties of the various
objects we introduced undergo a radical change as a consequence of
the interplay between the two chiralities which, in presence of a
mass scale, is no more trivial.

In the following we will first briefly review A.B. Zamolodchikov results
concerning integrable deformations of CFTs \cite{AlZa_89} and then
construct the $T$-operators for a particular class of these.

\subsection{Brief overview of CFT integrable deformations\label{sub:Brief-overview-of-1}}

Remember how in a CFT there exists an infinite set of integrals of
motion, which can be constructed from normal ordered products of the
energy momentum tensor $T(u)$ and its derivatives:
\[
\mathbf{I}_{2k-1}=\intop_{0}^{2\pi}\frac{dw}{2\pi}T_{2k}(w)\;,\qquad\overline{\mathbf{I}}_{2k-1}=\intop_{0}^{2\pi}\frac{d\overline{w}}{2\pi}\overline{T}(\overline{w})\;.
\]
All these IMs are in involution: they form an abelian subalgebra $\mathcal{I}$
of $\mathcal{U}\left(\textrm{Vir}\right)$ (same goes for the left
chirality, obviously).

Given a CFT with Hamiltonian $H_{\textrm{CFT}}$, we can think of
deforming it by a relevant field $\Phi$ (clearly belonging to that
same CFT), obtaining thus a massive field theory
\[
H_{\Phi}\doteq H_{\textrm{CFT}}+\hat{\mu}^{2}\int\Phi(x)dx^{2}\;.
\]
In general these theories are not integrable: the appearance of a
mass scale inevitably destroys the conformal symmetry and, along with
it, the abelian subalgebra containing the IMs. Only some of these
survive and can be written as
\[
\mathbb{I}_{s}=\intop_{0}^{2\pi}\left[T_{s+1}dw+\Theta_{s-1}d\overline{w}\right]\;,\qquad\overline{\mathbb{I}}_{s}=\intop_{0}^{2\pi}\left[\overline{\Theta}_{s}dw+\overline{T}_{s+1}d\overline{w}\right]\;,
\]
where $T_{s}$, $\overline{T}_{s}$, $\Theta_{s}$ and $\overline{\Theta}_{s}$
are some local fields satisfying the current conservation law
\[
\overline{\partial}T_{s+1}=\partial\Theta_{s-1}\;,\qquad\partial\overline{T}_{s+1}=\overline{\partial}\overline{\Theta}_{s-1}\;,
\]
and the index $s$ takes values in a finite set: $s\in\mathcal{X}\;,\;\mathfrak{C}\left(\mathcal{X}\right)<\infty$,
where $\mathfrak{C}$ denotes the cardinality of a set. It turns out,
however, that with the right choice of perturbing field this set $\mathcal{X}$
becomes infinite; in other words there exist particular perturbations
for which the abelian subalgebra $\mathcal{I}$ survives in its entirety%
\footnote{Actually this is not strictly a consequence of $\left\Vert \mathcal{X}\right\Vert =\infty$,
but for the models we are going to analyse it is conjectured to be
so.%
}. As a consequence the massive theory obtained through these deformations
is integrable inheriting \textit{in toto}, with suitable modifications,
the integrable structure of the corresponding CFT. In \cite{AlZa_89}
Al.B. Zamolodchikov showed%
\footnote{Actually he made conjectures based on strong physical assumptions.
His conjectures were never disproved since then and are assumed to
be true.%
} that this phenomenon happens in CFTs with $c<1$ if one chooses $\Phi_{(1,3)}$,
$\Phi_{(1,2)}$ or $\Phi_{(2,1)}$ as perturbations, where the bracketed
indices stands for $(r,s)$, identifying the fields on the Ka\v{c}
Table \cite{Gins_89}. In the following we will concentrate on $\Phi_{(1,3)}$
perturbations only.

\paragraph{The $\Phi_{(1,3)}$ perturbations of CFT}

Consider the massive field theory defined by the Hamiltonian
\[
H_{(1,3)}\doteq H_{\textrm{CFT}}+\hat{\mu}^{2}\int\Phi_{(1,3)}(x)dx^{2}\;,
\]
where $\Phi_{(1,3)}$ is a primary field of $H_{\textrm{CFT}}$ with
conformal dimensions
\[
h_{1,3}=\overline{h}_{1,3}=2\beta^{2}-1\equiv\frac{\xi-1}{\xi+1}\;,
\]
which satisfies the null-vector equations
\begin{eqnarray}
 &  & \left[2(1+2\beta^{2})L_{-3}-4L_{-1}L_{-2}+\frac{1}{\beta^{2}}L_{-1}^{3}\right]\Phi_{(1,3)}(w,\overline{w})=0\;,\label{eq:level3_null_vect}\\
 &  & \left[2(1+2\beta^{2})\overline{L}_{-3}-4\overline{L}_{-1}\overline{L}_{-2}+\frac{1}{\beta^{2}}\overline{L}_{-1}^{3}\right]\Phi_{(1,3)}(w,\overline{w})=0\;,\label{eq:Level3_null_vect_2}
\end{eqnarray}
and is assumed to have the following canonical normalisation
\[
\left\langle \Phi_{(1,3)}(w,\overline{w})\,\Phi_{(1,3)}(w',\overline{w}')\right\rangle _{\textrm{CFT}}\underset{(w,\overline{w})\rightarrow(w',\overline{w}')}{\sim}\Bigl|w-w'\Bigr|^{-4h_{1,3}}\;.
\]
From dimensional analysis one immediately sees that $\left[\hat{\mu}^{2}\right]=\left[\textrm{length}\right]^{h_{1,3}-1}$
and $\hat{\mu}$ can be thought of carrying an anomalous dimension
of $h_{\hat{\mu}}=\overline{h}_{\hat{\mu}}=\frac{1-h_{1,3}}{2}=\frac{1}{1+\xi}$.
Notice how this is exactly the opposite of the anomalous dimension
carried by the spectral parameter $\lambda$ in the analysis of the
above section.

Any field theory is completely characterised by the infinite-dimensional
vector space of local fields $\Omega=\left\{ \mathcal{O}_{j}\right\} _{j\in\mathbb{N}}$,
along with the totality of their correlation functions. Generically,
when taking a perturbation of a CFT, the fields of this last require
an infinite number of renormalisation parameters in order to cancel
the ultraviolet divergencies which arise in correlation functions.
In the case under consideration, however, the massive theory turns
out to be ``super-renormalisable'', meaning only a finite number
of counter-terms is needed (for $\beta^{2}<\frac{1}{2}$, that is
$c<-2$, there are actually no UV divergencies at all). If, furthermore,
we are in a finite size geometry, as we are, the infrared divergencies
are completely under control, thanks to the natural cutoff $R<\infty$.
For these reasons we can safely assume that the Hilbert space of the
perturbed theory has the same structure of that of the original CFT
\begin{equation}
\mathscr{H}_{(1,3)}\simeq\mathscr{H}_{\textrm{CFT}}\;,\label{eq:Hilbert_equivalence}
\end{equation}
and there exists a one-to-one correspondence between fields of the
two theories. In other words one can assign to fields in the massive
theory the roles they played in the CFT; in this sense there exists
in the massive theory a concept of primary fields, of descendants
and also of Virasoro operators. 

Consider the subspace $\Lambda_{\textrm{CFT}}\subset\Omega_{\textrm{CFT}}$,
consisting of all the composite fields built out of $T(w)$ and its
derivatives. Now take the quotient of this subspace by the action
of $L_{-1}$:
\[
\hat{\Lambda}_{\textrm{CFT}}\doteq\Lambda_{\textrm{CFT}}\Big\slash\left(L_{-1}\Lambda_{\textrm{CFT}}\right)\subset\Lambda_{\textrm{CFT}}\;.
\]
This further subset $\hat{\Lambda}_{\textrm{CFT}}$ consists of all
those composite fields which are not total derivatives (remember that
$L_{-1}$ acts on local fields as a derivative). Then it is immediate
to notice that all the integrals of motion are generated as integrals
of some element of $\hat{\Lambda}_{\textrm{CFT}}$; informally 
\[
\mathcal{I}=\int\hat{\Lambda}_{\textrm{CFT}}\;.
\]
Now, while clearly $\overline{\partial}\hat{\Lambda}_{\textrm{CFT}}=0$,
this is no more true when considering the corresponding subspace in
the deformed theory; in general one will have
\[
\overline{\partial}T_{s}=\sum_{n=1}^{\infty}\mu^{2n}R_{s-1}^{(n)}\;,\qquad T_{s}\in\Lambda\;,\qquad R_{s-1}^{(n)}\in\Omega\;.
\]
By simple dimensional analysis, we see that the fields $R_{s-1}^{(n)}$
must have dimensions $(h,\overline{h})=(s-n+nh_{1,3},1-n+nh_{1,3})$;
however, no field in $\Omega$ is allowed to have negative left conformal
dimension, meaning that the series above must truncate%
\footnote{Remember that $h_{1,3}=2\beta^{2}-1$ and, if $c<1$ then $h_{1,3}<1$.%
} for $n$ bigger than some integer $N\geq1$. Moreover, since $\Phi_{(1,3)}$
is the most relevant field in its OPE subalgebra, we conclude that
$N=1$
\[
\overline{\partial}T_{s}=\hat{\mu}^{2}R_{s-1}\;.
\]
Denoting as usual with $\mathcal{V}_{1,3}$ the Verma module with
highest weight $h_{1,3}$, we have 
\[
L_{0}\mathcal{V}_{1,3}^{(s)}=\left(h_{1,3}+s\right)\mathcal{V}_{1,3}^{(s)}\;,\quad\overline{L}_{0}\mathcal{V}_{1,3}^{(s)}=h_{1,3}\mathcal{V}_{1,3}^{(s)}\;,\qquad\mathcal{V}_{1,3}=\bigoplus_{s=0}^{\infty}\mathcal{V}_{1,3}^{(s)}\;,
\]
and, clearly%
\footnote{Note that we care-freely apply the CFT concept of Verma modules to
the deformed theory. This can actually be done thanks to the isomorphism
(\ref{eq:Hilbert_equivalence}), since we logically expect that the
decomposition of the CFT Hilbert space into Verma modules survives
to the deformation along with the other structures. Therefore it should
exist in the massive theory a decomposition of $\mathscr{H}_{1,3}=\bigoplus_{a}\left(\mathscr{V}_{a}\otimes\overline{\mathscr{V}_{a}}\right)$
into spaces $\mathscr{V}_{a}\simeq\mathcal{V}_{a}$. We use the same
notation as in the CFT, hoping this note will be sufficient to avoid
confusion.%
}
\[
R_{s-1}\in\mathcal{V}_{1,3}^{(s-1)}\;.
\]
We can thus interpret the anti-holomorphic partial derivative $\overline{\partial}$
as a map between subspaces of $\Omega$:
\[
\overline{\partial}\;:\quad\hat{\Lambda}_{s}\longrightarrow\mathcal{V}_{1,3}^{(s-1)}\;.
\]
By taking first-order corrections to correlation functions involving
he field $T_{s}$, one can show that
\[
\overline{\partial}T_{s}(w,\overline{w})=\left[T_{s}(w,\overline{w})\,,\,\hat{\mu}^{2}\intop_{0}^{2\pi}dw'\Phi_{(1,3)}(w',\overline{w})\right]\;,
\]
which is a usual formula of perturbation theory. As a consequence
\[
[\overline{\partial}\,,\, L_{-1}]=0\;,
\]
and we can define a set of operators $D_{n}\;:\;\Lambda\rightarrow\mathcal{V}_{1,3}$
from their action on the vectors of $\Lambda$, informally:
\[
D_{n}\Lambda\doteq\left[\Lambda\,,\,\hat{\mu}^{2}\intop_{0}^{2\pi}dw'e^{in(w'-w)}\Phi_{(1,3)}(w',\overline{w})\right]\;.
\]
Clearly we have $\overline{\partial}\equiv D_{0}$ and it is fairly
easy to prove the following relations
\begin{eqnarray*}
 &  & \left[L_{n}\,,\, D_{m}\right]=-\left[(1-h_{1,3})(n+1)+m\right]D_{n+m}\;,\\
 &  & D_{-n-1}\cdot1=\frac{1}{n!}L_{-1}^{n}\Phi_{(1,3)}(w,\overline{w})\;.
\end{eqnarray*}
Finally we are at a point where we can explicitly compute the action
of $\overline{\partial}$ on the elements of $\hat{\Lambda}$. Let
us begin with the energy-momentum tensor $T_{2}\equiv T=L_{-2}\cdot1$
itself
\[
\overline{\partial}T=\hat{\mu}^{2}D_{0}L_{-2}\cdot1=\hat{\mu}^{2}(h_{1,3}-1)D_{-2}\cdot1=\hat{\mu}^{2}(h_{1,3}-1)L_{-1}\Phi_{(1,3)}\;.
\]
Introducing the local field $\Theta_{0}\equiv\Theta\doteq\hat{\mu}^{2}(h_{1,3}-1)\Phi_{(1,3)}$,
we see that we can write
\[
\overline{\partial}T_{2}=\partial\Theta_{0}\;,
\]
which is a current conservation equation and, as such, define an integral
of motion $\mathbb{I}_{1}$.

Let us now try and see if there's a current conservation law also
for the next element of $\hat{\Lambda}$: $T_{4}\equiv L_{-2}^{2}\cdot1$.
In this case we obtain
\begin{eqnarray*}
 & \overline{\partial}T_{4} & =\hat{\mu}^{2}D_{0}L_{-2}^{2}\cdot1=\hat{\mu}^{2}(h_{1,3}-1)\left(D_{-2}L_{-2}+L_{-2}D_{-2}\right)\cdot1\\
 &  & =\hat{\mu}^{2}(h_{1,3}-1)\left(2L_{-2}L_{-1}+\frac{h_{1,3}-3}{6}L_{-1}^{3}\right)\Phi_{(1,3)}\;,
\end{eqnarray*}
and we cannot write $\overline{\partial}T_{4}$ as the holomorphic
derivative of a local field! As a consequence there seem to be no
conservation law.

\paragraph{Degenerate fields and integrals of motion}

We have seen that there seems to be no hope of recovering the continuity
equations $\overline{\partial}T_{2n}=\partial\Theta_{2n-2}$ for $n>1$
and, as such, we have only two integrals of motion: $\mathbb{I}_{1}$
and $\overline{\mathbb{I}}_{1}$, which is to say the energy $\mathbb{I}_{1}+\overline{\mathbb{I}}_{1}$
and the momentum $\mathbb{I}_{1}-\overline{\mathbb{I}}_{1}$ of our
system. However, we forgot that we have an ace in our sleeve: we chose
the perturbing field to be $\Phi_{(1,3)}$, a degenerate field which
satisfies the level-3 null vector equations (\ref{eq:level3_null_vect})
and (\ref{eq:Level3_null_vect_2})! Recalling that $L_{-2}L_{-1}=L_{-1}L_{-2}-L_{-3}$,
we can rewrite $\overline{\partial}T_{4}=\partial\Theta_{2}$ where
\[
\Theta_{2}\doteq\hat{\mu}^{2}\frac{h_{1,3}-1}{h_{1,3}+2}\left(2h_{1,3}L_{-2}+\left(h_{1,3}-2\right)\frac{\left(h_{1,3}-1\right)\left(h_{1,3}+3\right)}{6\left(h_{1,3}+1\right)}L_{-1}^{2}\right)\Phi_{(1,3)}\;,
\]
which is a proper conservation law and give rise to the integral of
motion $\mathbb{I}_{3}$.

The conjecture of Al. Zamolodchikov is that the phenomenon illustrated
just above, happens on every level subspace $\mathcal{V}_{(1,3)}^{(s-1)}$
with odd $s$. A nice way to see this for $s\leq7$ is the following.
Consider the operator $B_{s}$ defined as
\[
B_{s}\doteq\varPi_{s-1}\overline{\partial}\;:\quad\hat{\Lambda}_{s}\longrightarrow\hat{\mathcal{V}}_{(1,3)}^{(s-1)}\;,\qquad\hat{\mathcal{V}}_{(1,3)}^{(s-1)}\doteq\mathcal{V}_{(1,3)}^{(s-1)}\Big\slash\left(L_{-1}\mathcal{V}_{(1,3)}^{(s-1)}\right)\;,
\]
where $\varPi_{s}$ is the projector onto $\hat{\mathcal{V}}_{(1,3)}^{(s-1)}$:
\[
\varPi_{s}\;:\quad\mathcal{V}_{(1,3)}^{(s-1)}\longrightarrow\hat{\mathcal{V}}_{(1,3)}^{(s-1)}\;.
\]
By definition, if $B_{s+1}T_{s+1}=0$, then we are assured that there
exists a field $\Theta_{s-1}$ such that $\overline{\partial}T_{s+1}=\partial\Theta_{s-1}$.
This means that a conservation law is present at the level $s$ iff
$B_{s}$ has a non-vanishing kernel. The two conservation laws we
found above appear for a very simple reason
\[
\dim\left(\hat{\mathcal{V}}_{(1,3)}^{(1)}\right)=\dim\left(\hat{\mathcal{V}}_{(1,3)}^{(3)}\right)=0\;.
\]
This suggests that we can try to compare the dimensions of the spaces
$\hat{\Lambda}_{s}$ and $\hat{\mathcal{V}}_{(1,3)}^{(s-1)}$; we
can do this by using the character formulae \cite{DiFr_Math_Sene}:
\begin{eqnarray*}
 &  & \sum_{s=0}^{\infty}q^{s}\dim\left(\hat{\mathcal{V}}_{(1,3)}^{(s-1)}\right)=(1-q)\chi_{(1,3)}(q)\;,\qquad\chi_{(1,3)}(q)\doteq\prod_{\underset{n\neq3}{n=1}}^{\infty}(1-q^{n})^{-1}\;,\\
 &  & \sum_{s=0}^{\infty}q^{s}\dim\left(\hat{\Lambda}_{s}\right)=(1-q)\chi_{0}(q)+q\;,\qquad\chi_{0}(q)\doteq(1-q)\prod_{n=2}^{\infty}(1-q^{n})^{-1}+q\;.
\end{eqnarray*}

\begin{center}
In the following table we list the first few dimensions\\
\begin{tabular}{|c|c|c|c|c|c|c|c|c|c|c|c|}
\hline 
$s$ & 1 & 2 & 3 & 4 & 5 & 6 & 7 & 8 & 9 & 10 & 11\tabularnewline
\hline 
\hline 
$\dim\left(\hat{\mathcal{V}}_{(1,3)}^{(s-1)}\right)$ & 1 & 0 & 1 & 0 & 2 & 0 & 3 & 1 & 4 & 2 & 7\tabularnewline
\hline 
$\dim\left(\hat{\Lambda}_{s}\right)$ & 0 & 1 & 0 & 2 & 1 & 3 & 2 & 5 & 4 & 8 & 7\tabularnewline
\hline 
\end{tabular}
\par\end{center}

We see that for $s$ odd and $s\leq7$, $\dim\left(\hat{\Lambda}_{s}\right)=\dim\left(\hat{\mathcal{V}}_{(1,3)}^{(s-1)}\right)+1$,
meaning that $B_{s+1}$ has always at least a one-dimensional kernel
assuring the conservation laws. As $s$ grows over $7$, however,
this argument fails and the existence of higher spin conservation
laws is a conjecture which finds a partial justification in the classical
limit (where their existence is guaranteed to be true).

As a last note, one should pay particular attention when $c$ takes
values corresponding to the minimal models $\mathcal{M}_{p,p'}$,
as in those cases the structure of the spaces $\hat{\Lambda}_{s}$
and $\hat{\mathcal{V}}_{(1,3)}^{(s-1)}$ is warped by the infinite
ladder of null-vectors. Moreover in a minimal model, it not justified
to assume that $\overline{\partial}T_{s}=\hat{\mu}^{2}R_{s-1}$: terms
with higher powers of $\hat{\mu}^{2}$ might appear.

\subsection{T-operators in $\Phi_{(1,3)}$ deformed CFTs\label{sub:T-operators-in-}}

Let us consider the theory defined by the following formal action
\begin{equation}
\mathscr{A}_{(1,3)}\doteq\mathscr{A}_{\textrm{CFT}}+\hat{\mu}^{2}\int dudv\Phi_{(1,3)}(u,v)\;,\qquad\left[\hat{\mu}\right]=\left[\textrm{length}\right]^{2\beta^{2}-2}\;,\label{eq:Perturbed_CFT}
\end{equation}
defined on a cylinder of radius%
\footnote{We slightly change the geometry here by introducing the radius as
a new parameter of the system. We hope this will not be confusing.%
} $R$: $\left\{ (u,v)\;\Big\vert\; u+R=u\right\} $, with $u$ playing
the role of space and $v$ that of time. Here $\mathscr{A}_{\textrm{CFT}}$
is the formal action of a conformal field theory with $c<1$ and $\Phi_{(1,3)}$
is its primary field of conformal dimension $h_{1,3}=2\beta^{2}-1$.
We recall the relation between $\beta$ and the central charge:
\[
\beta=\sqrt{\frac{1-c}{24}}-\sqrt{\frac{25-c}{24}}\;\Longrightarrow\; c=13-6\left(\beta^{2}+\beta^{-2}\right)\;.
\]
As we already mentioned above, important facts of deformed CFTs are
the presence of a mass scale $\mathfrak{m}\propto\hat{\mu}^{\frac{1}{1-\beta^{2}}}$
and the breaking of conformal invariance, which translates into the
non-holomorphicity of the energy-momentum tensor. As we have seen
above, however, in a $\Phi_{(1,3)}$-deformed CFT, the local fields
$T_{2k}(w,\overline{w})$ satisfy the continuity equations
\[
\overline{\partial}T_{2k}(w,\overline{w})=\partial\Theta_{2k-2}(w,\overline{w})\;,\qquad\partial\overline{T}_{2k}(w,\overline{w})=\overline{\partial}\overline{\Theta}_{2k-2}(w,\overline{w})\;,
\]
meaning that the integrals
\begin{eqnarray}
 &  & \mathbb{I}_{2k-1}\doteq\intop_{0}^{R}\frac{du}{2\pi}\left[T_{2k}(w,\overline{w})+\Theta_{2k-2}(w,\overline{w})\right]\;,\label{eq:massive_right_IM}\\
 &  & \overline{\mathbb{I}}_{2k-1}\doteq\intop_{0}^{R}\frac{du}{2\pi}\left[\overline{T}_{2k}(w,\overline{w})+\overline{\Theta}_{2k-2}(w,\overline{w})\right]\;,\label{eq:massive_left_IM}
\end{eqnarray}
do not depend on the ``time'' $v$ and are, as such, integrals of
motion. It is not hard to verify, at least for small $k$, that these
integrals are in involution
\[
[\mathbb{I}_{2k-1},\mathbb{I}_{2l-1}]=[\mathbb{I}_{2k-1},\overline{\mathbb{I}}_{2l-1}]=[\overline{\mathbb{I}}_{2k-1},\overline{\mathbb{I}}_{2l-1}]=0\;.
\]
Notice that $\mathbb{H}\doteq\mathbb{I}_{1}+\overline{\mathbb{I}}_{1}$
and $\mathbb{P}=\mathbb{I}_{1}-\overline{\mathbb{I}}_{1}$ are, respectively,
the Hamiltonian and the momentum operators of $\mathscr{A}_{(1,3)}$.

\paragraph{The left chirality}

In order to describe the integrable structures of our deformed CFT,
we have to consider both chiralities at the same time. To this end,
let us construct the left-chiral integrable structure of the CFT,
which we will then join with the right-chiral part when moving to
the massive model.

Remember that the starting point of our construction was the Feigin-Fuchs
free field representation of the energy-momentum tensor (\ref{eq:Feigin_Fuchs}),
which allows us to express the right-chiral Hilbert space $\mathscr{H}_{ch}$
as a direct sum of Fock spaces $\mathscr{H}_{ch}=\bigoplus_{a}\mathscr{F}_{p_{a}}$,
with the latter generated by the free action of the Heisenberg operators
$\left\{ a_{-n}\right\} _{n=1}^{\infty}$ on the ``vacuum'' $\left|\, p\,\right\rangle $,
such that $P\left|\, p\,\right\rangle =p\left|\, p\,\right\rangle $
and $a_{n}\left|\, p\,\right\rangle =0\;,\;\forall n>0$. We will
need to repeat these steps in the left chirality, so let us introduce
the following free field%
\footnote{Note the appearance of the ratio $\frac{2\pi}{R}$, due to the rescaling
on a cylinder with radius $R$.%
}
\[
\overline{\varphi}(\overline{w})=\mathsf{i}\overline{Q}-\mathsf{i}\frac{2\pi}{R}\overline{w}\overline{P}+\sum_{n\neq0}\frac{\overline{a}_{-n}}{n}e^{-\mathsf{i}\frac{2\pi}{R}n\overline{w}}\;,
\]
where the operators $\left\{ \overline{Q},\overline{P};\overline{a}_{n}\right\} _{n\neq0}^{n\in\mathbb{Z}}$
span an Heisenberg algebra
\[
[\overline{Q},\overline{P}]=\frac{\mathsf{i}}{2}\beta^{2}\;,\qquad[\overline{a}_{n},\overline{a}_{m}]=\frac{n}{2}\beta^{2}\delta_{n+m,0}\;.
\]
The full Hilbert space can then be expressed as
\[
\mathscr{H}_{\textrm{CFT}}=\bigoplus_{a}\left(\mathscr{F}_{p_{a}}\otimes\overline{\mathscr{F}}_{-p_{a}}\right)\;,
\]
where $\mathscr{F}_{p_{a}}\otimes\overline{\mathscr{F}}_{-p_{a}}$
is an irreducible representation of $Vir\otimes\overline{Vir}$ with
highest weights $\left(h(p),h(p)\right)$ and highest-weight vector
$\left|\, p\,\right\rangle \otimes\overline{\left|\,-p\,\right\rangle }$.

Now we, most simply, define the left-chiral $L$-operator as
\[
\overline{\mathbf{L}}_{j}(\lambda)\doteq\pi_{j}\left[e^{-\pi\mathsf{i}\overline{P}H}\mathscr{P}\exp\left(\lambda\intop_{0}^{R}d\overline{w}\left(\overline{V}_{-}(\overline{w})q^{\frac{H}{2}}E+\overline{V}_{+}(\overline{w})q^{-\frac{H}{2}}F\right)\right)\right]\;,
\]
where $\left\{ H,E,F\right\} $ are $\mathcal{U}_{q}\left(sl(2)\right)$
generators and $\overline{V}_{\pm}(\overline{w})$ are the left-chiral
vertex operators:
\[
\overline{V}_{\pm}(\overline{w})\doteq:e^{\pm2\overline{\varphi}(\overline{w})}:\;.
\]
Just as for the right-chiral $L$-operators they satisfy a relation
with the $\mathcal{U}_{q}\left(sl(2)\right)$ trigonometric $R$-matrix
which we will call ``$\overline{L}\overline{L}R$ relation'':
\[
\left(\overline{\mathbf{L}}_{j}(\lambda)\otimes1\right)\left(1\otimes\overline{\mathbf{L}}_{j'}(\lambda')\right)\mathbf{R}_{j\, j'}(\lambda/\lambda')=\mathbf{R}_{j\, j'}(\lambda/\lambda')\left(1\otimes\overline{\mathbf{L}}_{j'}(\lambda')\right)\left(\overline{\mathbf{L}}_{j}(\lambda)\otimes1\right)\;.
\]
Notice the different ordering of this relation with respect to the
corresponding one for the right chirality (\ref{eq:RLL_relation}),
which can be traced down to the minus sign in front of $\overline{P}$
in the definition of $\overline{\varphi}$. From the knowledge of
the left-chiral $L$-operator, one can repeat exactly what has been
done for the right chirality and obtain the operators $\overline{\mathbf{T}}_{j}$,
$\overline{\mathbf{Q}}_{\pm}$, and so on. We will not go into details
as these constructions are essentially identical as the ones for the
right chirality.

\paragraph{The massive integrable structure}

Now that we have the $L$-operators of both right and left chiralities,
we need to fuse them into one single object. The ``vault key'' holding
the two pieces together will have to be the deformation parameter
$\hat{\mu}^{2}$ which, we recall, carries a dimension $[\textrm{length}]^{2\beta^{2}-2}$.
Since the spectral parameter carries a dimension of $[\lambda]=[\textrm{length}]^{\beta^{2}-1}$,
the most natural way (and, as it turns out, the correct one) to couple
the chiralities is
\[
\mathbb{L}_{j}(\mu\vert\lambda)\doteq\mathbf{L}_{j}(\lambda)\overline{\mathbf{L}}_{j}(\mu/\lambda)\;,
\]
where $\mu\propto\hat{\mu}$. As usual, the $T$-operators are obtained
taking the trace over the $\mathcal{U}_{q}\left(sl(2)\right)$ representation%
\footnote{Notice the missing exponential factor in front of the $L$-operator.
In fact $\overline{\mathbf{T}}_{j}(\lambda)=\textrm{tr}_{\pi_{j}}\left(e^{-\pi iPH}\overline{\mathbf{L}}_{j}(\lambda)\right)$
and, when combining the two chiralities, the exponential factors in
front of $\mathbf{L}_{j}$ and $\overline{\mathbf{L}}_{j}$ cancel
each other.%
}
\[
\mathbb{T}_{j}(\mu\vert\lambda)\doteq\textrm{tr}_{\pi_{j}}\left(\mathbb{L}_{j}(\mu\vert\lambda)\right)\;.
\]
The commutativity of these operators
\[
[\mathbb{T}_{j}(\mu\vert\lambda),\mathbb{T}_{j'}(\mu\vert\lambda)]=0\;,
\]
is easily inferred from the $RLL$ and $\overline{L}\overline{L}R$
relations, the commutativity of the chiralities
\[
[\mathbf{L}_{j}(\lambda),\overline{\mathbf{L}}_{j'}(\lambda')]=0\;,
\]
and from the unitarity of the $R$-matrices
\[
\mathbf{R}_{j\, j'}(\lambda)\mathbf{R}_{j\, j'}(\lambda^{-1})=1\;.
\]
The following two properties immediately descend from the definition
of $\mathbb{T}_{j}$ and from the properties of $\mathbf{T}_{j}$
and $\overline{\mathbf{T}}_{j}$:
\begin{itemize}
\item \textbf{Massless limit}: in the limit $\mu\rightarrow0$ we recover
the right and left chiral $T$-operators as
\begin{eqnarray*}
 &  & \mathbb{T}_{j}(\mu\vert\lambda)\underset{\mu\rightarrow0}{\longrightarrow}\mathbf{T}_{j}(\lambda)\otimes\overline{1}\;,\\
 &  & \mathbb{T}_{j}(\mu\vert\mu/\lambda)\underset{\mu\rightarrow0}{\longrightarrow}1\otimes\overline{\mathbf{T}}_{j}(\lambda)\;,
\end{eqnarray*}
where the tensor notation is employed here to specify that $\mathbb{T}$
acts as the identity in the left (right) chirality space.
\item \textbf{Analytic properties}: the operators $\mathbb{T}$ are single
valued functions of $\lambda^{2}$, regular everywhere except $\lambda^{2}=0,\infty$.
Moreover they inherit the asymptotic behaviour of the $\mathbf{T}$
and $\overline{\mathbf{T}}$ functions:
\[
\log\left(\mathbb{T}_{j}(\mu\vert\lambda)\right)\underset{\lambda\rightarrow\infty}{\sim}m_{j}\frac{R}{2\pi}\lambda^{1+\xi}\;,
\]
\[
\log\left(\mathbb{T}_{j}(\mu\vert\lambda)\right)\underset{\lambda\rightarrow0}{\sim}m_{j}\frac{R}{2\pi}\left(\frac{\mu}{\lambda}\right)^{1+\xi}\;,
\]
where
\[
m_{j}=\frac{\sin\left(j\pi\xi\right)}{\sin\left(\pi\frac{\xi}{2}\right)}m\;,
\]
and $m$ is given in (\ref{eq:m-parameter}). Clearly the massive
$T$-operators have two essential singularities: one at $\infty$
(like the chiral operators) and the other at $0$. It is natural to
expect the integrals of motion (\ref{eq:massive_right_IM}-\ref{eq:massive_left_IM})
to appear in the asymptotic expansion around these singular points.
\end{itemize}
As we already said above, all the purely algebraic relations that
we displayed in the previous section transfer directly here, without
any change in their appearance. So the $T$-system reads:
\[
\mathbb{T}_{j}(\mu\vert q^{\frac{1}{2}}\lambda)\mathbb{T}_{j}(\mu\vert q^{-\frac{1}{2}}\lambda)=1+\mathbb{T}_{j+\frac{1}{2}}(\mu\vert\lambda)\mathbb{T}_{j-\frac{1}{2}}(\mu\vert\lambda)\;,
\]
and, when $q$ is a root of unity, it truncates to a finite system,
exactly as in the massive case, which can then be recast in a $Y$-system,
from which a TBA equation can be extracted (under some suitable analyticity
assumptions).

The difference between the massive and the massless case emerges in
the analytic properties of the $T$-operators. As we have seen just
above, the $\mathbb{T}_{j}$ possess \uline{two} singularities
instead of just one, which corresponds to the two different chiralities
of the massive theory. Here below we list the conjectures about the
massive $T$-operators
\begin{itemize}
\item \textbf{Asymptotic expansion}: the integrals of motion (\ref{eq:massive_right_IM}-\ref{eq:massive_left_IM})
govern the asymptotic expansions of $\mathbb{T}=\mathbb{T}_{\frac{1}{2}}$
around its singular points $\lambda^{2}=0,\infty$ as follows
\[
\log\left(\mathbb{T}(\mu\vert\lambda)\right)=m\frac{R}{2\pi}\lambda^{1+\xi}-\sum_{n=1}^{\infty}C_{n}\lambda^{(1-2n)(1+\xi)}\mathbb{I}_{2n-1}\;,
\]
\[
\log\left(\mathbb{T}(\mu\vert\lambda)\right)=m\frac{R}{2\pi}\left(\frac{\mu}{\lambda}\right)^{1+\xi}-\sum_{n=1}^{\infty}C_{n}\left(\frac{\mu}{\lambda}\right)^{(1-2n)(1+\xi)}\overline{\mathbb{I}}_{2n-1}\;,
\]
where the constants $C_{j}$ are given in (\ref{eq:C-parameter}).
\item \textbf{Commutation with IMs}: The massive transfer matrices commute
with all the massive IMs
\[
[\mathbb{T}_{j}(\mu\vert\lambda),\mathbb{I}_{2k-1}]=[\mathbb{T}_{j}(\mu\vert\lambda),\overline{\mathbb{I}}_{2k-1}]=0\;.
\]

\item \textbf{Relation between $\mu$ and $\hat{\mu}$}: the parameter entering
in $\mathbb{T}$ is related to the deformation parameter $\hat{\mu}$
as follows
\[
\hat{\mu}^{2}=\frac{\Gamma^{2}\left(1-\beta^{2}\right)}{\pi\left(1-2\beta^{2}\right)\left(3\beta^{2}-1\right)}\left[\frac{\Gamma\left(3\beta^{2}\right)\Gamma\left(\beta^{2}\right)}{\Gamma\left(1-3\beta^{2}\right)\Gamma\left(1-\beta^{2}\right)}\right]^{\frac{1}{2}}\mu^{2}\;.
\]

\end{itemize}
The last of these conjectures might appear a bit outlandish, however
it becomes natural when one considers $\mathbb{T}$ to actually be
the transfer matrix of the sine-Gordon model. The reasons leading
us to conjecture that the operators $\mathbb{T}_{j}$ introduced above
can be interpreted as $T$-operators of sine-Gordon model is briefly
discussed in the following subsection.

\subsection{Moving on to sine-Gordon model}

The quantum sine-Gordon model on a cylinder of radius $R$ is a massive
integrable QFT described by the following action
\[
\mathscr{A}_{sG}=\intop_{0}^{R}du\intop_{-\infty}^{\infty}dv\left[\frac{1}{16\pi}\left(\partial_{\mu}\phi(u,v)\right)^{2}+2\tilde{\mu}\cos\left(\beta\phi(u,v)\right)\right]\;,
\]
where $\phi(w,\overline{w})$ is a scalar field, $\beta$ is the coupling
and $\tilde{\mu}$ is a parameter with dimensions $[\textrm{length}]^{2\beta^{2}-2}$.
It possesses an infinite set of integrals of motion whose form is
exactly the same as (\ref{eq:massive_right_IM}-\ref{eq:massive_left_IM}),
where now $T_{2k}$, $\Theta_{2k-2}$, $\overline{T}_{2k}$ and $\overline{\Theta}_{2k-2}$
are local fields of sine-Gordon model and $\hat{\mu}$ has to be replaced
by $\tilde{\mu}$. The spectrum of the model contains two topologically
charged particles (the soliton and the anti-soliton) with mass $\mathfrak{M}$
\begin{equation}
\mathfrak{M}=\frac{2}{\sqrt{\pi}}\frac{\Gamma\left(\frac{\xi}{2}\right)}{\Gamma\left(\frac{1}{2}+\frac{\xi}{2}\right)}\left[\pi\tilde{\mu}\frac{\Gamma\left(\frac{1}{1+\xi}\right)}{\Gamma\left(\frac{\xi}{1+\xi}\right)}\right]^{\frac{1+\xi}{2}}\;,\qquad\xi=\frac{\beta^{2}}{1-\beta^{2}}\;,\label{eq:soliton_mass}
\end{equation}
and a set of neutral particles (bound-states), whose number depends
on the coupling, with masses
\[
\mathfrak{m}_{j}=2\mathfrak{M}\sin\left(j\pi\xi\right)\;,\qquad j=\frac{1}{2},1,\ldots,n\;\textrm{s.t.}\; n<\frac{1}{2\xi}\;.
\]

The connection between sine-Gordon model and the perturbed CFTs we
considered above is not evident at first. However, as it was unveiled
in \cite{Resh_Smir_90,Bern_LeCl_90} these latter can be obtained
as ``quantum group reductions'' of the former, as we will recall
very briefly. When considered in infinite volume, sine-Gordon model
exhibits a symmetry with respect to the quantum group $\mathcal{U}_{\tilde{q}}\left(SL(2)\right)$,
where 
\[
\tilde{q}=e^{\mathsf{i}\frac{\pi}{\beta^{2}}}\;.
\]
This means that the soliton and anti-soliton transform in the two-dimensional
representation of this quantum group and that the local IMs and $S$-matrix
commute with the generators $\left\{ \tilde{H},\tilde{E},\tilde{F}\right\} $
of the associated quantum algebra $\mathcal{U}_{\tilde{q}}\left(sl(2)\right)$.
Now, the Hilbert space $\mathscr{H}_{sG}^{\infty}$ of sine-Gordon
model in infinite volume, contains a subspace $\mathscr{H}_{sG}^{\infty,\,\textrm{singlet}}$,
consisting of those states annihilated by the $\mathcal{U}_{\tilde{q}}\left(sl(2)\right)$
generators. What is remarkable is that this last Hilbert space can
be interpreted as the space of states of a certain local QFT, which
was called \textit{restricted sine-Gordon model}. As it turns out,
this model coincides exactly with the perturbed CFT (\ref{eq:Perturbed_CFT})
(considered in infinite volume!) where the parameters $\hat{\mu}$
and $\tilde{\mu}$ are related as
\[
\tilde{\mu}^{2}=\frac{\left(1-2\beta^{2}\right)\left(3\beta^{2}-1\right)}{\pi}\left[\frac{\Gamma^{3}\left(\beta^{2}\right)\Gamma\left(1-3\beta^{2}\right)}{\Gamma^{3}(1-\beta^{2})\Gamma\left(3\beta^{2}\right)}\right]^{\frac{1}{2}}\hat{\mu}^{2}\;.
\]
Although in finite volume the quantum group symmetry breaks down,
it is still possible to define singlet states and their Hilbert sub-space
$\mathscr{H}_{sG}^{R,\,\textrm{singlet}}$ and these still allow an
interpretation in terms of deformed CFTs. In particular
\[
\mathscr{H}_{sG}^{R,\,\textrm{singlet}}\simeq\mathscr{H}_{(1,3)}\simeq\mathscr{H}_{\textrm{CFT}}\;.
\]

It is not difficult to verify that the action of massive $T$-operators
and their properties, defined in the previous sub-section naturally
extend to the full sine-Gordon Hilbert space $\mathscr{H}_{sG}^{R}$
so that they can be interpreted as the transfer matrices of the unrestricted
sine-Gordon model. In this optics, the Feigin-Fuchs fields are naturally
identified with sine-Gordon one as
\[
\varphi(w)=\frac{\beta}{2}\phi(w,0)\;,\qquad\overline{\varphi}(\overline{w})=-\frac{\beta}{2}\phi(0,R-\overline{w})\;.
\]
It can moreover be shown that the truncation of the $T$-system happens
in sine-Gordon as well, meaning that it is possible employ the methods
of subsection \ref{sub:T-system,-Y-system-and} to obtain TBA-like
equations for the ground state of the system%
\footnote{Actually, there exists a method \cite{Klum_Pear_92} which, in theory,
allows one to recover non-linear integral equations from the $Y$-system
for all the eigenvalues. %
}. It is as well possible to proceed as in sub-sections \ref{sub:Baxter-Q-operators}
and \ref{sub:Destri-de-Vega-equation}, constructing the operators
$\mathbb{Q}$ and $\mathbb{A}$ and recovering the Destri-deVega equation.
This path is actually more rewarding than the $Y$-system one, since
it works for any value of the coupling $\beta^{2}$ (in the region
$(0,1)$, as discussed for the CFTs). In order to obtain the $Q$-operators,
one proceeds in the same exact way as for the operators $\mathbb{T}$,
that is by combining the right and left-chiral $L$-operators in the
$q$-oscillator representation into a single operator $\mathbb{L}_{\pm}$
and then taking its trace. We will not delve in the detail of this
construction as, really, it is a simple variation on the theme of
what has been done in the CFT case. We wish instead to present a different
approach to the integrability structure of sine-Gordon model, which
relies on a surprising and still not completely understood connection
between classical and quantum worlds.

\section{Sine-Gordon model and the massive ODE/IM correspondence\label{sec:Sine-Gordon-model-and}}

In this section we wish to briefly present an approach to quantum
sine-Gordon%
\footnote{This approach can actually be extended to sinh-Gordon model without
too many difficulties.%
} model in finite geometry which was proposed in \cite{Luky_AZam_10}
by S.L. Lukyanov and A.B. Zamolodchikov. This approach relies on older
studies on the so-called ODE/IM correspondence \cite{Dore_Tate_99,Bazh_Luky_AZam_98,Dore_Dunn_Glio_Tate_08}
(see \cite{Dore_Dunn_Tate_07} for a review) which related the integrals
of motion of certain CFTs to the spectral properties of certain ordinary
differential equations (ODE). This setting was extended by Lukyanov
and Zamolodchikov to the massive case \cite{Luky_AZam_10} with the
study of sine- and sinh-Gordon cases and later this method was generalised,
first to the Tzitzéica-Bullough-Dodd model \cite{Dore_Fald_Negr_Tate_13},
corresponding to the affine Lie algebra $A_{2}^{(2)}$, then to the
Toda theories associated to the affine algebras $A_{n}^{(1)}$ \cite{Adam_Dunn_14}.
Finally, the ODE/IM was applied to the whole set Toda field theories,
associated both to simply-laced \cite{Negro_14,Maso_Raim_Vale_15-1}
and non-simply-laced \cite{Maso_Raim_Vale_15-2,Ito_Lock_15} algebras.

\subsection{Quantum sine-Gordon $T$ and $Q$ operators}

Before venturing in the description of the ODE/IM correspondence for
the sine-Gordon model, we wish to recapitulate the properties (proved
or conjectured) satisfied by the operators $\mathbb{T}$ and $\mathbb{Q}$.
These properties will allow us to identify certain particular objects
in the ODE side of the correspondence with the vacuum eigenvalues
of $T$ and $Q$ operators.

Consider the quantum sine-Gordon model as defined by the Lagrangian%
\footnote{Note that here we set the parameter in the Lagrangian as $\mu$, while
before it was $\tilde{\mu}$. We hope this will not create too much
confusion.%
}
\begin{equation}
\mathscr{L}_{sG}=\frac{1}{16\pi}\left[\left(\partial_{v}\phi\right)^{2}-\left(\partial_{u}\phi\right)^{2}\right]+2\mu\cos\left(\beta\phi\right)\;,\label{eq:sG_Lagrangian}
\end{equation}
which is obviously invariant under shifts of the field $\phi\longrightarrow\phi+2\frac{\pi}{\beta}$.
As a consequence, the Hilbert space $\mathscr{H}_{sG}$ splits into
orthogonal subspaces $\mathscr{H}_{k}$ characterised by the \textit{quasi-momentum}
$k$. Denote $\mathbb{U}$ the operator performing the shift of $\phi$,
then:
\[
\mathbb{U}\;:\qquad\begin{array}{c}
\phi\longrightarrow\phi+2\frac{\pi}{\beta^{2}}\\
\\
\left|\,\Phi_{k}\,\right\rangle \longrightarrow e^{2\pi\mathsf{i}k}\left|\,\Phi_{k}\,\right\rangle 
\end{array}\;,\qquad\forall\left|\,\Phi_{k}\,\right\rangle \in\mathscr{H}_{k}\;.
\]
Let us also introduce the charge and parity operators as:
\[
\mathbb{C}\phi(u,v)\mathbb{C}=-\phi(u,v)\;,\qquad\mathbb{P}\phi(u,v)\mathbb{P}=\phi(-u,v)\;.
\]
This model possesses an infinite set of $T$-operators $\left\{ \mathbb{T}_{\frac{j}{2}}\right\} _{j=1}^{\infty}$
and two Baxter $Q$-operators $\mathbb{Q}_{\pm}$. Their properties,
which are mostly conjectured on the basis of massless, classical and
discrete limits analysis, are listed below. From this moment on we
will drop the explicit dependence of $\mathbb{T}$ and $\mathbb{Q}$
on the parameter $\mu$. We will also use the variable $\theta=\log\left(\lambda^{1+\xi}\right)$
instead of the spectral parameter.

\paragraph{Properties of $T$-operators}
\begin{itemize}
\item \textbf{Mutual commutativity}:
\[
[\mathbb{T}_{j}(\theta),\mathbb{T}_{j'}(\theta')]=0\;,
\]

\item \textbf{Invariance under discrete shift}:
\[
[\mathbb{T}_{j}(\theta),\mathbb{U}]=0\;,
\]

\item \textbf{Invariance under charge conjugation}:
\[
[\mathbb{T}_{j}(\theta),\mathbb{C}]=0\;,
\]

\item \textbf{Parity conjugation}:
\[
\mathbb{P}\mathbb{T}_{j}(\theta)\mathbb{P}=\mathbb{T}_{j}(-\theta)\;,
\]

\item \textbf{Analytic properties}: the functions $\mathbb{T}_{j}(\theta)$
are entire functions of the variable $\theta$ with essential singularities
at $\theta\rightarrow\pm\infty$,
\item \textbf{Hermiticity}:
\[
\mathbb{T}_{j}^{\dagger}(\theta)=\mathbb{T}_{j}(\theta^{\ast})\;,
\]

\item \textbf{Periodicity}:
\[
\mathbb{T}_{j}\left(\theta+\mathsf{i}\pi(1+\xi)\right)=\mathbb{T}_{j}(\theta)\;,
\]
note that this property is the translation in terms of $\theta$ of
property of single-valuedness of $\mathbb{T}_{j}$ as a function of
$\lambda^{2}$.
\item \textbf{Fusion relation ($T$-system)}:
\[
\mathbb{T}_{\frac{1}{2}}(\theta)\mathbb{T}_{j}\left(\theta+\mathsf{i}\pi\xi\frac{2j+1}{2}\right)=\mathbb{T}_{j+\frac{1}{2}}\left(\theta+\mathsf{i}\pi\xi\frac{2j+2}{2}\right)+\mathbb{T}_{j-\frac{1}{2}}\left(\theta+\mathsf{i}\pi\xi\frac{2j}{2}\right)\;,
\]
or, equivalently
\[
\mathbb{T}_{j}\left(\theta+\mathsf{i}\pi\frac{\xi}{2}\right)\mathbb{T}_{j}\left(\theta-\mathsf{i}\pi\frac{\xi}{2}\right)=1+\mathbb{T}_{j+\frac{1}{2}}(\theta)\mathbb{T}_{j-\frac{1}{2}}(\theta)\;,
\]

\item \textbf{Asymptotic behaviour on the real line}:
\[
\log\left(\mathbb{T}_{\frac{1}{2}}(\theta)\right)\underset{\theta\rightarrow+\infty}{\sim}\sum_{n=0}^{\infty}2(-1)^{n}\sin\left(\pi\xi\frac{2n-1}{2}\right)C_{n}\mathbb{I}_{2n-1}e^{(1-2n)\theta}\;,
\]
\[
\log\left(\mathbb{T}_{\frac{1}{2}}(\theta)\right)\underset{\theta\rightarrow-\infty}{\sim}\sum_{n=0}^{\infty}2(-1)^{n}\sin\left(\pi\xi\frac{2n-1}{2}\right)C_{n}\overline{\mathbb{I}}_{-2n+1}e^{(2n-1)\theta}\;,
\]
where we set $\mathbb{I}_{-1}=\overline{\mathbb{I}}_{-1}\doteq\frac{R}{2\pi}$
and $C_{0}=m$.
\end{itemize}

\paragraph{Properties of $Q$-operators}
\begin{itemize}
\item \textbf{Commutativity}:
\[
[\mathbb{Q}_{\pm}(\theta),\mathbb{T}_{j}(\theta')]=[\mathbb{Q}_{\pm}(\theta),\mathbb{Q}_{\pm}(\theta')]=[\mathbb{Q}_{+}(\theta),\mathbb{Q}_{-}(\theta')]=0\;,
\]

\item \textbf{Invariance under discrete shift}:
\[
[\mathbb{Q}_{\pm}(\theta),\mathbb{U}]=0\;,
\]

\item \textbf{Charge conjugation}:
\[
\mathbb{C}\mathbb{Q}_{\pm}(\theta)\mathbb{C}=\mathbb{Q}_{\mp}(\theta)\;,
\]

\item \textbf{Parity conjugation}:
\[
\mathbb{P}\mathbb{Q}_{\pm}(\theta)\mathbb{P}=\mathbb{Q}_{\mp}(-\theta)\;,
\]

\item \textbf{Analytic properties}: the functions $\mathbb{Q}_{\pm}(\theta)$
are entire functions of the variable $\theta$ with essential singularities
at $\theta\rightarrow\pm\infty$,
\item \textbf{Hermiticity}:
\[
\mathbb{Q}_{\pm}^{\dagger}(\theta)=\mathbb{Q}_{\pm}(\theta^{\ast})\;,
\]

\item \textbf{Baxter $T$-$Q$ relation}:
\[
\mathbb{T}_{\frac{1}{2}}(\theta)\mathbb{Q}_{\pm}(\theta)=\mathbb{Q}_{\pm}(\theta+\mathsf{i}\pi\xi)+\mathbb{Q}_{\pm}(\theta-\mathsf{i}\pi\xi)\;,
\]

\item \textbf{Shift property}:
\[
\mathbb{Q}_{+}\left(\theta+\mathsf{i}\pi(\xi+1)\right)=\mathbb{U}\mathbb{Q}_{+}(\theta)\;,
\]
\[
\mathbb{Q}_{-}\left(\theta+\mathsf{i}\pi(\xi+1)\right)=\mathbb{U}^{-1}\mathbb{Q}_{-}(\theta)\;,
\]
this property, along with the $T$-$Q$ relation, can be regarded
as defining the $Q$-operators as ``Bloch-wave''%
\footnote{This is a consequence of the periodicity of $\mathbb{T}_{j}$.%
} solutions to a second order finite difference equation,
\item \textbf{Quantum Wronskian}:
\[
\mathbb{Q}_{+}\left(\theta+\mathsf{i}\pi\frac{\xi}{2}\right)\mathbb{Q}_{-}\left(\theta-\mathsf{i}\pi\frac{\xi}{2}\right)-\mathbb{Q}_{+}\left(\theta-\mathsf{i}\pi\frac{\xi}{2}\right)\mathbb{Q}_{-}\left(\theta+\mathsf{i}\pi\frac{\xi}{2}\right)=\mathbb{U}^{-1}-\mathbb{U}\;,
\]

\item \textbf{Wronskian expression of operators $\mathbb{T}_{j}$}:
\begin{eqnarray*}
 &  & \mathbb{T}_{j}(\theta)\left(\mathbb{U}^{-1}-\mathbb{U}\right)=\left[\mathbb{Q}_{+}\left(\theta+\mathsf{i}\pi\xi\frac{2j+1}{2}\right)\mathbb{Q}_{-}\left(\theta-\mathsf{i}\pi\xi\frac{2j+1}{2}\right)\right.+\\
 &  & \qquad\qquad\qquad\qquad\quad-\left.\mathbb{Q}_{+}\left(\theta-\mathsf{i}\pi\xi\frac{2j+1}{2}\right)\mathbb{Q}_{-}\left(\theta+\mathsf{i}\pi\xi\frac{2j+1}{2}\right)\right]\;,
\end{eqnarray*}

\item \textbf{Leading asymptotic}:
\[
\mathbb{Q}_{+}(\theta)\underset{\Re(\theta)\rightarrow\infty}{\sim}\mathbb{U}^{\pm\frac{1}{2}}\mathbb{S}^{\frac{1}{2}}\exp\left[\mathfrak{M}R\frac{e^{\theta\mp\mathsf{i}\pi\frac{\xi+1}{2}}}{4\cos\left(\pi\frac{\xi}{2}\right)}\right]\;,\qquad\theta\in H_{\pm}\;,
\]
\[
\mathbb{Q}_{+}(\theta)\underset{\Re(\theta)\rightarrow-\infty}{\sim}\mathbb{U}^{\pm\frac{1}{2}}\mathbb{S}^{-\frac{1}{2}}\exp\left[\mathfrak{M}R\frac{e^{-\theta\pm\mathsf{i}\pi\frac{\xi+1}{2}}}{4\cos\left(\pi\frac{\xi}{2}\right)}\right]\;,\qquad\theta\in H_{\pm}\;,
\]
where $H_{\pm}\doteq\left\{ \theta\in\mathbb{C}\;\Big\vert\;0<\pm\Im(\theta)<\pi(\xi+1)\right\} $
and $\mathbb{S}$ is some operator such that
\[
[\mathbb{S},\mathbb{P}]=[\mathbb{S},\mathbb{U}]=0\;,\quad\mathbb{C}\mathbb{S}\mathbb{C}=\mathbb{S}^{-1}\;,\quad\mathbb{S}^{\dagger}=\mathbb{S}\;.
\]

\end{itemize}

\subsection{The modified sinh-Gordon equation and its linear problem}

We begin our study of quantum sine-Gordon starting from an apparently
far away point. Indeed let us consider the following \uline{classical}
partial differential equation
\begin{equation}
\partial\overline{\partial}\eta(z,\overline{z})-e^{2\eta(z,\overline{z})}+p(z)p(\overline{z})e^{-2\eta(z,\overline{z})}=0\;,\quad p(z)=z^{2\alpha}-s^{2\alpha}\;,\label{eq:MshG_equation}
\end{equation}
where $z$ and $\overline{z}$ are formal complex variables in no
way related to the space-time of the quantum model. On the other hand,
the real and positive parameters $\alpha$ and $s$ will be, later,
related to parameters of the quantum model. This equation, whose name
is \textit{modified sinh-Gordon} (MshG) equation, arise in the context
of differential geometry (see e.g. \cite{Bobe_91}) where it describes
the conformal metric of certain surfaces with smooth constant mean
curvature immersed in $\mathbb{R}^{3}$. The recent interest in this
equation was sparked by its appearance in the computation of gluon
scattering amplitudes in $\mathcal{N}=4$ super Yang-Mills at strong
coupling \cite{Alda_Gaio_Mald_11}; these can be analysed in terms
of classical strings in $AdS_{5}$ which, in turn, lead to the study
of minimal surfaces, whence the MshG equation arises. This equation
is integrable, as we will see, for any choice of the function $p$.
In our case, the MshG equation (\ref{eq:MshG_equation}) possesses
an evident discrete symmetry
\begin{equation}
(z,\overline{z})\longrightarrow(e^{\mathsf{i}\frac{\pi}{\alpha}}z,e^{-\mathsf{i}\frac{\pi}{\alpha}}\overline{z})\;,\label{eq:Discrete_symmetry}
\end{equation}
and we will restrict our attention to solutions which respect this
symmetry. More in detail, it is not difficult to verify that there
exists a family of solutions to (\ref{eq:MshG_equation}), parametrised
by the real number $l\in[-\frac{1}{2},\frac{1}{2}]$, satisfying the
following properties (here we sit on the real slice of $\mathbb{C}^{2}$,
by setting $z=\rho e^{\mathsf{i}\phi}$ and $\overline{z}=\rho e^{-\mathsf{i}\phi}$)%
\footnote{Note that here $\phi$ denotes the argument of $z$ and not the sine-Gordon
field! Hopefully this will not cause confusion.%
}:
\begin{enumerate}
\item \textbf{Periodicity}:
\[
\eta(\rho,\phi)=\eta(\rho,\phi+\frac{\pi}{\alpha})\;,
\]
or, in other words, we consider the MshG equation restricted on the
cone of apex angle $\frac{\pi}{\alpha}$
\[
\mathscr{C}_{\frac{\pi}{\alpha}}\doteq\left\{ (\rho,\phi)\in\mathbb{R}^{+}\times[-\frac{\pi}{2\alpha},\frac{\pi}{2\alpha})\;\Big\backslash\;\phi+\frac{\pi}{\alpha}\sim\phi\right\} \;,
\]

\item \textbf{Analyticity}: the solution $\eta(\rho,\phi)$ is single-valued,
real and finite everywhere on $\mathscr{C}_{\frac{\pi}{\alpha}}$
with the sole exception of $\rho=0,\infty$,
\item \textbf{Large-$\rho$ asymptotics}:
\[
\eta(\rho,\phi)\underset{\rho\rightarrow\infty}{\sim}\alpha\log\rho+o(1)\;,
\]

\item \textbf{Small-$\rho$ asymptotics}:
\[
\eta(\rho,\phi)\underset{\rho\rightarrow0}{\sim}\begin{cases}
2l\,\log\rho+O(1) & \left|l\right|<\frac{1}{2}\\
\pm\log\rho+O\left(\log(-\log\rho)\right) & l=\pm\frac{1}{2}
\end{cases}\;.
\]

\end{enumerate}
Starting from the above small-$\rho$ asymptotics we can iteratively
construct a $(z,\overline{z})\rightarrow(0,0)$ expansion of the following
form
\begin{eqnarray*}
 &  & \eta(z,\overline{z})=l\,\log\left(z\overline{z}\right)+\eta_{0}+\sum_{k=1}^{\infty}\gamma_{k}\left(z^{2\alpha k}+\overline{z}^{2\alpha k}\right)-s^{4\alpha}\frac{e^{-2\eta_{0}}}{(1-2l)^{2}}\left(z\overline{z}\right)^{1-2l}\\
 &  & \qquad\qquad+\frac{e^{2\eta_{0}}}{(1+2l)^{2}}\left(z\overline{z}\right)^{1+2l}+\cdots\;,
\end{eqnarray*}
where $\eta_{0}$ and $\left\{ \gamma_{k}\right\} _{k=1}^{\infty}$
are integration constants that are to be determined by imposing the
properties listed above on this expansion. The utility of this expression
is that it remains valid on the whole $\mathbb{C}^{2}$, which means
that we can safely fix $z$ to some finite value and send only $\overline{z}\rightarrow0$,
obtaining
\[
\eta(z,\overline{z})\underset{\overline{z}\rightarrow0}{\sim}l\,\log\left(z\overline{z}\right)+\eta_{0}+\gamma(z)\;,\qquad\gamma(z)\doteq\sum_{k=1}^{\infty}\gamma_{k}z^{2\alpha k}\;.
\]

As hinted at above, the MshG equation is integrable and, as such,
possesses a Lax pair $\left\{ \mathscr{D},\overline{\mathscr{D}}\right\} $
and an associated linear problem (from here on, we will omit the explicit
dependence on the complex variables unless necessary and denote $\overline{p}\equiv p(\overline{z})$)
\begin{equation}
\begin{cases}
\mathscr{D}\Psi=0 & ,\qquad\mathscr{D}\doteq\partial+\frac{1}{2}\partial\eta\,\sigma^{3}-e^{\tilde{\theta}}\left(\sigma^{+}e^{\eta}+\sigma^{-}pe^{-\eta}\right)\\
\overline{\mathscr{D}}\Psi=0 & ,\qquad\overline{\mathscr{D}}\doteq\overline{\partial}-\frac{1}{2}\overline{\partial}\eta\,\sigma^{3}-e^{-\tilde{\theta}}\left(\sigma^{-}e^{\eta}+\sigma^{+}\overline{p}\, e^{-\eta}\right)
\end{cases}\;,\label{eq:Linear_Problem}
\end{equation}
where $\Psi$ is a $2D$ vector function and $\left\{ \sigma^{3},\sigma^{\pm}\right\} $
are the usual Pauli matrices
\[
\sigma^{3}=\begin{pmatrix}1 & 0\\
0 & -1
\end{pmatrix}\;,\quad\sigma^{+}=\begin{pmatrix}0 & 1\\
0 & 0
\end{pmatrix}\;,\quad\sigma^{-}=\begin{pmatrix}0 & 0\\
1 & 0
\end{pmatrix}\;.
\]
The parameter $\tilde{\theta}=\log\tilde{\lambda}$ is called spectral
parameter, just as $\theta=\log\lambda$; even though the two will
turn out to be one and the same thing, we prefer to denote them differently
to stress the difference in their origin.

\paragraph{Analysis of the linear problem around $(z,\bar{z})=0$}

Clearly the general solution to linear problem are not known, however
the knowledge of the asymptotic behaviour of the function $\eta$
allow us to perform a local analysis at the singular points which,
as easily verified, are the same as those of MshG equation. Let us
start by analysing the behaviour of solutions around zero. Notice
how the linear problem (\ref{eq:Linear_Problem}) is not invariant
under the discrete symmetry (\ref{eq:Discrete_symmetry}); instead
it is invariant under a joint rotation of $\phi$ and $\tilde{\theta}$%
\footnote{The rotation in $\tilde{\theta}$ is actually an hyperbolic rotation%
}:
\[
\hat{\Omega}\;:\qquad(\phi,\tilde{\theta})\longrightarrow(\phi+\frac{\pi}{\alpha},\tilde{\theta}-\mathsf{i}\frac{\pi}{\alpha})\;.
\]
This property suggests that a solution $\Psi$ of the linear problem
should be represented, in the neighborhood of zero%
\footnote{This argument works around zero, since this is a Fuchsian singularity
of the linear problem. The same reasoning does not hold around $\infty$,
since the presence of the potential $p$ causes the \textit{Stokes
phenomenon} to arise. We will return on this later, but, in short
words, this means that solutions to (\ref{eq:Linear_Problem}) can
be represented correctly as asymptotic expansions only inside wedges
of the complex plane; as one tries to move outside these, the control
on the asymptotic behavior gets lost.%
}, by a function of $\mathsf{i}\phi+\tilde{\theta}$. Another easily
verified symmetry is the parity
\[
\hat{\Pi}\;:\qquad\begin{array}{l}
\tilde{\theta}\longrightarrow\tilde{\theta}-\mathsf{i}\pi\\
\mathscr{D}\longrightarrow\sigma^{3}\mathscr{D}\sigma^{3}\\
\overline{\mathscr{D}}\longrightarrow\sigma^{3}\overline{\mathscr{D}}\sigma^{3}
\end{array}\;,
\]
under which the solution should behave as
\[
\hat{\Pi}\;:\qquad\Psi\longrightarrow e^{\mathsf{i}\pi\tau}\sigma^{3}\Psi\;,\;\textrm{for some}\;\tau\in\mathbb{R}\;.
\]
With these facts in mind, we define two particular solutions $\Psi_{\pm}(\rho,\phi\vert\tilde{\theta})$
to the linear problem specifying their asymptotic behaviour (note
that we assume $\left|l\right|<\frac{1}{2}$)
\[
\Psi_{+}\underset{\rho\rightarrow0}{\sim}\frac{1}{\sqrt{\cos\left(\pi l\right)}}\begin{pmatrix}0\\
e^{(\mathsf{i}\phi+\tilde{\theta})l}
\end{pmatrix}\;,\qquad\Psi_{-}\underset{\rho\rightarrow0}{\sim}\frac{1}{\sqrt{\cos\left(\pi l\right)}}\begin{pmatrix}e^{-(\mathsf{i}\phi+\tilde{\theta})l}\\
0
\end{pmatrix}\;.
\]
Here follows a list of easily proven properties of these two solutions
\begin{itemize}
\item Analyticity: the solutions $\Psi_{\pm}(\rho,\phi\vert\tilde{\theta})$
are entire functions of $\tilde{\theta}$ for any $\phi\in[-\frac{\pi}{2\alpha},\frac{\pi}{2\alpha})$
and any $\rho>0$,
\item $\hat{\Omega}$-invariance
\[
\Psi_{\pm}\left(\rho,\phi+\frac{\pi}{2\alpha}\left\vert\tilde{\theta}-\mathsf{i}\frac{\pi}{2\alpha}\right.\right)=\Psi_{\pm}\left(\rho,\phi-\frac{\pi}{2\alpha}\left\vert\tilde{\theta}+\mathsf{i}\frac{\pi}{2\alpha}\right.\right)\;,
\]

\item $\hat{\Pi}$-transformation
\[
\Psi_{+}(\rho,\phi\vert\tilde{\theta}\pm\mathsf{i}\pi)=-e^{\pm\pi\mathsf{i}l}\sigma^{3}\Psi_{+}(\rho,\phi\vert\tilde{\theta})\;,
\]
\[
\Psi_{-}(\rho,\phi\vert\tilde{\theta}\pm\mathsf{i}\pi)=e^{\mp\pi\mathsf{i}l}\sigma^{3}\Psi_{-}(\rho,\phi\vert\tilde{\theta})\;,
\]

\item Linear independence
\[
\det\left|\left(\Psi_{+},\Psi_{-}\right)\right|=-\frac{1}{\cos\left(\pi l\right)}\;,
\]
where $\left(\Psi_{+},\Psi_{-}\right)$ is the matrix with $\Psi_{+}$
and $\Psi_{-}$ as columns. This determinant joins the role of Wronskian
of solutions,
\item Complex-conjugation:
\[
\begin{array}{l}
\Psi_{\pm}^{\ast}(\rho,\phi\vert\tilde{\theta})=\sigma^{1}\Psi_{\mp}(\rho,\phi\vert-\tilde{\theta})\\
\\
\Psi_{\pm}^{\ast}(\rho,0\vert\tilde{\theta})=\Psi_{\pm}(\rho,0\vert\tilde{\theta})
\end{array}\;,\qquad\forall\tilde{\theta}\in\mathbb{R}\;.
\]

\end{itemize}

\paragraph{Large $\rho$ analysis of the linear problem}

Now we move to the study of large-$\rho$ asymptotic of solutions
to the linear problem. For this task we can employ the WKB method
\cite{Bend_Orsz} and see that there exists a solution $\Xi_{-}(\rho,\phi\vert\tilde{\theta})$
which is uniquely specified by its asymptotic behaviour
\begin{equation}
\Xi_{-}\underset{\rho\rightarrow\infty}{\sim}\begin{pmatrix}e^{-\mathsf{i}\phi\frac{\alpha}{2}}\\
-e^{\mathsf{i}\phi\frac{\alpha}{2}}
\end{pmatrix}\exp\left[-2\frac{\rho^{\alpha+1}}{\alpha+1}\cosh\left(\tilde{\theta}+\mathsf{i}(\alpha+1)\phi\right)\right]\;,\quad\left|\phi\right|<\frac{\pi}{2(\alpha+1)}\;.\label{eq:Xi_funct}
\end{equation}
Note that this behaviour is limited to the wedge $\left|\phi\right|\leq\frac{\pi}{2(\alpha+1)}$;
when crossing one of the lines $\phi=\pm\frac{\pi}{2(\alpha+1)}$,
the solution moves from a decaying behaviour to a growing one. This
particular fact is called \textit{Stokes phenomenon} and the lines
$\phi=\pm\frac{\pi}{2(\alpha+1)}$ are known as \textit{Stokes lines}.
The reason why we have to limit the asymptotic behaviour to a given
sector of the complex plane is that only there we have full control
over the decaying solution. In fact, when rotating this last across
a Stokes line, exponentially decaying factors might appear in the
expansion without us noticing; these will then, when crossing another
Stokes line, become exponentially growing, depriving us completely
of the control on the asymptotic expansion. We will not delve further
in this fascinating subject and direct the interested reader to the
book \cite{Cost}, which present this subject in a modern way, underscoring
its connections with the theory of transseries, Borel summability
and resurgent functions.

We have seen that a basis of solutions of our linear problem is given
by $\left\{ \Psi_{\pm}\right\} $, which are entire functions of $\theta$.
This immediately means that $\Xi_{-}$ as well is entire in $\theta$
and we can thus perform analytic continuation in it. In particular
we can exploit the existence of the symmetry $\hat{\Omega}$, which
allows us to generate a new solution starting from $\Xi_{-}$:
\[
\Xi_{+}(\rho,\phi\vert\tilde{\theta})\doteq\hat{\Omega}\left[\Xi_{-}(\rho,\phi\vert\tilde{\theta})\right]\equiv\Xi_{-}\left(\rho,\phi+\frac{\pi}{\alpha}\left\vert\tilde{\theta}-\mathsf{i}\frac{\pi}{\alpha}\right.\right)\;.
\]
One immediately obtain the asymptotic expansion of this function
\[
\Xi_{+}\underset{\rho\rightarrow\infty}{\sim}-\mathsf{i}\begin{pmatrix}e^{-\mathsf{i}\phi\frac{\alpha}{2}}\\
e^{\mathsf{i}\phi\frac{\alpha}{2}}
\end{pmatrix}\exp\left[2\frac{\rho^{\alpha+1}}{\alpha+1}\cosh\left(\tilde{\theta}+\mathsf{i}(\alpha+1)\phi\right)\right]\;,\qquad\left|\phi\right|<\frac{\pi}{2(\alpha+1)}\;,
\]
which allows us to compute the determinant
\[
\det\left|\left(\Xi_{-},\Xi_{+}\right)\right|=-2\mathsf{i}\;.
\]
Thus $\left\{ \Xi_{\pm}\right\} $ is another basis of the space of
solutions to the linear problem (\ref{eq:Linear_Problem}). This,
however, is not the end as we can in fact repeatedly apply the transformation
$\hat{\Omega}$ on $\Xi_{-}$ and generate an infinite set of solutions
\[
\Xi_{n}(\rho,\phi\vert\tilde{\theta})\doteq\hat{\Omega}^{n}\left[\Xi_{-}(\rho,\phi\vert\tilde{\theta})\right]\equiv\Xi_{-}\left(\rho,\phi+\pi\frac{n}{\alpha}\left\vert\tilde{\theta}-\mathsf{i}\pi\frac{n}{\alpha}\right.\right)\;.
\]
Note that these solutions can be interpreted as decaying solutions
in the wedge $-\frac{\pi}{2}\frac{1+2n}{\alpha-1}<\phi<\frac{\pi}{2}\frac{1-2n}{\alpha-1}$

\paragraph{Spectral determinants}

Since the two solutions $\Psi_{\pm}$ are linearly independent, they
form a basis of the space of solutions of the linear problem (\ref{eq:Linear_Problem}),
meaning that we can expand the solution $\Xi_{-}$ as
\[
\Xi_{-}(\rho,\phi\vert\tilde{\theta})=Q_{-}(\tilde{\theta})\Psi_{+}(\rho,\phi\vert\tilde{\theta})+Q_{+}(\tilde{\theta})\Psi_{-}(\rho,\phi\vert\tilde{\theta})\;,
\]
where the \textit{connection coefficients}
\[
Q_{\pm}\equiv\pm\cos\left(\pi l\right)\det\left|\left(\Xi_{-},\Psi_{\pm}\right)\right|\;,
\]
are functions of $\tilde{\theta}$ and $l$ only. They are also known
as \textit{spectral determinants} for the central problem of (\ref{eq:Linear_Problem}).
This last denomination simply means that the zeroes of these functions
are precisely the eigenvalues of the linear problem considered on
functions in $L^{2}(0,\infty)$: those values of the spectral parameters
for which the function $\Psi_{\pm}$ decays at infinity. The notation
chosen for these functions is not random: in the following sub-section
we will see how these functions can be interpreted as the ground state
eigenvalues of quantum sine-Gordon $Q$-functions. Similarly we can
exploit the existence of the basis $\left\{ \Xi_{\pm}\right\} $ and
expand the solutions $\Xi_{n}$ defined above as linear combinations:
\[
\Xi_{n}(\rho,\phi\vert\tilde{\theta})=-T_{\frac{n-2}{2}}\left(\tilde{\theta}-\mathsf{i}\pi\frac{n+1}{2\alpha}\right)\Xi_{-}(\rho,\phi\vert\tilde{\theta})+T_{\frac{n-1}{2}}\left(\tilde{\theta}-\mathsf{i}\pi\frac{n}{2\alpha}\right)\Xi_{+}(\rho,\phi\vert\tilde{\theta})\;.
\]
Here too the coefficients
\begin{align}
T_{n}\left(\tilde{\theta}-\mathsf{i}\pi\frac{2n+1}{2\alpha}\right)= & \frac{1}{2\mathsf{i}}\det\left|\left(\Xi_{2n+1},\Xi_{-}\right)\right|\label{eq:T-spect_det}\\
T_{n+\frac{1}{2}}\left(\tilde{\theta}-\mathsf{i}\pi\frac{2n+2}{2\alpha}\right)= & \frac{1}{2\mathsf{i}}\det\left|\left(\Xi_{2n+1},\Xi_{+}\right)\right|\;,\nonumber 
\end{align}
are spectral determinants, this time for the lateral problems of of
our linear system. These problems consist in determining solutions
to (\ref{eq:Linear_Problem}) decaying in both wedges
\begin{align*}
 & -\frac{\pi}{2}\frac{1}{\alpha-1}<\phi<\frac{\pi}{2}\frac{1}{\alpha-1}\;\textrm{and}\;-\frac{\pi}{2}\frac{4n+3}{\alpha-1}<\phi<-\frac{\pi}{2}\frac{4n+1}{\alpha-1}\;,\quad\textrm{for}\, n\in\mathbb{Z}\;,\\
 & -\frac{\pi}{2}\frac{3}{\alpha-1}<\phi<\frac{\pi}{2}\frac{-1}{\alpha-1}\;\textrm{and}\;-\frac{\pi}{2}\frac{4n+1}{\alpha-1}<\phi<-\frac{\pi}{2}\frac{4n-1}{\alpha-1}\;,\quad\textrm{for}\, n\in\mathbb{Z}+\frac{1}{2}\mathbb{Z}\;,
\end{align*}
and its particular eigenvalues corresponds to the zeroes of the functions
$T_{j}$. Again, the choice of notation for these spectral determinants
hints at the fact that they can be interpreted as eigenvalues of the
operators $\mathbb{T}_{j}$, as it will be shown later.

\subsection{From spectral determinants to the quantum $Q$-operators}

Before beginning to unveil the correspondence between the linear problem
(\ref{eq:Linear_Problem}) and the quantum world, let us list the
properties of the spectral determinants $Q_{\pm}$ introduced just
above
\begin{itemize}
\item Analyticity: $Q_{\pm}(\tilde{\theta})$ are entire functions of $\tilde{\theta}$,
note also that the functions are defined for $l=\pm\frac{1}{2}$ by
continuity
\[
\lim_{l\rightarrow\pm\frac{1}{2}}\left(Q_{+}(\tilde{\theta})-Q_{-}(\tilde{\theta})\right)=0
\]

\item Quasi-periodicity:
\[
Q_{\pm}\left(\tilde{\theta}+\mathsf{i}\pi\frac{\alpha+1}{2\alpha}\right)=e^{\pm\mathsf{i}\pi(l+\frac{1}{2})}Q_{\pm}\left(\tilde{\theta}-\mathsf{i}\pi\frac{\alpha+1}{2\alpha}\right)\;,
\]

\item Complex conjugation:
\[
Q_{\pm}^{\ast}(\tilde{\theta})=Q_{\pm}(\tilde{\theta}^{\ast})\;,\qquad\forall\tilde{\theta}\in\mathbb{R}\;,
\]

\item Parity symmetry:
\[
Q_{\pm}(\tilde{\theta})=Q_{\mp}(-\tilde{\theta})\;,\qquad\forall\tilde{\theta}\in\mathbb{R}\;,
\]

\item Quantum Wronskian:
\begin{equation}
Q_{+}\left(\tilde{\theta}+\mathsf{i}\frac{\pi}{2\alpha}\right)Q_{-}\left(\tilde{\theta}-\mathsf{i}\frac{\pi}{2\alpha}\right)-Q_{-}\left(\tilde{\theta}+\mathsf{i}\frac{\pi}{2\alpha}\right)Q_{+}\left(\tilde{\theta}-\mathsf{i}\frac{\pi}{2\alpha}\right)=-2\mathsf{i}\,\cos\left(\pi l\right)\;.\label{eq:Quantum_Wronskian-1}
\end{equation}

\end{itemize}
In order to proceed, it is convenient to define the following single
function of two variables
\[
Q(\tilde{\theta},\tilde{k})\doteq\begin{cases}
Q_{+}(\tilde{\theta})\Big\vert_{l=2\tilde{k}-\frac{1}{2}} & 0<\tilde{k}<\frac{1}{2}\\
Q_{-}(\tilde{\theta})\Big\vert_{l=-2\tilde{k}-\frac{1}{2}} & -\frac{1}{2}<\tilde{k}<0
\end{cases}\;,
\]
where $\tilde{k}=0$ is treated by continuity. Since, obviously, $Q(\tilde{\theta},\tilde{k})=Q(\tilde{\theta},\tilde{k}+1)$,
this function admits an analytic extension to all $\tilde{k}\in\mathbb{R}$.
The properties above become in term of this function
\[
Q\left(\tilde{\theta}+\mathsf{i}\pi\frac{\alpha+1}{\alpha},\tilde{k}\right)=e^{2\mathsf{i}\pi k}Q(\tilde{\theta},\tilde{k})\;,
\]
\[
Q^{\ast}(\tilde{\theta},\tilde{k})=Q(-\tilde{\theta},\tilde{k})\;,\qquad Q(-\tilde{\theta},\tilde{k})=Q(\tilde{\theta},-\tilde{k})\;,
\]
\[
Q\left(\tilde{\theta}+\mathsf{i}\frac{\pi}{2\alpha},\tilde{k}\right)Q\left(\tilde{\theta}-\mathsf{i}\frac{\pi}{2\alpha},-\tilde{k}\right)-Q\left(\tilde{\theta}-\mathsf{i}\frac{\pi}{2\alpha},\tilde{k}\right)Q\left(\tilde{\theta}+\mathsf{i}\frac{\pi}{2\alpha},-\tilde{k}\right)=-2i\sin\left(2\pi\tilde{k}\right)\;.
\]

The similarity of the above properties with the one listed at the
beginning of this section for the operators $\mathbb{Q}_{\pm}$ are
striking, however, in order to univocally fix the function $Q$ we
still need to determine its asymptotic behaviour and the distribution
of its zeroes. Thanks to the property of periodicity we can concentrate
on the strip $\tilde{H}\doteq\tilde{H}_{+}\cup\tilde{H}_{-}$ where
$\tilde{H}_{\pm}\doteq\left\{ \tilde{\theta}\in\mathbb{C}\;\Big\backslash\;0<\pm\Im(\tilde{\theta})<\pi\frac{\alpha+1}{\alpha}\right\} $.
With a careful and thorough WKB analysis of the solutions of the linear
problem, it is possible to establish the following behaviours
\[
Q\underset{\Re(\tilde{\theta})\rightarrow\infty}{\sim}e^{\pm\mathsf{i}\pi\tilde{k}}\mathscr{S}^{\frac{1}{2}}(\tilde{k})\exp\left[r\frac{e^{\tilde{\theta}\mp\mathsf{i}\pi\frac{\alpha+1}{2\alpha}}}{4\cos\left(\frac{\pi}{2\alpha}\right)}\right]\;,\qquad\tilde{\theta}\in H_{\pm}\;,
\]
\[
Q\underset{\Re(\tilde{\theta})\rightarrow-\infty}{\sim}e^{\pm\mathsf{i}\pi\tilde{k}}\mathscr{S}^{-\frac{1}{2}}(\tilde{k})\exp\left[r\frac{e^{-\tilde{\theta}\pm\mathsf{i}\pi\frac{\alpha+1}{2\alpha}}}{4\cos\left(\frac{\pi}{2\alpha}\right)}\right]\;,\qquad\tilde{\theta}\in H_{\pm}\;,
\]
where we introduced $r=B\, s^{1+\alpha}$ with $B=2\sqrt{\pi}\frac{\Gamma\left(1+\frac{1}{2\alpha}\right)}{\Gamma\left(\frac{3}{2}+\frac{1}{2\alpha}\right)}$
and the function
\[
\mathscr{S}(\tilde{k})\doteq\frac{\Gamma\left(2\tilde{k}\right)}{\Gamma\left(1-2\tilde{k}\right)}2^{4\tilde{k}-1}e^{\eta_{0}}\;,\qquad0\leq\tilde{k}\leq\frac{1}{2}\;,
\]
which enjoys the symmetries
\[
\mathscr{S}(\tilde{k})\mathscr{S}(-\tilde{k})=1\;,\qquad\mathscr{S}(\tilde{k}+1)=\mathscr{S}(\tilde{k})\;.
\]

Now that we verified that the asymptotic of the function $Q$ has
exactly the same form of the operator $\mathbb{Q}_{+}$ (for $\tilde{k}>0$),
the fact that we are actually dealing with an eigenvalue of this last
in disguise is becoming more than a simple suspect. What is left to
do is to identify which is this specific eigenvalue by studying the
pattern of the zeroes of $Q(\tilde{\theta})$: this is done again
by thorough WKB analysis (remember that they are the eigenvalues of
a central problem for the system (\ref{eq:Linear_Problem})). As it
turns out, for any $k\in\mathbb{R}$, these zeroes are real, simple,
symmetrically disposed with respect to the origin and accumulating
at the singularities $\tilde{\theta}\rightarrow\pm\infty$, which
identifies the spectral determinants $Q(\tilde{\theta},\pm\tilde{k})$
with the eigenvalues $Q_{\pm}^{(\textrm{vac})}(\theta)$ of the operator
$\mathbb{Q}_{\pm}(\theta)$ on the vacuum state with quasi-momentum
$k$. The parameters on the two sides of this correspondence have
to be identified as
\begin{eqnarray}
 &  & \alpha=\frac{1}{\xi}=\frac{1}{\beta^{2}}-1\;,\qquad\tilde{k}=k\;,\label{eq:parameter1}\\
 &  & r=\mathfrak{M}R\;\Longrightarrow s=\left(\frac{R}{\pi\beta^{2}}\right)^{\beta^{2}}\left[\mu\pi\frac{\Gamma\left(1-\beta^{2}\right)}{\Gamma\left(\beta^{2}\right)}\right]^{\frac{\beta^{2}}{2-2\beta^{2}}}\;,\label{eq:parameter2}
\end{eqnarray}
where $\mathfrak{M}$ is the soliton mass of quantum sine-Gordon (\ref{eq:soliton_mass})
and $\mu$ is the parameter appearing in the Lagrangian (\ref{eq:sG_Lagrangian}).

\begin{framed}%

\paragraph*{The NLIE equation}

Just as we did in the previous section, we wish to use the analytic
properties of the spectral determinants/$Q$-functions to derive a
NLIE equation, as a further verification of the identification we
performed. From now on we will drop the tildas on $\theta$ and $k$
and use equivalently $\alpha$, $\xi$ or $\beta$, depending on notational
reasons. Consider the following function
\[
\varepsilon(\theta)\doteq\mathsf{i}\,\log\left[\frac{Q(\theta+\mathsf{i}\pi\xi,k)}{Q(\theta-\mathsf{i}\pi\xi,k)}\right]\;,
\]
with the branch of the logarithm fixed so that
\[
\varepsilon(\theta)-r\frac{e^{\theta}}{2\cos\left(\pi\frac{\xi}{2}\right)}\underset{\Re(\theta)\rightarrow\infty}{\sim}-2\pi k\;,\qquad\left|\Im(\theta)\right|<\frac{\pi}{2}\;.
\]
Thanks to this function we can label univocally the zeroes of $Q$
by integers $n\in\mathbb{Z}$ in such a way that
\[
\theta_{n}<\theta_{n+1}\;,\qquad\varepsilon(\theta_{n})=\pi(2n+1)\;,
\]
where the last relation descends from the Quantum Wronskian relation.
The zeroes $\left\{ \theta_{n}\right\} _{n\in\mathbb{Z}}$ are more
conveniently represented in the following form
\[
e^{2\theta_{n}\frac{\alpha}{\alpha+1}}=\begin{cases}
s^{-2\alpha}E_{n}(k) & ,\; n\geq0\\
s^{2\alpha}E_{-n-1}^{-1}(-k) & ,\; n<0
\end{cases}\;,\quad0\leq k\leq\frac{1}{2}\;,
\]
which makes more explicit the symmetry $n\rightarrow-n-1$. The ``zeroes''
$\left\{ E_{n}\right\} _{n=0}^{\infty}$ are functions of $k$ and
satisfy the following relations
\[
E_{n}(k+1)=E_{n+1}(k)\;,\qquad E_{0}(-k-1)E_{0}(k)=s^{4\alpha}\;,
\]
and have the following asymptotic behaviour
\[
E_{n}(\pm k)\underset{n\rightarrow\infty}{\sim}\left[\frac{2\pi}{B}\left(2n\pm2k+1\right)\right]^{2\frac{\alpha}{\alpha+1}}\;,
\]
which, again, can be obtained by thorough WKB analysis of the linear
problem. We have now all the informations needed to express the function
$Q$ as a Hadamard product
\[
Q(\theta,k)=\mathscr{C}(k)e^{2k\theta\frac{\alpha}{\alpha+1}}\prod_{n=0}^{\infty}\left(1-s^{2\alpha}\frac{e^{2\theta\frac{\alpha}{\alpha+1}}}{E_{n}(k)}\right)\left(1-s^{2\alpha}\frac{e^{-2\theta\frac{\alpha}{\alpha+1}}}{E_{n}(-k)}\right)\;,
\]
which converges only for $\alpha>1$; if one wishes to extend the
above product to the region $0<\alpha\leq1$, then a Weierstrass prime
multiplier has to be added in order to regulate the divergency of
the product \cite{Conw}. The normalisation in front of the product
satisfies the following relations
\[
\mathscr{C}(k)=\mathscr{C}(-k)\;,\qquad\mathscr{C}(k)=-s^{-2\alpha}E_{0}(k)\mathscr{C}(k+1)\;.
\]

We can finally write down the NLIE equation for the function $\varepsilon$
by combining the Hadamard representation of the $Q$, its analytic
properties, its asymptotic behaviour and the equation $\varepsilon(\theta_{n})=\pi(2n+1)$,
obtaining
\[
\varepsilon(\theta)=-2\pi k+r\sinh\theta-2\intop_{-\infty}^{\infty}d\theta'G(\theta-\theta')\Im\left[\log\left(1+e^{-\mathsf{i}\varepsilon(\theta'-i0)}\right)\right]\;,
\]
where we introduced the kernel
\[
G(\theta)\doteq\intop_{-\infty}^{\infty}\frac{d\nu}{2\pi}\frac{\sinh\left(\pi\nu\frac{1-\alpha}{2\alpha}\right)}{2\cosh\left(\pi\frac{\nu}{2}\right)\sinh\left(\pi\frac{\nu}{2\alpha}\right)}e^{\mathsf{i}\theta\nu}\;.
\]
As was expected, this equation coincides exactly with the NLIE equation
for the ground state of quantum sine-Gordon model \cite{Dest_deVe_95},
given the identifications (\ref{eq:parameter1}-\ref{eq:parameter2})
are made. Once $\varepsilon$ is known, one can then recover the function
$Q$ from the following formula
\begin{eqnarray}
 &  & \log\left(Q(\theta+\mathsf{i}\pi\frac{\alpha+1}{2\alpha},k)\right)=\frac{r}{2}\frac{\cosh\theta}{\cos\frac{\pi}{2\alpha}}+\mathsf{i}\pi k+\frac{1}{2}\log\left(\mathscr{S}(k)\right)+\nonumber \\
 &  & \qquad\qquad\qquad+2\mathsf{i}\intop_{-\infty}^{\infty}d\theta'\Im\left[F(\theta-\theta'-\mathsf{i}0)\log\left(1+e^{-\mathsf{i}\varepsilon(\theta'-i0)}\right)\right]\;,\label{eq:Q_from_DDV}
\end{eqnarray}
where the following kernel was introduced
\[
F(\theta)\doteq\intop_{-\infty}^{\infty}\frac{d\nu}{2\pi}\frac{e^{\mathsf{i}\nu\theta}}{4\cosh\left(\pi\frac{\nu}{2}\right)\sinh\left(\pi\frac{\nu-\mathsf{i}0}{2\alpha}\right)}\;.
\]
The formula (\ref{eq:Q_from_DDV}) is actually valid for $\Im(\theta)=0$
only; however it is possible to suitably modify it so that it provides
$Q$ in the whole strip $H_{+}$. The function $\mathscr{S}(k)$ can
also be recovered from the solution to the NLIE equation
\[
\log\left(\mathscr{S}(k)\right)=\alpha\intop_{-\infty}^{\infty}\frac{d\theta}{\pi}\Im\left[\log\left(1+e^{-\mathsf{i}\varepsilon(\theta-i0)}\right)\right]\;.
\]

Before closing this in-depth box we ``close the circle'' by recovering
the integrals of motion from the asymptotic expansion of the function
$Q$. For this task the equation (\ref{eq:Q_from_DDV}) is perfect,
since it allows strightforward evaluation of the large-$\left|\theta\right|$
expansion. This most simply involves writing $F(\theta)$ as an sum
over his residues in the correct half plane and plugging it in the
formula for the function $Q$. The result is
\begin{eqnarray*}
 &  & \log\left(Q(\theta+\mathsf{i}\pi\frac{\alpha+1}{2\alpha},k)\right)\underset{\Re(\theta)\rightarrow\pm\infty}{\sim}\frac{re^{\pm\theta}}{4\cos\left(\frac{\pi}{2\alpha}\right)}+\mathsf{i}\pi k+\frac{1}{2}\log\left(\mathscr{S}(k)\right)+\\
 &  & \qquad\qquad\qquad\qquad\qquad\qquad-\sum_{n=1}^{\infty}\left[\mathscr{I}_{\pm(2n-1)}e^{\mp(2n-1)\theta}-\mathscr{G}_{\pm2n}e^{\mp2n\alpha\theta}\right]\;,
\end{eqnarray*}
where we defined the following objects
\[
\mathscr{I}_{\pm(2n-1)}\doteq-\frac{r}{4\cos\left(\frac{\pi}{2\alpha}\right)}\delta_{n,1}\pm\frac{(-1)^{n+1}}{\sin\left(\pi\frac{2n-1}{\alpha}\right)}\intop_{-\infty}^{\infty}\frac{d\theta}{\pi}\Im\left[e^{\pm(2n-1)(\theta-\mathsf{i}0)}\log\left(1+e^{-\mathsf{i}\varepsilon(\theta-\mathsf{i}0)}\right)\right]\;,
\]
\[
\mathscr{G}_{\pm2n}\doteq\pm\frac{\alpha(-1)^{2n}}{\cos\left(2\pi\alpha n\right)}\intop_{-\infty}^{\infty}\frac{d\theta}{\pi}\Im\left[e^{\pm4\alpha n(\theta-\mathsf{i}0)}\log\left(1+e^{-\mathsf{i}\varepsilon(\theta-\mathsf{i}0)}\right)\right]\;.
\]
which are related to the local and non-local integrals of motion,
e.g.
\[
\begin{array}{c}
\mathscr{I}_{2n-1}=\mathfrak{C}_{n}I_{2n-1}(k)\\
\\
\mathscr{I}_{1-2n}=\mathfrak{C}_{n}\overline{I}_{2n-1}(k)
\end{array}\;,\quad\forall n>0\;,\qquad\begin{array}{c}
\mathbb{I}_{2n-1}\left|\,\Phi_{k}^{(\textrm{vac})}\,\right\rangle =I_{2n-1}\left|\,\Phi_{k}^{(\textrm{vac})}\,\right\rangle \\
\\
\overline{\mathbb{I}}_{2n-1}\left|\,\Phi_{k}^{(\textrm{vac})}\,\right\rangle =\overline{I}_{2n-1}\left|\,\Phi_{k}^{(\textrm{vac})}\,\right\rangle 
\end{array}\;,
\]
with
\[
\mathfrak{C}_{n}\doteq\left(-\frac{\alpha^{2}}{\alpha+1}\right)^{n-1}\frac{\Gamma\left(-\frac{2n-1}{2\alpha}\right)\Gamma\left((2n-1)\frac{\alpha+1}{2\alpha}\right)}{2\sqrt{\pi}n!}\left[\frac{2\mathfrak{M}\sin\left(\frac{\pi}{2\alpha}\right)}{8\sqrt{\pi}}\Gamma\left(\frac{\alpha+1}{2\alpha}\right)\Gamma\left(-\frac{1}{2\alpha}\right)\right]^{1-2n}\;.
\]
Similar formulae exist for the non-linear integrals of motion.\end{framed}

\subsection{The $T$-functions}

We wish to conclude this review by pointing out the connection of
the spectral determinants (\ref{eq:T-spect_det}) to the $T$-functions
of Quantum sine-Gordon. In order to bring it into light, we can expand
the $\Xi$ functions entering the definition of $T_{j}$ in terms
of the basis $\Psi_{\pm}$, remembering that $\hat{\Omega}\left[\Psi_{\pm}\right]=\Psi_{\pm}$
and then use the quantum Wronskian relation (\ref{eq:Quantum_Wronskian-1}).
The result is, as expected
\begin{eqnarray}
 &  & T_{j}(\theta)=\frac{\mathsf{i}}{2\cos\left(\pi l\right)}\left[Q_{+}\left(\theta+\mathsf{i}\pi\frac{2j+1}{2\alpha}\right)Q_{-}\left(\theta-\mathsf{i}\pi\frac{2j+1}{2\alpha}\right)\right.+\nonumber \\
 &  & \qquad\quad-\left.Q_{+}\left(\theta-\mathsf{i}\pi\frac{2j+1}{2\alpha}\right)Q_{-}\left(\theta+\mathsf{i}\pi\frac{2j+1}{2\alpha}\right)\right]\;,\label{eq:T_ODE_Wronskian}
\end{eqnarray}
\[
T_{-\frac{1}{2}}(\theta)=0\;,\qquad T_{0}(\theta)=1\;,
\]
which is precisely the Wronskian expression for $T$-functions of
the quantum sine-Gordon model. We easily see that this relation directly
implies the validity of the $T$-system
\[
T_{\frac{1}{2}}(\theta)T_{j}\left(\theta+\mathsf{i}\pi\frac{2j+1}{2\alpha}\right)=T_{j-\frac{1}{2}}\left(\theta+\mathsf{i}\pi\frac{2j+2}{2\alpha}\right)+T_{j+\frac{1}{2}}\left(\theta+\mathsf{i}\pi\frac{2j}{2\alpha}\right)\;.
\]
To perform the precise identification between a spectral determinant
$T$ and the $k$-vacuum eigenvalue of the operator $\mathbb{T}_{j}$,
one has to analyse the analytic properties of the first and compare
them with the last's ones. At this point it is not a surprise anymore
to find out that these properties match exactly, allowing us a perfect
identification of the results given in this section with those presented
in Sec.s \ref{sec:Integrable-conformal} and \ref{sec:Integrable-massive}.
We will not present these calculations here, but encourage the interested
reader to go through them.

\section*{Acknowledgements}

It is for me a great pleasure to thank the GATIS network and the organisers
of YRIS summer school for giving me the opportunity of participating
there as a lecturer. A particular thanks goes to A. Cagnazzo, R. Frassek,
A. Sfondrini, I.M. Szécsényi and S. van Tongeren for their work as
editors of the special issue this review is part of. I would like
also to thank the other lecturer at YRIS school, D. Bombardelli, F.
Levkovich-Maslyuk, F. Loebbert, A. Torrielli and S. van Tongeren as
well as A. Cavagliá, R. Tateo and P. Dorey for useful discussions
on the topics addressed here and I.M. Szécsényi for helpful comments
on the manuscript. I also wish to thank Durham University and, in
particular, its Department of Mathematical Sciences, for providing
excellent hospitality. Finally I wish to thank all the students which
attended to the YRIS school, their interest in the lectures was comforting
and their questions and comments have been precious.

The work of the author was supported by the European Research Council
(Programme \textquotedblleft Ideas\textquotedblright{} ERC- 2012-AdG
320769 AdS-CFT-solvable).

\end{document}